\documentclass[NET,biber]{nowfnt} 

\usepackage[utf8]{inputenc}
\usepackage{graphicx}
\usepackage{amsmath}
\usepackage{amssymb}
\usepackage{mathtools}

\usepackage[nolist]{acronym}
\usepackage{MathDefinitions} 
\usepackage{tabularx}
\usepackage{array}
\usepackage{tikz}
\usepackage{latexsym}
\usepackage{multirow}
\usepackage{makecell}
\usepackage{subfig} 
\usepackage{booktabs}
\usepackage{amsthm}
\usepackage{comment}
\theoremstyle{definition}

\usepackage{ragged2e}
\newcolumntype{L}{>{\RaggedRight}X} 
\usepackage{enumitem}
\newlist{tabitemize}{itemize}{1}
\setlist[tabitemize,1]{label=\textbullet, left=0pt, nosep,
    before={\begin{minipage}[t]{1.35\linewidth}},
        after ={\end{minipage}}}

\title{Machine Learning for Spectrum Sharing: A Survey}

\maintitleauthorlist{
Francisco R. V. Guimar\~{a}es \\[-1pt]
Fed. Inst. of Education, Sci. and Tech. of Ceara (IFCE), Brazil \\[-1pt]
rafael.vasconcelos@ifce.edu.br
\and
Jos\'{e} Mairton B. da Silva Jr. \\[-1pt]
Uppsala University, Sweden \\[-1pt]
mairton.barros@it.uu.se
\and 
Charles Casimiro Cavalcante \\[-1pt]
Federal University of Ceara (UFC), Brazil \\[-1pt]
charles@ufc.br
\and
Gabor Fodor \\
KTH Royal Institute of Technology, Sweden\\[-1pt]
Ericsson Research, Sweden \\[-1pt]
gabor.fodor@ericsson.com
\and
Mats Bengtsson \\[-1pt]
KTH Royal Institute of Technology, Sweden \\[-1pt]
matben@kth.se
\and
Carlo Fischione \\[-1pt]
KTH Royal Institute of Technology, Sweden \\[-1pt]
carlofi@kth.se
}

\issuesetup
{%
 copyrightowner={F.R.V.~Guimar\~{a}es \emph{et al}},
 volume        = 14,
 issue         = 1-2,
 pubyear       = 2024,
 isbn          = 978-1-63828-430-7,
 eisbn         = 978-1-63828-431-4,
 doi           = 10.1561/1300000037,
 firstpage     = 1, 
 lastpage      = 159
 }

\usepackage{mwe}

\author[1]{Guimar\~{a}es, Francisco R. V.}
\author[2]{Silva Jr., Jos\'{e} Mairton B.}
\author[3]{Cavalcante, Charles Casimiro}
\author[4]{Fodor, Gabor}
\author[5]{Bengtsson, Mats}
\author[6]{Fischione, Carlo}

\affil[1]{IFCE, Brazil; rafael.vasconcelos@ifce.edu.br}
\affil[2]{Uppsala University, Sweden; mairton.barros@it.uu.se}
\affil[3]{UFC, Brazil; charles@ufc.br}
\affil[4]{KTH and Ericsson, Sweden; gabor.fodor@ericsson.com}
\affil[5]{KTH, Sweden; matben@kth.se}
\affil[6]{KTH, Sweden; carlofi@kth.se}

\articledatabox{\nowfntstandardcitation}

\newcommand{\tabitem}{~~\llap{\textbullet}~~}

\DeclareUnicodeCharacter{0301}{\'{e}}

\begin{acronym}
  \acro{2G}{Second Generation}
  \acro{3G}{3$^\text{rd}$~generation}
  \acro{3GPP}{3$^\text{rd}$~Generation Partnership Project}
  \acro{4G}{4$^\text{th}$~Generation}
  \acro{5G}{5$^\text{th}$~generation}
  \acro{6G}{6$^\text{th}$~generation}
  \acro{A2C}{advantage actor-critic}
  \acro{AE}{auto-encoder}
  \acro{AI}{artificial inteligence}
  \acro{ANN}{artificial neural network}
  \acro{AoA}{angle of arrival}
  \acro{AR}{augmented reality}
  \acro{AWGN}{additive white Gaussian noise}
  \acro{BER}{bit error rate}
  \acro{BF}{beamforming}
  \acro{BS}{base station}
  \acro{CAP}{Combinatorial Allocation Problem}
  \acro{CAPEX}{Capital Expenditure}
  \acro{CBF}{coordinated beamforming}
  \acro{CDF}{cumulative distribution function}
  \acro{CSI}{channel state information}
  \acro{CSIT}{channel state information at the transmitter}
  \acro{CNN}{convolutional neural network}
  \acro{CDI}{channel distribution information}
  \acro{CODIPAS}{combined fully distributed payoff and strategy learning}
  \acro{CR}{cognitive radio}
  \acro{CRN}{cognitive radio network}
  \acro{CSS}{cooperative spectrum sensing}
  \acro{D2D}{device-to-device}
  \acro{DBF}{Digital beamforming}
  \acro{DCA}{dynamic channel allocation}
  \acro{DDPG}{deep deterministic policy gradient}
  \acro{DDQN}{double deep Q network}
  \acro{DFT}{Discrete Fourier Transform}
  \acro{DL}{deep learning}
  \acro{DPGMM}{Dirichlet process GMM}
  \acro{D2DM}{D2D Mode}
  \acro{DMS}{D2D Mode Selection}
  \acro{DNN}{deep neural network}
  \acro{DPC}{Dirty Paper Coding}
  \acro{DQN}{deep Q network}
  \acro{DRA}{Dynamic Resource Assignment}
  \acro{DRL}{deep reinforcement learning}
  \acro{DSA}{dynamic spectrum access}
  \acro{DSS}{dynamic spectrum sharing}
  \acro{DTDD}{dynamic time division duplexing}
  \acro{EE}{energy efficiency}
  \acro{eMBB}{enhanced mobile broadband}
  \acro{eNB}{evolved Node B}
  \acro{EM}{expectation maximization}
  \acro{FDA}{Fisher discriminant analysis}
  \acro{FDD}{frequency division duplex}
  \acro{GMM}{Gaussian mixture model}
  \acro{HBF}{hybrid beamforming}
  \acro{HD}{half duplex}
  \acro{HMM}{hidden Markov model}
  \acro{IBFD}{in-band full-duplex}
  \acro{IC}{interference channel}
  \acro{IDS}{intrusion detection system}
  \acro{IOPSS}{inter-operator proximal spectrum sharing }
  \acro{IoT}{Internet of Things}
  \acro{IMT}{International Mobile Telecommunication}
  \acro{IPNP}{interference plus noise power}
  \acro{ISM}{industrial, scientific and medical}
  \acro{ITU}{International Telecommunication Union}
  \acro{k-NN}[$k$-NN]{$k$-nearest neighbor}
  \acro{KPI}{key performance indicator}
  \acro{LLC}{local learning-based clustering}
  \acro{LoS}{line-of-sight}
  \acro{LTE}{long-term evolution}
  \acro{LTE-U}{long-term evolution on unlicensed spectrum}
  \acro{LRI}{linear reward inaction}
  \acro{LRMM}{log-Rayleigh mixture model}
  \acro{LSTM}{long short-term memory}
  \acro{MAB}{multi-armed bandit}
  \acro{MARL}{multi-agent reinforcement learning}
  \acro{MDP}{Markov decision process}
  \acro{METIS}{Mobile Enablers for the Twenty-Twenty Information Society}
  \acro{ML}{machine learning}
  \acro{MIMO}{multiple-input multiple-output}
  \acro{MIMO-IC}{MIMO-interference channel}
  \acro{MISO}{multiple-input single-output}
  \acro{MISO-IC}{MISO-interference channel}
  \acro{MM}{mixture model}
  \acro{mmWave}{millimeter-wave}
  \acro{MRC}{maximum ratio combining}
  \acro{MRT}{maximum ratio transmission}
  \acro{MS}{mode selection}
  \acro{MSE}{mean square error}
  \acro{MMSE}{minimum mean square error}
  \acro{MNO}{mobile network operator}
  \acro{MSVM}{multiclass SVM}
  \acro{MTC}{machine type communications}
  \acro{NLoS}{non-line-of-sight}
  \acro{NN}{neural network}
  \acro{NOMA}{non-orthogonal multiple access}
  \acro{NR}{New Radio}
  \acro{NSPS}{national security and public safety}
  \acro{NWC}{network coding}
  \acro{PC}{power control}
  \acro{PHY}{physical layer}
  \acro{PPO}{proximal policy optimization}
  \acro{PU}{primary user}
  \acro{PR}{primary receiver}
  \acro{PRB}{Physical Resource Block}
  \acro{PSO}{particle swarm optimization}
  \acro{QoE}{quality of experience}
  \acro{QoS}{quality of service}
  \acro{RAN}{radio access network}
  \acro{RA}{Resource Allocation}
  \acro{RAT}{radio access technology}
  \acro{RB}{resource block}
  \acro{ReLU}{rectified linear unit}
  \acro{RF}{random forest}
  \acro{RNN}{recurrent neural network}
  \acro{RL}{reinforcement learning}
  \acro{ROC}{receiver operating characteristics}
  \acro{RSS}{received signal strength}
  \acro{SAc}{spectrum access}
  \acro{SAl}{spectrum allocation}
  \acro{SAE}{stacked autoencoder}
  \acro{SARSA}{state–action–reward–state–action}
  \acro{SINR}{signal-to-interference-plus-noise ratio}
  \acro{SISO}{single-input single-output}
  \acro{SGD}{stochastic gradient descent}
  \acro{SH}{spectrum handoff}
  \acro{SOCP}{second order cone programming}
  \acro{SON}{self-optimizing network}
  \acro{SPBA}{sub-optimal pricing-based algorithm}
  \acro{SNR}{signal-to-noise ratio}
  \acro{SSe}{spectrum sensing}
  \acro{SU}{secondary user}
  \acro{SR}{secondary receiver}
  \acro{STC}{space-time coding}
  \acro{SVM}{support vector machine}
  \acro{SVR}{support vector regression}
  \acro{TDD}{time division duplexing}
  \acro{TTI}{Transmission Time Interval}
  \acro{UAV}{unmanned aerial vehicle}
  \acro{UE}{user equipment}
  \acro{UL}{uplink}
  \acro{ULPB}{uplink Power Boosting}
  \acro{URLLC}{ultra-reliable low-latency communication}
  \acro{VUE}{vehicular user equipment}
  \acro{V2I}{vehicle-to-infrastructure}
  \acro{V2V}{vehicle-to-vehicle}
  \acro{V2X}{vehicle-to-vehicle and vehicle-to-infrastructure}
  \acro{VR}{virtual reality}
  \acro{ZF}{zero-forcing}
\end{acronym}

\begin{document}

\makeabstracttitle

\begin{abstract}
The 5th generation (5G) of wireless systems is being deployed with the aim to provide many sets of wireless
communication services, such as low data rates for a massive amount of devices, broad band, low latency, and industrial wireless access. Such an aim is even more complex in the next generation wireless systems (6G) where wireless connectivity is expected to serve any connected intelligent unit, such as software robots and
humans interacting in the meta verse, autonomous vehicles, drones, trains, or smart sensors monitoring cities, buildings and the environment. Because of the wireless devices will be orders of magnitude denser than in 5G cellular systems, and because of their complex quality of service requirements, the access to the wireless spectrum will have to be appropriately shared to avoid congestion, poor quality of service, or unsatisfactory communication delays. Spectrum sharing methods have been the objective of intense study through model based approaches, such as optimization or game theories. However, these methods may fail when facing the complexity of the communication environments in 5G, 6G and beyond. Recently, there has been significant interest into the application and development of data-driven methods, namely machine learning methods, to handle the complex operation of spectrum sharing. In this survey paper, we provide a complete overview of the state-of-the-art of machine learning for spectrum sharing. First, we make a mapping of the most prominent  methods that we encounter in spectrum sharing. Then, we show how these machine learning methods are applied to the numerous dimensions and sub-problems of spectrum sharing, such as: spectrum sensing, spectrum allocation, spectrum access, and spectrum handoff. We also highlight the open questions and future trends.

\end{abstract}

\chapter{Introduction}\label{Sec:Intro}
Due to the rapidly increasing number of mobile data subscriptions and the continuous increase in the average data volume per mobile broadband subscription,
the demand for wireless services and applications has been experiencing a large growth in recent years.
Users of \ac{eMBB}, \ac{IoT}, smart factory, remote health care,  connected \ac{UAV} (drone), and urban air mobility applications as well intelligent transportation and smart home services,
demand high functional safety and rely on the exchange of large amount of data with low latency and often with high reliability.
To meet these requirements, \ac{5G} systems are deployed to support 10-100 times more connected devices, transmit 100 times more data,
and support 1000 times the capacity compared with the capabilities \ac{4G} systems \cite{Osseiran2014}.
For \ac{6G} systems,  meeting new requirements on data volumes, coverage and capacity as well as
on the massive number of connected devices, spectrum management will be even more challenging and important \cite{DOCOMO2021, 
Akyildiz2020}.

Recognizing the increasing demands for wireless services, and thereby for spectrum resources in cellular and local area networks,
several previous works have suggested that the static assignment of spectrum to \acp{MNO} and/or specific wireless
technologies confine the utilization of spectrum resources. The key observation of these works is that a certain geographical area,
such as a single cell of a cellular network, may occasionally be populated by users
-- including connected vehicles, drones or \ac{IoT} devices -- belonging to different \acp{MNO} \cite{ShokriGhadikolaei2016}.
In such scenarios, spectrum sharing among multiple players is a flexible and efficient paradigm, which enables to better utilize
the spectrum, avoid spectrum shortage in sub \ac{mmWave} bands, and enhance the return-of-investment in spectrum resources by \acp{MNO} \cite{Papadias:20}, \cite{Wang:11}.
Following these early works on spectrum sharing, several technical and economical aspects of spectrum sharing have been discussed in 
the literature \cite{Dosch:11,ECC:18,Frascolla:16,ETSI:18,FCC:15}.
One of the practical results of these ideas is the protocols and mechanisms standardized by the 3GPP and implemented by \acp{MNO}
for sharing spectrum between 4G and 5G networks \cite{Papadias:20, Boccardi:16}.

Massive \ac{MTC}, \ac{eMBB} enablers and \ac{URLLC} are technology components that
aim to fulfill the aforementioned \ac{5G} and emerging 6G requirements \cite{Popovski2014,Popovski2018}. The \ac{MTC} and a part of \ac{eMBB} implementation should be deployed in sub-6 GHz band due to cost reduction, since sub-6 GHz bands have favourable propagation characteristics \cite{Hayashida2019}. 
However, this spectrum is heavily used by other wireless systems,
including cellular and local area wireless networks using licensed and unlicensed spectrum bands.

\begin{figure}[!t]
	\begin{center}		
		{\includegraphics[width=0.95\textwidth]{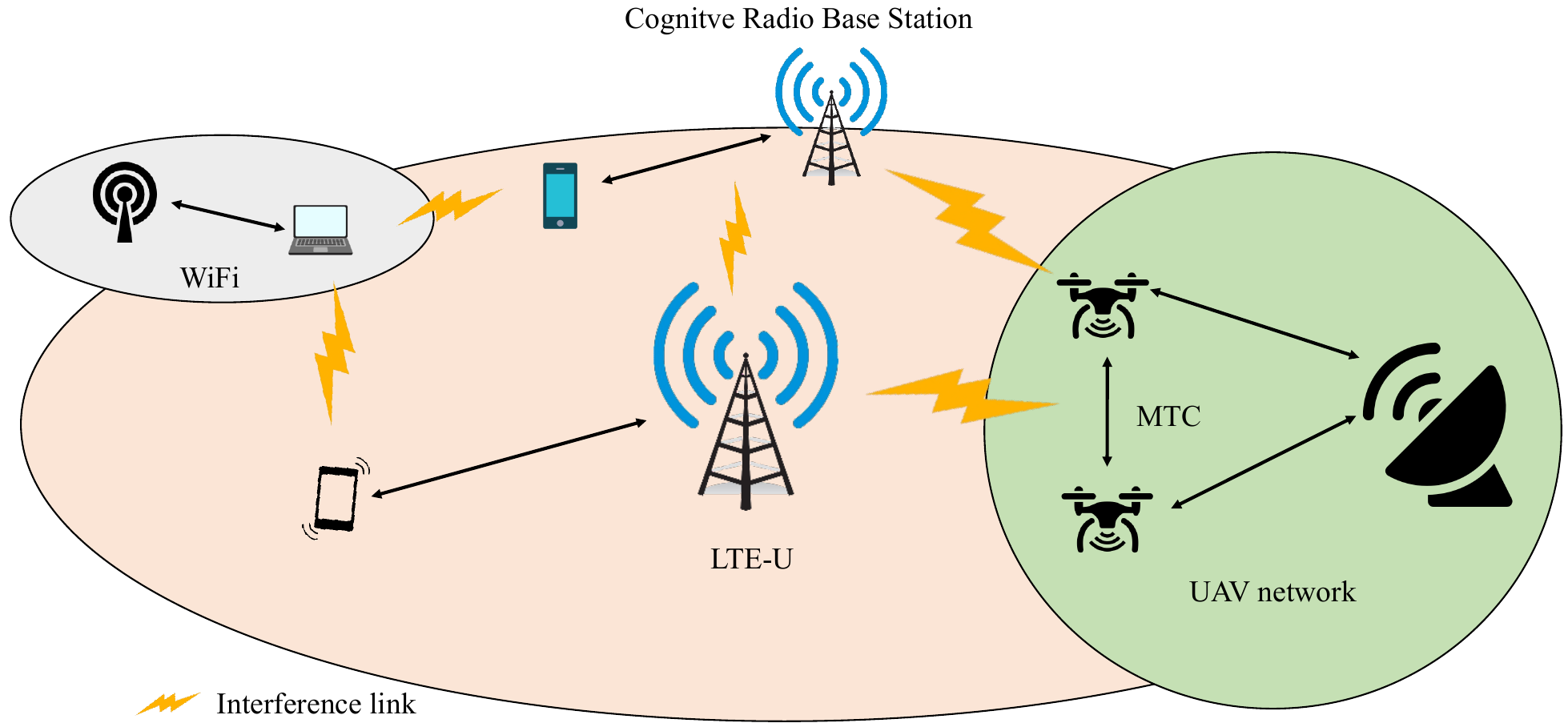}}
		\caption{Coexistence of different technologies in a spectrum sharing scenario.}
		\label{fig:spectrumsharing}
	\end{center}
\end{figure}

To accommodate the emerging 5G and the upcoming 6G services an appealing alternative is to utilize \ac{mmWave} frequencies, which operate between 10 and 300 GHz.
Unfortunately, even this spectrum range has availability problems,
due to other service requirements, which are already allocated in these frequencies \cite{ShokriGhadikolaei2016}.
Due to the pressing demand for efficient ways to allocate and access spectrum,
the concept of \ac{DSS} has attracted significant research attention \cite{Yang2016,Khan2014,Goldsmith2009,Sharma2015}.
Currently, \acp{MNO} have to refarm
their available cellular frequency bands either to enable exclusive \ac{5G} operations or to support shared operations of 4G and 5G infrastructures
in the same or overlapping frequency bands \cite{Candal:21, Agiwal:21}. 
As a natural step beyond currently available spectrum sharing solutions designed for 4G and 5G systems,
the more general concept of \ac{DSS} facilitates the coexistence of cellular and others technologies such as WiFi, \ac{UAV} networks and \acp{CRN}, as illustrated in Fig. \ref{fig:spectrumsharing}.
Indeed, \ac{DSS} will enable to share the same spectrum resources across multiple radio access technologies
allowing to gradually deploy new services that are best served by different access technologies.

To serve a growing number of users and applications by spectrum  sharing between 4G and 5G systems -- while maintaining
high spectrum utilization and meeting capital and operational expenditure constraints -- comes at the cost of
considerable complexity.
While operating 4G and 5G systems in dedicated bands allows to use a wide range of \ac{SON} functionalities,
introducing \ac{DSS} between 4G and 5G systems increases the number of parameters to tune considerably.
However, this increasing complexity makes it difficult to continue using the current resource allocation and
optimization techniques.
To cope with such complexity, the 3GPP and the research community have started to explore
the use of \ac{ML} and \ac{AI} for spectrum sharing.

With an \ac{ML}-based SON, the network self-adjusts and fine-tunes a range of parameters according to the prevailing radio and traffic conditions,
alleviating the burden of manual optimization by the \acp{MNO}.
While \ac{SON} algorithms are not standardized in 3GPP, \ac{SON} implementations may be assisted by various \ac{ML} algorithms, including
those employing supervised learning, unsupervised learning and \ac{RL}-based schemes \cite{Bonati:21}.

\section{Spectrum Sharing State-of-the-Art Surveys}
Spectrum sharing can be performed either in a centralized  or distributed manner. The former is characterized by a central unity, often 
called spectrum  server, which is responsible for optimizing the spectrum usage among all users. 
In the latter, all network users participate in the spectrum optimization process. It is a more practical solution for high spectral demand since the computational complexity at the central unity increases with the number of spectrum requests \cite{Lin2010}.

In a spectrum sharing scenario, the coexistence of different wireless systems are supported by four  mechanisms:
\begin{enumerate}
	\item \textit{Spectrum sensing}: in this mechanism, signal features are extracted from the environment to determine the radio frequency occupancy condition, i.e., which channels are in use and which ones are free.
	\item \textit{Spectrum allocation}: receives the channel characterization from sensing mechanism or directly from the environment in case of frequency planing. The main goal is to assign users on available channels for data transmission.
	\item \textit{Spectrum access}: the user assignment is used in this stage to provide channel access for allocated users in order to guarantee the data transmission.
	\item \textit{Spectrum handoff}: responsible for user channel switching whenever necessary. It sends a request to the spectrum allocation mechanism to check and to assign a new channel to the user so it can continue to access the medium sending its data.
\end{enumerate}

\begin{figure}[!t]
	\begin{center}
	{\includegraphics[width=0.95\textwidth]{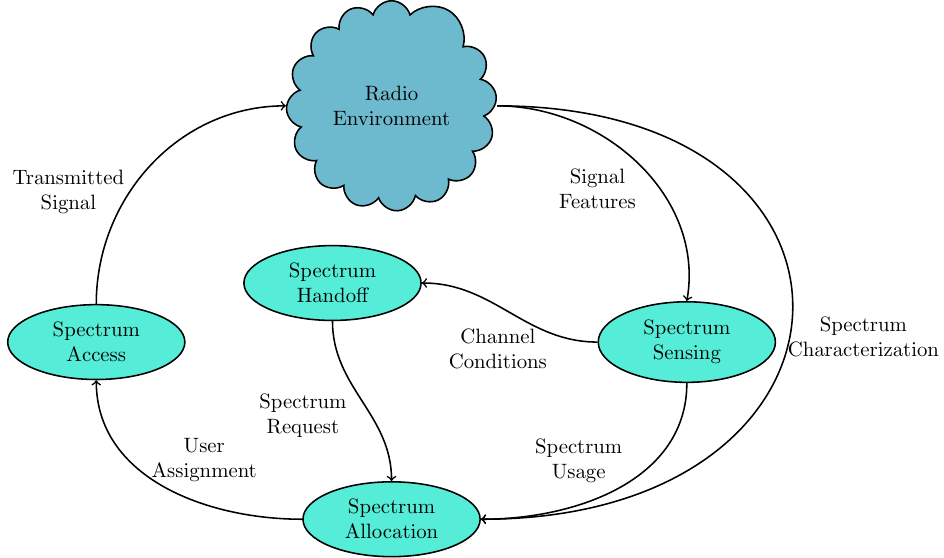}}
	\caption{Relationship among spectrum sharing mechanisms.}
	\label{fig:spectrumsharingscenario}
	\end{center}
\end{figure}

This relationship between the four spectrum sharing mechanisms is showed in Fig \ref{fig:spectrumsharingscenario}.

The use of \ac{ML} solutions as a tool for spectrum sharing has been investigated by recent surveys \cite{Arjoune2019,Syed2023,Agrawal2022,Fernando2023,Wang2018,Zhang2017,Zhang2019a,Tehrani2016,Puspita2019,Janu2021,Samanta2022,Hu2018,Kaur2022,ElHalawany2022,Lu2023,Wang2022,Dangi2022}.

References \cite{Arjoune2019,Syed2023,Agrawal2022,Fernando2023} cover the state-of-the-art of spectrum sensing for \ac{CR}.
The main focus of \cite{Arjoune2019} is the classification and review of different sensing techniques using traditional and \ac{ML} schemes, while  \cite{Syed2023} provides a \ac{DL} detailed survey for spectrum sensing. Reference \cite{Agrawal2022} discusses recent spectrum sensing and \ac{DSA} schemes and topics related to \ac{CR} including \ac{ML} solutions.
The authors highlight the efficiency, limitations and implementation challenges of both narrowband and wideband sensing approaches. On the other hand, reference \cite{Fernando2023} presents spectrum sensing in \ac{IoT} context, giving a brief discussion of recent papers in the area. These works also discuss the open issues related to spectrum sensing and the way how \ac{CR} can be used to solve spectrum sharing problems in next generation networks. However, these references do not discuss \ac{ML} issues in details.
To bridge this research gap, in this paper we provide an overview of spectrum sensing \ac{ML} works that address narrowband and 
wideband spectrum sharing schemes, and present a mathematical formulation of the \ac{ML}-assisted spectrum sensing problem.

The work in \cite{Wang2018} presents a survey on spectrum allocation using \ac{RL} algorithms for \acp{CRN}.
The authors analyze the advantages and disadvantages of each \ac{RL} algorithm by dividing them into two groups:
minor and major implementation improvements.
They also address challenges and open issues related to spectrum allocation for \acp{CRN} and \ac{RL} algorithms.
However, the usage of \ac{ML} methods for other spectrum sharing networks
such as \ac{LTE-U}, \ac{UAV} networks and \ac{NOMA} systems, are not investigated in that paper.

Reference \cite{Zhang2017} presents a survey on spectrum sharing techniques that address
the basic principles and state-of-the-art for \ac{CR}, \ac{D2D}, \ac{IBFD}, \ac{NOMA} and \ac{LTE-U} technologies.
The authors also discuss the challenges related to deploying each of these techniques,
as well as how they can be integrated into \ac{5G} networks.
Spectrum access is also in the focus of reference \cite{Zhang2019a}.  
It presents the basic principles and spectrum sharing solutions for the most popular \ac{IoT}
technologies applicable in both licensed and unlicensed spectrum.
That paper also identifies future challenges of \ac{IoT} systems and suggests research directions
for next generation technologies.
However, none of these references address \ac{ML} solutions for spectrum access.
Differently from both works, in this paper
we describe the \ac{DSA} problem and discuss \ac{ML} solutions by surveying the most relevant recent works in this area.

The authors of \cite{Tehrani2016} study various scenarios on licensed cellular networks
with different topologies in order to demonstrate the importance of spectrum sharing for future networks.
That paper provides an analysis of spectrum sharing involving \acp{MNO} using
licensed shared access for wide area broadband services.
The main concepts of spectrum sharing are explained,
and open issues for future research are suggested.
However, such a paper does not discuss the potential and challenges related to \ac{ML} schemes for spectrum sharing.

Reference \cite{Puspita2019} focuses on recent \ac{RL}-based surveys for \acp{CRN}.
The work discusses how \ac{ML} algorithms can be used to solve spectrum sharing problems for \ac{CR}.
It also presents future research directions and networks solutions for upcoming \ac{CR} technologies.
The authors, however, dedicate only a small section to discuss \ac{RL} for \ac{CRN}.
Other \ac{ML} schemes such as supervised and unsupervised learning are not addressed.

The work  in \cite{Janu2021} addresses the usage of \ac{ML} for cooperative spectrum sensing and \ac{DSS}. The authors characterized the surveyed papers based on the applied \ac{ML} methods (supervised, unsupervised or \ac{RL}) and on the evaluation performance metrics of the adopted approaches, showing their advantages and limitations. It also addresses \ac{DSS} scenarios providing useful discussion on spectrum allocation and spectrum access. However, the authors did not survey \ac{ML} papers on these topics, which are covered in \cite{Samanta2022}. In this work, the authors provide an overview of \ac{ML} techniques focusing on address \ac{5G} network issues such as resource allocation, spectrum access and security aspects. Although, relevant spectrum sharing topics are discussed in this work, a spectrum sensing discussion is missing.

References \cite{Hu2018} and \cite{Kaur2022} provide an extensive review of works related to spectrum sensing, allocation, access and handoff in the context of \acp{CRN}.
They also present a summary of existing survey works on \ac{CRN}
and discuss design aspects of \ac{CR} control mechanisms and energy efficiency.
Although the former includes a large set of spectrum sharing works,
\ac{ML} papers are out of the scope of that survey. On the other hand, the latter presents a comprehensive review of \ac{ML} works for 
spectrum sharing, however beamforming and security are not addressed.

Since \ac{mmWave} has arisen as a key technology to accommodate new services in next generation systems, authors in \cite{ElHalawany2022} presented an extensive survey on \ac{ML}-based beamforming for \ac{mmWave} scenario. The authors provided an overview and applicability of \ac{ML} techniques, summarized \ac{mmWave} beamforming strategies and provided insightful discussion about \ac{ML} usage for \ac{mmWave} beamforming. Although sub 6GHz frequencies were out of the scope, there are important recent references not covered by the authors. In our work, we cover relevant \ac{ML} works for beamforming design in all range of frequencies. 

 The exponential growth on data traffic in next generation networks motivates recent surveys to explore spectrum sharing security. Reference \cite{Lu2023} surveyed \ac{RL} strategies for the physical layer, focusing on jammers, eavesdroppers, spoofers and inference attackers. Although the authors provided a large overview of security techniques and defense strategies, unsupervised and supervised learning classification strategies were not considered. Falsification attacks, for example, rely on camouflaging the attacker as an authorized node. Classification methods were proved to be efficient to combat this strategy \cite{Wang2022,Dangi2022}. In \cite{Wang2022} the authors review spectrum sharing for various types of network frameworks. They also investigate the state-of-the-art \ac{ML} of security threats and defensive strategies in different network layers. Instead of considering all network aspects, the work in \cite{Dangi2022} address security issues focusing on network slice lifecycle. The authors present insightful discussions on \ac{ML} strategies for network slicing and an existing related surveys mapping. Although \cite{Wang2022} and  \cite{Dangi2022}  have many contributions in the security field, they did not survey works related with spectrum sharing mechanisms.

To summarize the above discussion on recent related works, Table \ref{Tab:Surveys} presents the main aspects  and Table \ref{Tab:Surveys2} summarizes the main contributions covered by each work. Differently from other surveys, our work covers the fundamentals of \ac{ML} methods,
which are prevalent in the topic of spectrum sharing and are expected to play a key role in emerging 6G systems.
The main reason for this is that 6G systems will cope with the increasing traffic demands, complexity and scalability
requirements by employing cognitive and learning technologies, as inherent parts of both lower and upper layers of the system.
Also, we provide a mathematical description of \ac{ML} methods, highlight the conceptual differences among them,
and discuss spectrum sharing applications for which \ac{ML} techniques have already been successfully applied.
We also provide an in-depth comparison of the proposals available in the literature,
identify research gaps in the existing solutions, 
and discuss open questions related to spectrum sharing that will be important in the upcoming generation of wireless
systems.

\begin{table}
	\centering
	\caption{Spectrum sharing surveys aspects overview.}
	{\scriptsize \begin{tabularx}{\linewidth}{
				|>{\hsize=1.54\hsize}X|
				>{\hsize=0.59\hsize}X|
				>{\hsize=0.79\hsize}X|
				>{\hsize=0.59\hsize}X|
				>{\hsize=0.59\hsize}X|
				>{\hsize=0.3\hsize}X|
				>{\hsize=1\hsize}X|
				>{\hsize=0.6\hsize}X|
			}
		\hline
		
		\multirow{2}{*}{\centering \hspace*{0.6cm}Work} & \multicolumn{4}{c|}{Spectrum Sharing}  & \multicolumn{3}{c|}{Additional Aspects}  \\
		& Sensing   & Allocation  & Access   &   Handoff        & \centering ML           &   Beamforming  &           Security        \\ \hline
		\parbox[c]{2cm}{\centering\cite{Arjoune2019}} &  \centering  $\checkmark$ & \centering $-$ & \centering $-$  & \centering $-$ & \centering $\checkmark$ & \centering $-$ & \centering\arraybackslash$-$ \\
		\hline
		\parbox[c]{2cm}{\centering\cite{Syed2023}}	&  \centering  $\checkmark$ & \centering $-$  & \centering $-$  & \centering $-$ & \centering $\checkmark$ & \centering $-$  & \centering\arraybackslash $-$ \\
		\hline
		\parbox[c]{2cm}{\centering\cite{Agrawal2022}}	&  \centering  $\checkmark$ & \centering $-$  & \centering $\checkmark$  & \centering $-$  & \centering $\checkmark$  & \centering $-$ & \centering\arraybackslash $-$ \\
		\hline
		\parbox[c]{2cm}{\centering\cite{Fernando2023}}	&  \centering  $\checkmark$ & \centering $-$  & \centering $\checkmark$  & \centering $-$  & \centering $\checkmark$  & \centering $-$ & \centering\arraybackslash $-$ \\
		\hline
		\parbox[c]{2cm}{\centering\cite{Wang2018}}	&  \centering  $-$ & \centering $\checkmark$  & \centering $-$  & \centering $-$   & \centering $\checkmark$   & \centering $-$ & \centering\arraybackslash $-$  \\  
		\hline	
		\parbox[c]{2cm}{\centering\cite{Zhang2017}}	&  \centering  $-$ & \centering $-$  & \centering $\checkmark$ & \centering $-$  & \centering $-$  & \centering $-$ & \centering\arraybackslash $-$   \\  
		\hline	 		
		\parbox[c]{2cm}{\centering\cite{Zhang2019a}} &  \centering  $-$ & \centering $-$  & \centering $\checkmark$  & \centering $-$  & \centering $-$  & \centering $-$ & \centering\arraybackslash $-$\\  
		\hline	
		\parbox[c]{2cm}{\centering\cite{Tehrani2016}} &  \centering  $\checkmark$ & \centering $\checkmark$  & \centering $\checkmark$  & \centering $-$  & \centering $-$  & \centering $-$ & \centering\arraybackslash $-$  \\  
		\hline	
		\parbox[c]{2cm}{\centering\cite{Puspita2019}} &  \centering   $-$ & \centering  $-$  & \centering $\checkmark$  & \centering $-$  & \centering $\checkmark$ & \centering\arraybackslash $-$ & \centering\arraybackslash $-$  \\ 
		\hline				
		\parbox[c]{2cm}{\centering\cite{Janu2021}} &  \centering  $\checkmark$ & \centering $-$  & \centering $-$ & \centering $-$   & \centering $ \checkmark$  & \centering $-$ & \centering\arraybackslash $-$ \\  
		\hline	
		\parbox[c]{2cm}{\centering\cite{Samanta2022}} &  \centering  $-$ & \centering  $ \checkmark$  & \centering  $ \checkmark$ & \centering $-$   & \centering $ \checkmark$  & \centering $-$ & \centering\arraybackslash $ \checkmark$ \\  
		\hline				
		\parbox[c]{2cm}{\centering\cite{Hu2018}} &  \centering  $\checkmark$ & \centering $\checkmark$  & \centering $\checkmark$  & \centering $\checkmark$   & \centering $-$  & \centering $-$ & \centering\arraybackslash $-$ \\ 	
		\hline							
		\parbox[c]{2cm}{\centering\cite{Kaur2022}} &  \centering  $\checkmark$ & \centering $\checkmark$  & \centering $\checkmark$  & \centering $\checkmark$   & \centering $\checkmark$  & \centering $-$ & \centering\arraybackslash $-$   \\ 	
		\hline		
		\parbox[c]{2cm}{\centering\cite{ElHalawany2022}} &  \centering  $-$ & \centering $-$  & \centering $-$  & \centering $-$   & \centering $\checkmark$  & \centering $\checkmark$ & \centering\arraybackslash $-$  \\ 
		\hline
		\parbox[c]{2cm}{\centering\cite{Lu2023}} &  \centering   $-$ & \centering $\checkmark$  & \centering  $-$  & \centering  $-$   & \centering  $\checkmark$  & \centering $-$ & \centering\arraybackslash  $\checkmark$  \\ 	
		\hline	
		\parbox[c]{2cm}{\centering\cite{Wang2022}} &  \centering   $-$ & \centering $\checkmark$  & \centering  $-$  & \centering  $-$   & \centering  $\checkmark$  & \centering $-$ & \centering\arraybackslash  $\checkmark$  \\ 	
		\hline	
	    \parbox[c]{2cm}{\centering\cite{Dangi2022}} &  \centering   $-$ & \centering $-$  & \centering  $-$  & \centering  $-$   & \centering  $\checkmark$  & \centering $-$ & \centering\arraybackslash  $\checkmark$     \\ 	
	    \hline	
		
		\parbox[c]{2cm}{\centering\textbf{Our work}} &  \centering  $\checkmark$ & \centering $\checkmark$  & \centering $\checkmark$  & \centering $\checkmark$   & \centering $\checkmark$  & \centering $\checkmark$  & \centering\arraybackslash $\checkmark$ \\
		\hline
	\end{tabularx}}
\label{Tab:Surveys}
\end{table}

\begin{table}
	\centering
	\caption{Spectrum sharing surveys key contributions summary.}
	{\footnotesize  \begin{tabularx}{\linewidth}{
				|>{\hsize=0.5\hsize}X|
				>{\hsize=1.5\hsize}X|
			}
			\hline
			\parbox[c]{2.5cm}{\centering\vspace*{0.05cm}Work}   &        Key Contribution        \\ \hline
			\vspace{0.0005cm}\parbox[c]{2.5cm}{\centering\cite{Arjoune2019}} & A survey on narrowband and wideband spectrum sensing schemes for \acp{CRN}. \\ 	
			\hline	
			\parbox[c]{2.5cm}{\centering\vspace*{0.05cm}\cite{Syed2023}}	 & A survey on \ac{DL} spectrum sensing schemes for \acp{CRN}. \\ 
			\hline	
			\vspace*{0.15cm}\parbox[c]{2.5cm}{\centering\vspace*{0.05cm}\cite{Agrawal2022}}	&  A review on spectrum sensing and \ac{DSA} for cognitive radar networks. It provides a detailed spectrum sensing classification and a spectrum management framework.\\ 
			\hline			
			\vspace*{0.15cm}\parbox[c]{2.5cm}{\centering\vspace*{0.05cm}\cite{Fernando2023}}	 & A systematic review on the relationship between spectrum sensing, clustering algorithms, and energy-harvesting for \ac{CRN}s in \ac{IoT} context.\\ 	
			\hline	
			\vspace{0.0005cm}\parbox[c]{2.5cm}{\centering\cite{Wang2018}}	 &   An overview of the state-of-the-art of \ac{RL} algorithms for spectrum allocation on \acp{CRN}. \\ 
			\hline	
			\vspace{0.0005cm}\parbox[c]{2.5cm}{\centering\cite{Zhang2017}} & A brief discussion on spectrum sharing techniques for \ac{CR}, \ac{D2D}, \ac{IBFD}, \ac{NOMA} and \ac{LTE-U} technologies. \\  \hline	 		
			\vspace{0.0005cm}\parbox[c]{2.5cm}{\centering\cite{Zhang2019a}} & Discussion of spectrum sharing solutions for popular \ac{IoT} technologies. \\  
			\hline	
			\vspace{0.0005cm}\parbox[c]{2.5cm}{\centering\cite{Tehrani2016}} & Study of potential scenarios that can benefit from spectrum sharing. \\  
			\hline	
			\vspace*{0.15cm}\parbox[c]{2.5cm}{\centering\vspace*{0.05cm}\cite{Puspita2019}}   & Brief survey on \ac{RL} works for spectrum sharing on \acp{CRN}, including the discussion of efficient spectrum management on \ac{5G} technology.\\ 	
			\hline				
			\vspace{0.0005cm}\parbox[c]{2.5cm}{\centering\cite{Janu2021}} &  A survey on \ac{ML} algorithms in the \ac{CSS} and \ac{DSS} domain for \acp{CRN}. \\  
			\hline	
			\vspace{0.0005cm}\parbox[c]{2.5cm}{\centering\cite{Samanta2022}} &   A deep learning discussion to tackle \ac{5G} and beyond wireless systems issues. \\  
			\hline				
			\vspace*{0.15cm}\parbox[c]{2.5cm}{\centering\vspace*{0.05cm}\cite{Hu2018}} &  A survey on spectrum sharing for \ac{CR} towards \ac{5G} networks, including a taxonomy from the perspective of Wider-Coverage, Massive-Capacity, Massive-Connectivity, and Low-Latency. \\ 	\hline							
			\vspace{0.0005cm}\parbox[c]{2.5cm}{\centering\cite{Kaur2022}} &  Provides a classification and a survey for \ac{ML} techniques on spectrum sharing scenario. \\ 	
			\hline		
			\vspace{0.0005cm}\parbox[c]{2.5cm}{\centering\cite{ElHalawany2022}} &  Provides an overview of \ac{mmWave} beamforming design with \ac{ML}. \\ 	
			\hline
			\parbox[c]{2.5cm}{\centering\vspace*{0.4cm}\cite{Lu2023}} &  Surveys \ac{RL} techniques for  physical layer attacks in \ac{6G} systes. \\ 	
			\hline	
			\vspace*{0.15cm}\parbox[c]{2.5cm}{\centering\vspace*{0.05cm}\cite{Wang2022}} &  Investigates the state-of-the-art \ac{ML} defensive strategies, such as primary user emulation, spectrum sensing data falsification, jamming and eavesdropping attacks. \\ 	
			\hline	
			\parbox[c]{2.5cm}{\centering\vspace*{0.05cm}\cite{Dangi2022}} & A survey on security issues for \ac{5G} networking slicing. \\ 	
			\hline	
			
			\vspace*{1.5cm}\parbox[c]{2.5cm}{\centering\vspace*{0.05cm}\textbf{Our work}} &  An extensive up to date survey on \ac{ML} for spectrum sharing. Particularly,
			
			\parbox[c]{8cm}{\vspace*{0.05cm} \tabitem We present several works that use \ac{ML} as a tool for  spectrum sharing problem, 
			including beamforming and security aspects.\\  \hspace*{0.005cm} \tabitem We include a ML review section and summary tables  
			that  provide useful insights on \ac{ML} techniques for spectrum  sharing.  \\ \hspace*{0.005cm} \tabitem We also highlight 
			spectrum sharing  challenges and future research directions. }  \\ 	
			\hline
	\end{tabularx}}
	\label{Tab:Surveys2}
\end{table}

\begin{figure*}[!tb]
  \centering
  \includegraphics[width=1\textwidth]{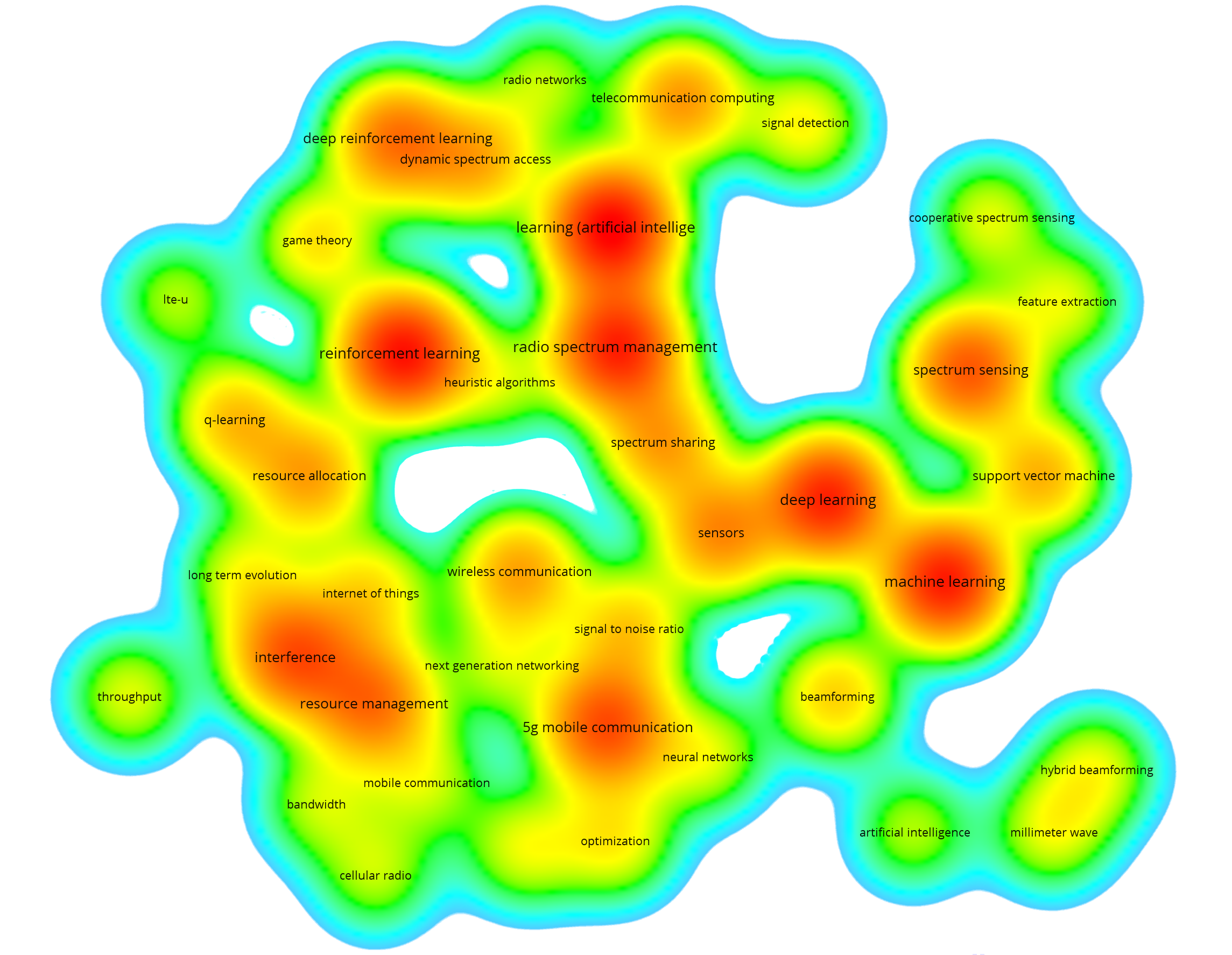}
  \caption{Density of the keywords presented in the cited references of this survey.}\label{Fig:Landscape}
\end{figure*}

Another point also provided by our survey is the evaluation of the most active keywords in the recent literature. We provide in Fig.~\ref{Fig:Landscape} a density illustration of the works cited in this survey, showing the most relevant topics (keywords) considered in the surveyed literature. The darker the color where the keyword is being shown the more frequent the keyword is in the considered database. The neighborhood of the keywords is related to their joint occurrence in the references and therefore the figure allow us to see which topics are more correlated. Finally, only the keywords which are being mentioned in at least 5 (five) references are being displayed in the density map.

The major contributions of the present paper are summarized as follows:
\begin{itemize}
\item We cover the recent spectrum sharing surveys state-of-the-art addressing the strong points and pointing the main gaps of each work, providing a comparison among various \ac{ML} papers on spectrum sensing, allocation, access and handoff scenarios highlighting the main contributions of each work.
\item We outline \ac{ML} methods providing a general discussion and a mathematical formulation in the context of spectrum sharing networks describing the benefits of these approaches.
\item We discuss the contributions of \ac{ML} to fundamental aspects on spectrum sharing security and beamforming applications.
\item We identify existing challenges on spectrum sharing and we point how \ac{ML} can be used as a potential solution to overcome those issues.
We also point future open researches on spectrum sharing using \ac{ML} applications.
\end{itemize}

This work is structured as follows.
Chapter~\ref{Sec:ML} introduces \ac{ML} schemes which can be used in the context of spectrum sharing. More than just a recall about the main \ac{ML} methods, the goal is to provide a more suitable description of the methods for the applications of spectrum sharing. Although the literature has a high number of introductory texts about machine learning and the topics covered in Chapter~\ref{Sec:ML} could be just assumed to be known by the reader, the section is intended to be a self-contained introduction to the most important \ac{ML} methods. This will allow the unfamiliar reader to see the details of  some of the strategies available in the literature to be able to understand the underlying concepts that are used for the solution of the spectrum sharing problems. Hence, the reader already familiar with the \ac{ML} strategies and models can skip Chapter~\ref{Sec:ML} without any loss of continuity.
Chapters \ref{Sec:Sensing}, \ref{Sec:Allocation} and \ref{Sec:Access}  review the most relevant works in the literature covering \ac{ML} solutions
for spectrum sensing, allocation and access, respectively. Chapter~\ref{Sec:Aspects} addresses \ac{ML} usage on spectrum handoff, beamforming and spectrum sharing security.
Subsequently, Chapter~\ref{Sec:FutureResearch} discusses
the main issues and challenges on spectrum sharing and highlights important points on spectrum sharing for future research.
Finally, Chapter~\ref{Sec:Conclusion} concludes this paper.
A list of key acronyms and abbreviations used throughout
the paper is given in Table \ref{Table:Acro}.

\begin{table*}
	\caption{List of key acronyms.}
	\centering
	{\scriptsize  \begin{tabular}{r|l||r|l}
			\hline
			\textbf{Acronym} & \textbf{Definition} & \textbf{Acronym} & \textbf{Definition} \\
			\hline
			{AE} & {Autoencoder} & {LRMM} & {Log-Rayleigh Mixture Model} \\
			{AI} & {Artificial Intelligence} &	{LSTM} & {Long Short-Term Memory} \\
			{ANN} & {Artificial Neural Network} & {MAB} & {Multi-Armed Bandit} \\  
			{BS} & {Base Station} & {MARL} & \makecell{Multi-Agent Reinforcement \\ Learning} \\
			{BF} & {Beamforming} & {MDP} & {Markov Decision Process} \\
			{CBF} & {Coordinated Beamforming} & {ML} & {Machine Learning} \\
			{CSI} & {Channel State Information} & {MM} & {Mixture Model} \\
			{CSIT}& \makecell{Channel State Information \\ at the Transmitter} & {mmWave} & {Millimeter-Wave}  \\
			{CNN} &{Convolutional Neural Network} & {NOMA} & {Non-Orthogonal Multiple  Access} \\
			{CRN} & {Cognitive Radio Network} & {NR} & {New Radio} \\
			{CSS} & {Cooperative Spectrum Sensing} & {PU} & {Primary User}  \\
			{DDQN} & {Double Deep Q Network} & {PR} & {Primary Receiver} \\
			{DL} & {Deep Learning} & {PSO} & {Particle Swarm Optimization} \\
			{DNN} & {Deep Neural Network} & {QoE} & {Quality of Experience} \\
			{DRL} & {Deep Reinforcement Learning} & {QoS} & {Quality of Service}   \\
			{DSA} & {Dynamic Spectrum Access} & {RAN} & {Radio Access Network} \\
			{DSS} & {Dynamic Spectrum Sharing} &  {RAT} & {Radio Access Technology} \\
			{eMBB} & {Enhanced Mobile Broadband} & {RF} & {Random Forest} \\
			{eNB} & {Evolved Node B} & {RNN} & {Recurrent Neural Network} \\
			{FDA} & {Fisher Discriminant Analysis} & {RL} & {Reinforcement Learning} \\
			{GMM} & {Gaussian Mixture Model} & {ROC} & {Receiver Operating  Characteristics} \\
			{HBF} & {Hybrid Beamforming} & {RSS} & {Received Signal Strength}  \\
			{HMM} & {Hidden Markov Model} &  {SAE} & {Stacked Autoencoder}  \\
			{IBFD} & {In-Band Full-Duplex}  & {SGD} & {Stochastic Gradient Descent} \\
			{IDS} & {Intrusion Detection System} &  {SINR} & \makecell{Signal-to-Interference-plus-Noise \\ Ratio} \\
			{IoT} & {Internet of Things} & {SU} & {Secondary User}  \\ 
			{ITU} & \makecell{International Telecommunication\\  Union} & {SVM} & {Support Vector Machine} \\
			{k-NN} & {k-Nearest Neighbor} & {UAV} & {Unmanned Aerial Vehicle} \\
			{KPI} & {Key Performance Indicator} & {UE} & {User Equipment} \\
			{LTE} & {Long-Term Evolution} & {URLLC} & \makecell{Ultra-Reliable Low-Latency \\ Communication}  \\
			{LTE-U} & \makecell{Long-Term Evolution on \\ Unlicensed  Spectrum} & {VUE} & {Vehicular User Equipment} \\
			\hline
		\end{tabular}\label{Table:Acro}}
\end{table*}

\section{Summary}\label{sub:Chap_Intro:summary}
In this chapter, we introduced the spectrum sharing problematic. Specifically, we contextualized the need for the use of spectral sharing in \ac{5G} and beyond networks and we pointed out \ac{ML} as one of the enables to do it efficiently. We also presented the state of the art of recent spectrum sharing surveys, along with the contributions of our work.

In the next chapter we will discuss the \ac{ML} approaches and common algorithms used by spectrum sharing \ac{ML} works in literature.

\chapter{Preliminary: Introduction to Machine Leaning}\label{Sec:ML}
In this chapter, we discuss the fundamentals of the three major \ac{ML} categories, and the commonly used algorithms
within each category that are commonly used specifically for spectrum sharing. 

\begin{figure}[!t]
	\centering
	\subfloat[Supervised learning~\label{fig:classification_ex}]
	{\includegraphics[width=0.55\textwidth]
		{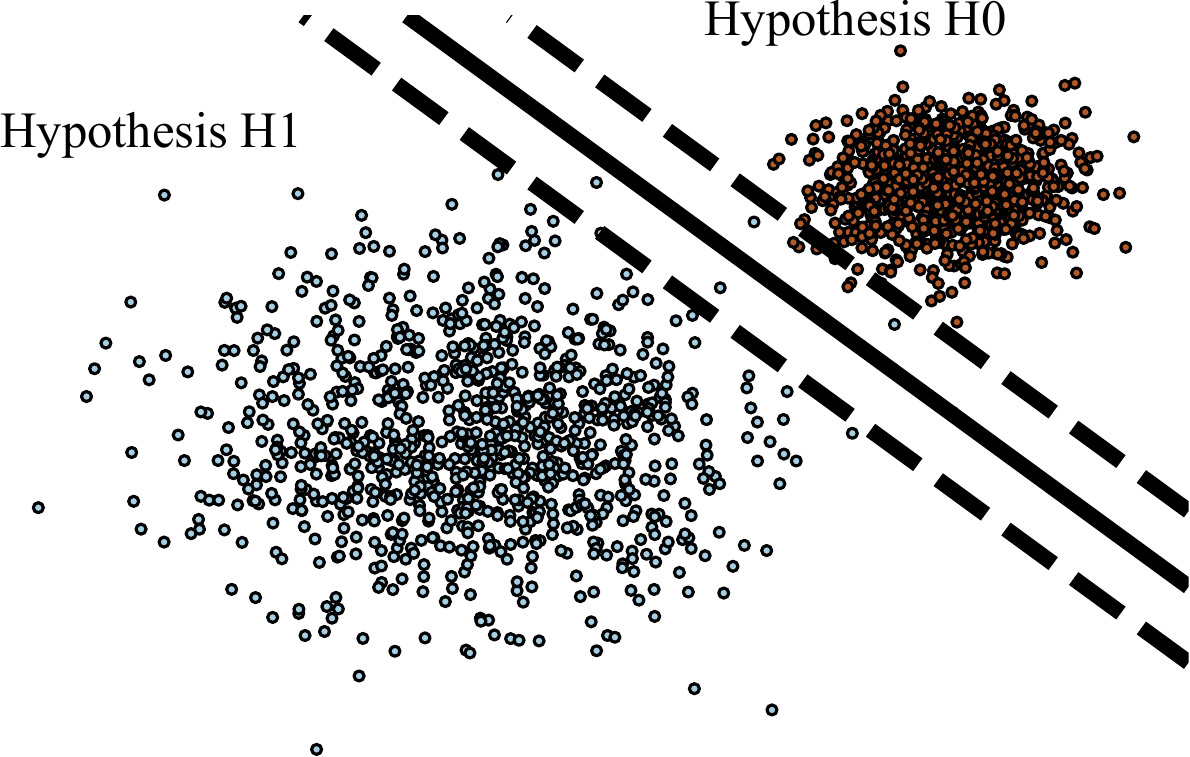}} \hfill
	\subfloat[Unsupervised learning~\label{fig:clustering_ex}]
	{\includegraphics[width=0.55\textwidth]
		{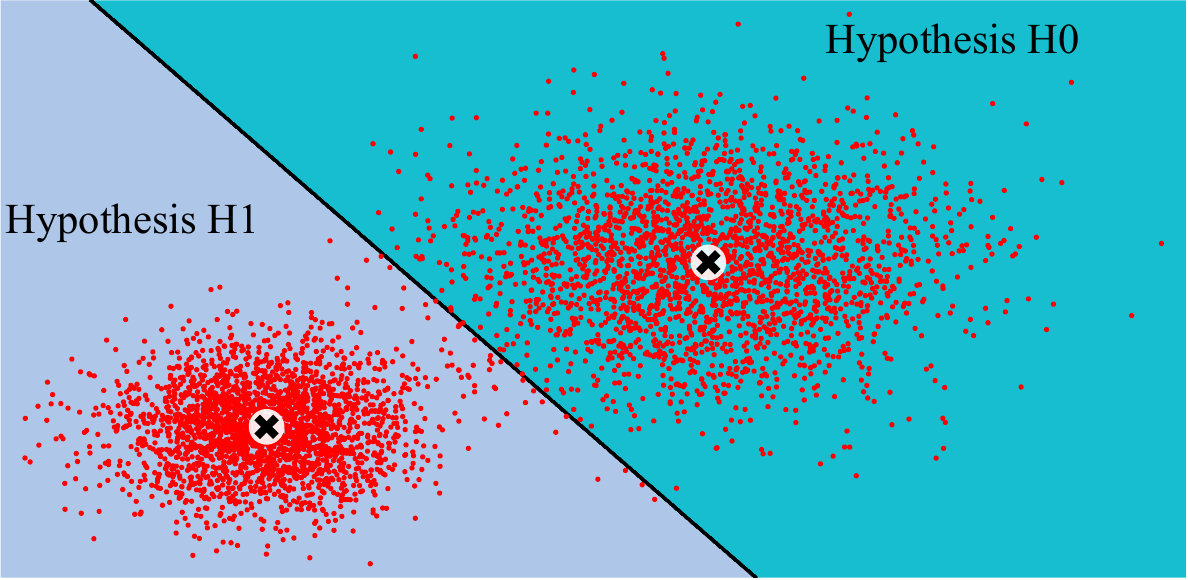}}\hfill\hfill
	\subfloat[Reinforcement learning~\label{fig:reinforcement_ex}]
	{\includegraphics[width=0.55\textwidth]
		{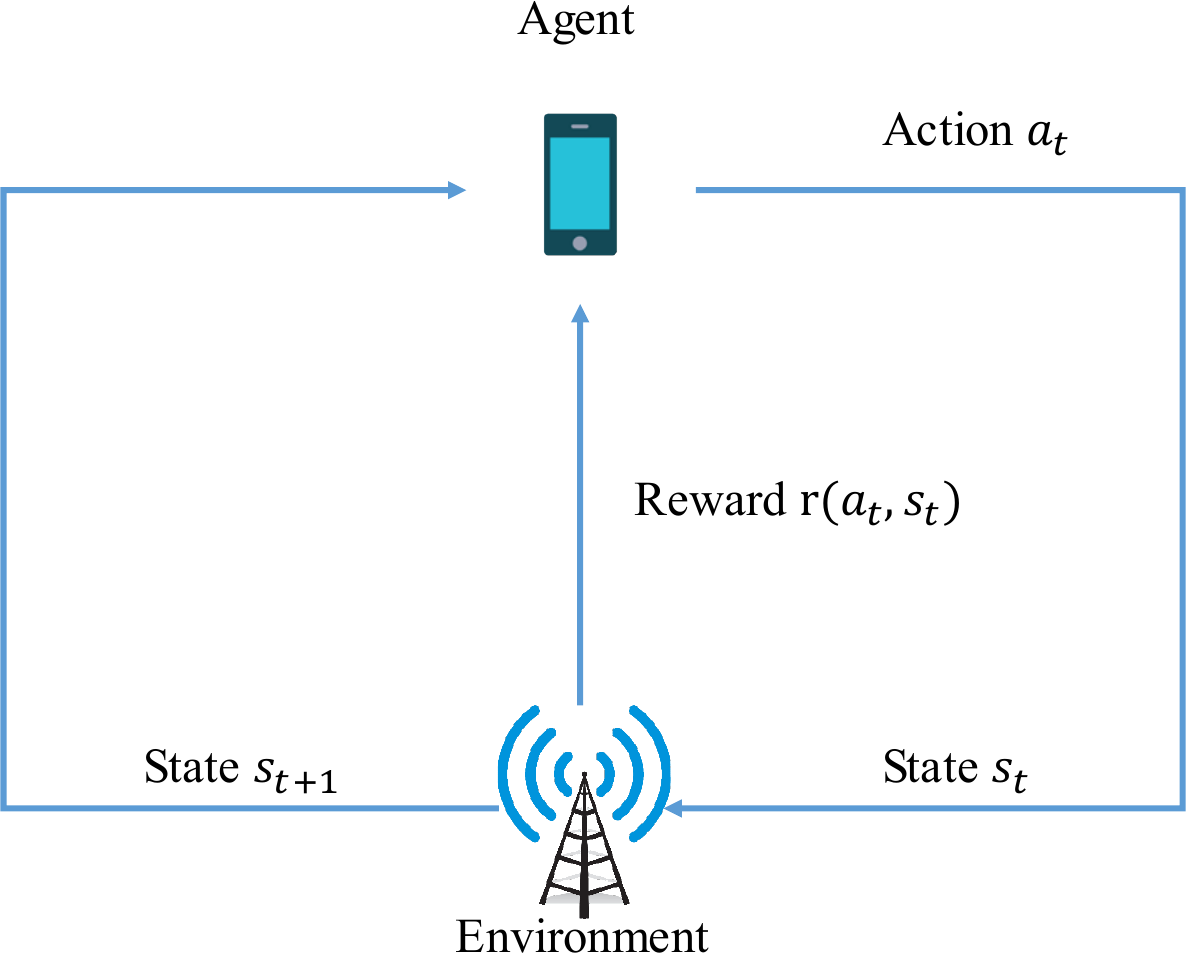}}
	\caption{Examples of supervised, unsupervised, and reinforcement learning problems. First, we have a classification
	problem solved using supervised and unsupervised learning, respectively. Then, we show an example of the
	reinforcement learning process between an agent with the environment in a cellular networks.}
	\label{fig:ML_examples}
\end{figure}

With the ongoing world-wide deployment of \ac{5G} systems and the initial discussions on \ac{6G}~\citep{Letaief2019}, 
\ac{ML} has a key role to substantially improve the spectrum utilization and network throughput, thus reducing the 
latency and improving the data rates.
Hence, it is also expected that spectrum sharing techniques, which have been evolving since the \ac{3G}, also benefit 
from the gains brought by \ac{ML} techniques.
In general, \ac{ML} techniques can be divided into three major categories~\citep{Simeone2018}: supervised, unsupervised,
and reinforcement learning.
Figure~\ref{fig:ML_examples} shows one example for each category.

\pdfsuppresswarningpagegroup=1

Figure~\ref{fig:classification_ex} shows an example of supervised learning, specifically the classification
problem between hypothesis $\mathcal{H}_0$ and $\mathcal{H}_1$.
In spectrum sharing, hypothesis $\mathcal{H}_0$ and $\mathcal{H}_1$ may represent the presence or absence of a transmitting signal,
respectively.
In supervised learning~\citep{Simeone2018}, there are $N$ labelled training pieces of data $\calD = 
\{(x_i,y_i)\}_{i=1}^N$, where
$x_i$ represents the explanatory variable (or input) and $y_i$ is the label (or output).
The goal of supervised learning is to predict the true output $y_{i \geq N}$ given an input $x_{i \geq N}$ that the
learning algorithm has not seen before.
Notice that in Figure~\ref{fig:classification_ex}, some blue dots are in the region of the red dots (i.e., Hypothesis 
$\mathcal{H}_0$), indicating that they are categorized incorrectly.

Conversely, Figure~\ref{fig:clustering_ex} shows an example of unsupervised learning, specifically the clustering of
points in two dimensions, which are based on hypothesis $\mathcal{H}_0$ and $\mathcal{H}_1$.
In unsupervised learning~\citep{Simeone2018}, there are only $N$ unlabelled training data $\calD = \{x_i\}_{i=1}^N$, in
which the label $\{y_i\}_{i=1}^N$ needs to be determined or does not exist.
Thus, the goal of unsupervised learning is to learn the properties of the mechanism that generates and helps to explain
the dataset, such as in dimensionality reduction and feature extraction problems; and in some cases, determines the 
correct
label for the data, such as clustering problems.
Notice that Figure~\ref{fig:clustering_ex} shows the points in two regions with the same color, i.e., without the
labels, while the clustering algorithm provides the label of the dataset in the hypothesis $\mathcal{H}_0$ and $\mathcal{H}_1$.

Finally, Figure~\ref{fig:reinforcement_ex} shows an example of reinforcement learning, specifically the actions of a
mobile user in a cellular network.
In reinforcement learning~\citep{Sutton2018}, the agent is on a certain state $s_t$ and interacts with the environment at 
time $t$ through a series of actions $a_t$. 
Once action $a_t$ is selected, the agent receives the reward $r(a_t,s_t)$.
With the reward, the agent moves to state $s_{t+1}$ and the decision making problem starts again, but now at state
$s_{t+1}$ and time $t+1$.
The goal of reinforcement learning is to develop a sequential decision making solution to maximize the reward given to
the agent by the environment.
Notice that reinforcement learning is neither supervised because the learning algorithm does not have the optimal
action $a_t$ to select in state $s_t$, nor unsupervised because the reward is given when selecting action $a_t$ in
state $s_t$~\citep{Simeone2018}.
Moreover, reinforcement learning allows the environment to influence future states and rewards based on the previous
actions; a feature that is present in neither supervised nor unsupervised learning.
Figure~\ref{fig:reinforcement_ex} shows an example of a user interacting with the environment at state $s_t$,
which represents the resource currently used, through a series of actions $a_t$, which represents the series of
resources used until time $t$, and the reward $r(a_t,s_t)$, which may represent the achievable spectral efficiency at
state $s_t$.

\section{Supervised Learning}\label{sub:sup_learning}

We discuss herein problems in the context of supervised learning, i.e., the problem of labelling unseen
data based on information from a set of labelled training data~\citep{bottou2018}.
The common learning goal is to represent a prediction function $h:\calX\rightarrow \calY$ from an input space $\calX$
to an output space $\calY$ such that, given $\vtX\in\calX$, the value $h(\vtX)$ offers an accurate prediction, also
termed as hypothesis, about the true output $y \in \calY$.
Hence, the prediction function $h$ should minimize a risk measure over an adequately selected family of prediction
functions, termed $\calH$.
Rather than optimizing over a generic family of prediction functions, it is commonly assumed that the prediction
function $h$ has a fixed form and it is parameterized by a real vector $\vtW\in\bbR^d$.
Then, for some $h(\cdot;\cdot):\bbR^{d_x}\times\bbR^d\rightarrow\bbR^{d_y}$, the family of prediction functions is
$\calH\triangleq \{h(\cdot;\vtW): \vtW\in\bbR^d\}$.

The aim is then to obtain the prediction function in the family $\calH$ that minimizes the losses due to inaccurate
predictions.
To this end, we assume a given loss function $l:\bbR^{d_y}\times\bbR^{d_y}\rightarrow\bbR$ that given an
input-output pair $(\vtX,\vtY)$, yields the loss $l\big(h(\vtX;\vtW),\vtY\big)$~\citep{bottou2018}.
Notice that $h(\vtX;\vtW)$ and $\vtY$ represent the predicted and true outputs, respectively.
The parameter $\vtW$ is chosen such that the expected loss incurred from any input-output pair is minimized.
The loss functions $l(\cdot;\vtW)$ can be either convex on $\vtW$, such as when used for linear regression and binary
classification (linear \ac{SVM}), or nonconvex, such as when used for image classification using neural networks with
several layers.

Let us assume that the losses are measured with respect to a probability distribution $\Pr(\vtX,\vtY)$ in the
input-output space $\bbR^{d_x}\times\bbR^{d_y}$, i.e., $P:\bbR^{d_x}\times\bbR^{d_y}\rightarrow[0,1]$.
Then, the objective function we want to minimize is
\begin{equation}\label{eq:expected_risk}
R(\vtW) = \int_{\bbR^{d_x}\times\bbR^{d_y}} l(h(\vtX;\vtW),\vtY) d \Pr(\vtX,\vtY) = \bbE[l(h(\vtX;\vtW),\vtY)],
\end{equation}
in which $R: \bbR^d\rightarrow \bbR$ is the expected risk given a parameter vector $\vtW$ with respect to the
probability distribution $\Pr(\vtX,\vtY)$.
The minimum expected risk, denoted by $R(\vtW^\star)$ with $\vtW^\star \coloneqq \ArgMin{\vtW}{R(\vtW)}$, is also
known as the \emph{test} or \emph{generalization error}.
To minimize the expected risk in Eq.~\eqref{eq:expected_risk}, it is necessary to have complete information about the
probability distribution $P$ of the input-output pair.
However, such minimization is not possible in most situations because complete information of $P$ is not available.
Due to this reason, the practical goal becomes the minimization of an estimation of the expected risk $R$.
To this end, we assume that there are $N\in\bbN$ independently drawn input-output samples $\{(\vtX_i,\vtY_i)\}_{i=1}^N
\subseteq \bbR^{d_x}\times\bbR^{d_y}$, and we define the empirical risk function $R_N:\bbR^d\rightarrow\bbR$ as
\begin{equation}\label{eq:empirical_risk}
R_N(\vtW) = \frac{1}{N} \sum\nolimits_{i=1}^{N} l( h(\vtX_i;\vtW),\vtY_i ).
\end{equation}
With the empirical risk, the optimization problem is written as follows:
\begin{align}\label{eq:problem_emp_risk}
\underset{\vtW}{\text{minimize}}\quad
& \frac{1}{N} \sum_{i=1}^{N} l( h(\vtX_i;\vtW),\vtY_i ),
\end{align}
in which the minimization of $R_N$ is the practical optimization problem that needs to be solved when performing
supervised learning.
The minimum empirical risk is also known as the \emph{training error} and can be understood as an estimation of the
\emph{test error}~\citep{Hastie2009}.

To solve the optimization problem in~\eqref{eq:problem_emp_risk}, several optimization algorithms have been proposed using
stochastic optimization methods, such as \ac{SGD}, with or without the use of data partition in
batches~\citep{bottou2018}.
A general \ac{SGD} method solves iteratively the optimization problem in~\eqref{eq:problem_emp_risk}, with iterations given by
\begin{equation}\label{eq:general_sgd}
\vtW_{t+1} \gets \vtW_t -\eta \sum_{i_t=1}^{n} \nabla f_{i_t}(\vtW_t), \forall t\in\bbN,
\end{equation}
where $\vtW_t\in\bbR^d$, $\eta$ is the learning rate, and $f_i(\vtW)$ is the composition of the loss function $l$
and $h$ evaluated at sample $i$.
The sum in~\eqref{eq:general_sgd} may represent pure \ac{SGD}, batch gradient descent, or a joint approach with
mini-batch \ac{SGD}~\citep{bottou2018}.
For $n=1$, Eq.~\eqref{eq:general_sgd} is generally called \emph{model update} and represents the pure \ac{SGD} method,
and the index $i_t$, which corresponds to the
seed $\xi_{[i_t]}$ of the sample pair $(\vtX_{i_t},\vtY_{i_t})$, is chosen randomly from $\{1,\ldots,N\}$.
For $n=N$, Eq.~\eqref{eq:general_sgd} represents the batch gradient descent method, in which the gradient is
evaluated for all samples $N$ and taken into account at each iteration $t$.
For $1 < n < N$, Eq.~\eqref{eq:general_sgd} represents the mini-batch \ac{SGD} method, in which a set $\calS_t 
\subseteq
\{1,\ldots,N\}$ of cardinality $n$ is chosen randomly at each iteration $i_t$.
The iterations are evaluated until they reach a minimizer of the empirical risk $R_N$.

In Table~\ref{tab:loss_func}, we present some loss functions that are commonly used in the spectrum sharing
literature, such as the \ac{MSE}, cross-entropy, and hinge loss.
One common aspect to the different models in Table~\ref{tab:loss_func} is the regularization function $r(\vtW)$, which
is intended to reduce the overfitting, which is when the model is overly trained to the given dataset and does not
generalize well when using data not part of the training dataset.
To promote sparsity and reduce model dimensionality, the regularization function is often assumed as $r(\vtW) =
\lambda\NormOne{\vtW}$, where $\lambda$ is the regularization constant; whereas to reduce the large variations of
vector $\vtW$, the regularization is often assumed as $r(\vtW) = \lambda\NormTwo{\vtW}^2$, where $\NormOne{\cdot}$ and
$\NormTwo{\cdot}$ are one-norm and two-norm, respectively.
Similarly, Table~\ref{tab:pred_func} shows the prediction functions commonly used in spectrum sharing problems
across different \ac{ML} models.
In the following, we present the most used \ac{ML} models in spectrum sharing problems.
\begin{table*}
\caption{Common loss functions $l(h(\vtX_i;\vtW),\vtY_i)$ in spectrum sharing problems.}\label{tab:loss_func}
\centering
{\footnotesize \begin{tabular}{|c|c|}%
	\hline
	\textbf{Loss Function} 					  & \textbf{Definition} \\
	\hline
	\ac{MSE}                                  & $(y_i - h(\vtX_i;\vtW))^2 + r(\vtW)$ \\
	\hline
	Cross-entropy (binary) 					  & $-y_i\log(h(\vtX_i;\vtW)) - (1-y_i)\log(1-h(\vtX_i;\vtW)) +
	r(\vtW)$\\
	\hline
	Hinge loss                                & $\max \left(0, 1-y_i(h(\vtX_i;\vtW))\right) + r(\vtW)$\\
	\hline
	Rayleigh quotient                         & $\frac{\vtW^T \mtA \vtW }{\vtW^T \mtB \vtW}$\\
	\hline
	Misclassification error                   & $1-\hat{p}_{mk},$ with $\hat{p}_{mk}=\frac{1}{N_m}\sum_{x_i\in\calR_m}
	I(y_i=k)$\\
	\hline
	Gini index                   			  & $\sum_{k\neq k'} \hat{p}_{mk} \hat{p}_{mk'}$\\
	\hline
\end{tabular}}
\end{table*}

\subsection{$k$-Nearest Neighbour}\label{subsub:k-nn}
The \ac{k-NN} is among the simplest and most popular supervised learning techniques. 
The goal of \ac{k-NN} is to predict both qualitative (discrete) and quantitative (continuous) outputs using simply the 
inputs from the neighbour samples.
The prediction function is assumed as $h(\vtX,\vtY;\vtW)=\frac{1}{k}\sum_{\vtX_i\in \calN_k(\vtX)} \vtY_i$, where the 
scalar $k$ specifies the $k$ closest points to $\vtX_i$ that will be considered. 
Moreover, the set $\calN_k(\vtX)$ is the neighbourhood of $\vtX$ defined by the $k$ closest points $\vtX_i$ in the 
training set.
Note that the metric to define the closest points can vary, but the most popular one is the Euclidean distance with 
$d(\vtX_i,\vtX_j)=\NormTwo{\vtX_i - \vtX_j}$.
Moreover, the weight parameter $\vtW$ can be understood as the scalar $k$, which is usually a design choice and not 
trained.
Therefore, this implies that the \ac{k-NN} method does not have a clear definition of loss function.

\begin{table*}
    \caption{Common prediction function $h(\vtX_i,\vtW)$ in spectrum sharing problems.}\label{tab:pred_func}
    \centering
    {\footnotesize \begin{tabular}{|c|c|c|c|}
            \hline
            \textbf{\ac{ML} Model} & \textbf{Loss Function}  & \textbf{Prediction Function}\\
            \hline
            $k$-Nearest Neighbour  & -                       & $\frac{1}{k}\sum_{\vtX_i\in \calN_k(\vtX)} \vtY_i$\\
            \hline
            Linear Regression      & \ac{MSE}                & $\vtW^T\vtX_i$\\
            \hline
            \parbox[c]{2.5cm}{\centering\vspace*{0.05cm}Logistic Regression (binary)} & Cross-entropy       & $\vtW^T\vtX_i$\\
            \hline
            \parbox[c]{2.5cm}{\centering\vspace*{0.05cm}Fisher discriminant analysis (binary)} & Rayleigh quotient & $\vtW^T\vtX_i$\\
            \hline
            Bayes classifier & Rayleigh quotient & $\vtW^T\vtX_i$\\	
            \hline		
            \ac{SVM} (binary) & Hinge loss & $\vtW^T\vtX_i-b$ \\
            \hline
            Polynomial \ac{SVM} & Hinge loss  & $(1+\vtW^T\vtX_i)^d$ \\
            \hline
            Radial \ac{SVM} & Hinge loss  & $\exp (-\gamma\Norm{\vtW-\vtX_i}^2)$ \\
            \hline
            Decision Trees (regression)   & \ac{MSE} & $\sum_{m=1}^{M} c_m I(\vtX_i\in \calR_m)$\\
            \hline
            \parbox[c]{2.5cm}{\centering\vspace*{0.05cm}Decision Trees (classification)}   & \parbox[c]{3cm}{\centering\vspace*{0.05cm} 
            Misclassification error, Gini index
                and cross-entropy} & $\ArgMax{k}{\hat{p}_{mk}}$\\
            \hline
            \acs{HMM} (decoding)   & Misclassification error & $\ArgMax{Q_1,\ldots,Q_{T-1}}{\Pr(Q|O,\lambda)}$\\
            \hline
            \acs{ANN}  & \ac{MSE} and cross-entropy & $\vtR_{L}$ (see Eq.~\eqref{eq:predic_func_ann}) \\
            \hline
            \acs{CNN}  & \ac{MSE} and cross-entropy & $\vtR_{L}$ (see Eq.~\eqref{eq:predic_func_cnn})\\
            \hline
            \acs{RNN}  & \ac{MSE} and cross-entropy & See Eqs.~\eqref{eq:predic_func_rnn}\\
            \hline
            \acs{LSTM} & \ac{MSE} and cross-entropy & See Eqs.~\eqref{eq:predic_func_lstm}\\
            \hline
    \end{tabular}}
\end{table*}
\subsection{Linear and Logistic Regression}\label{subsub:lin_log_regre}
In linear regression, the objective is to predict quantitative (continuous) outputs and to understand how the inputs
affect the output~\citep{Hastie2009}.
The loss function is the \ac{MSE} between the real output $y_i$ and the output provided by the
prediction function $h(\vtX;\vtW)$.
Note that the prediction function is assumed as $h(\vtX;\vtW) = \vtW^T\vtX$, which converts the minimization of
the loss function into a least squares estimation problem.
Differently from linear regression, the logistic regression is heavily used for classification problems.
Its main objective is to predict qualitative (discrete) outputs and to understand the role of the input variables in
explaining the outcome.
The loss function is the cross-entropy between the real output $y_i$ and the output provided by the
prediction function $h(\vtX;\vtW)$, which similar to the linear regression, is assumed as linear on $\vtW$.
Moreover, the logistic regression in Table~\ref{tab:pred_func} represents a binary classification, whereas a multiclass
classification problem can be modeled similarly and the problem can be solved iteratively as a binary classification
problem with one class being the selected one and the other class representing all the others.

\subsection{Fisher Discriminant Analysis}\label{subsub:fisher_analysis}
Similarly, Fisher linear discriminant analysis considers a linear prediction function $\vtW^T\vtX_i$,
focuses on classification tasks, and aims at separating different classes using a hyperplane.
Instead of separating the classes via \ac{MSE} as linear regression, Fisher discriminant analysis aims at maximizing
the class separation via the centroids of the classes while also giving a small variance within each
class~\citep{Bishop2006,Hastie2009}.
Thus, the overlap between classes is also minimized.

For a binary classification problem, let us define the centroids as $\vtM_0 = \frac{1}{N_0} \sum_{i\in\calC_0} \vtX_i$ 
and $\vtM_1 = \frac{1}{N_1} \sum_{i\in\calC_1} \vtX_i$, where $N_0, N_1$ are the number of samples in classes $0$ 
and $1$, and $\calC_0,\calC_1$ are the set of samples for classes $0$ and $1$, respectively.
Let us define the between-class covariance matrix, $\mtS_B\in\fdR^{d\times d}$, and the total within-class
covariance matrix, $\mtS_W\in\fdR^{d\times d}$, as follows
\begin{align}\label{eq:fisher_cov_mat}
\mtS_B &= \left(\vtM_1 - \vtM_1 \right) \left(\vtM_1 - \vtM_1 \right)^T,\\
\mtS_W &= \sum_{i\in\calC_0} \left(\vtX_i - \vtM_0 \right) \left(\vtX_i - \vtM_0 \right)^T +\sum_{j\in\calC_1} \left(\vtX_j - \vtM_1 \right) 
\left(\vtX_j - \vtM_1 \right)^T.
\end{align}
The objective of Fisher linear discriminant analysis can be formulated as the following Rayleigh quotient problem
\begin{align}\label{eq:fisher_problem_form}
\underset{\vtW}{\text{maximize}}\quad
& \frac{\vtW^T \mtS_B \vtW }{\vtW^T \mtS_W \vtW}.
\end{align}
Problem~\eqref{eq:fisher_problem_form} is a generalized eigenvalue problem whose optimal solution, $\vtW^{\star}$, is
the eigenvector with the largest eigenvalue of matrix $\mtS_W^{-1}\mtS_B$.
Fisher linear discriminant analysis can also be applied in multiclass classification problems, and we refer the
interested reader to~\cite[Section 4.1.6]{Bishop2006}.

\subsection{Bayes Classifier}\label{subsub:bayes_class}
Another classical classification technique is the Bayes classifier, also called Naive Bayes
classifier~\citep{Hastie2009,Jurafsky2020}.
The Bayes classifier is probabilistic, which means that for a sample $\vtX_i$ to be classified within all classes
$c\in\calC=\{1,\ldots,C\}$, the classifier returns the class $\hat{c}$ which has the maximum posterior
probability in the sample as
\begin{align}\label{eq:gen_Bayes}
\hat{c} = \arg\!\max_{c\in\calC} \Pr(y_i=c|\vtX_i).
\end{align}
To obtain the posterior probability, we can use Bayes' rule as follows
\begin{align}\label{eq:Bayes_rule}
\arg\!\max_{c\in\calC}\Pr(y_i=c|\vtX_i) &= \arg\!\max_{c\in\calC} \frac{\Pr(\vtX_i|y_i=c)\Pr(y_i=c)}{\Pr(\vtX_i)},\\
&= \arg\!\max_{c\in\calC} \Pr(\vtX_i|y_i=c)\Pr(y_i=c),
\end{align}
where we dropped the term in the denominator because $\Pr(\vtX_i)$ is common to all the terms to be maximized.
The probabilities are unknown, but at least $\Pr(y_i=c)$ can be estimated from all samples in the dataset.
However, $\Pr(\vtX_i|y_i=c)$ remains too complicated to be estimated directly.
To overcome this problem, Bayes classifier considers that all the $d$ dimensions of the feature space of $\vtX_i$ are
independent given $y_i$, which means that the probability $\Pr(\vtX_i|y_i=c)$ can be \emph{naively} considered as
follows
\begin{equation}\label{eq:naive_Bayes_indep}
\Pr(\vtX_i|y_i=c) = \prod_{l=1}^{d} \Pr(x_{il}|y_i=c),
\end{equation}
where $\Pr(x_{il}|y_i=c)$ is the probability that sample $i$ from feature $l$ belongs to class $c$.
To obtain the probabilities, we can use a parameter estimation based on the frequency of occurrence in the training set
such that $\Pr(y_i=c)$ and $\Pr(x_{il}|y_i=c)$ are estimated as
\begin{align}\label{eq:param_Bayes_estim}
\Pr(y_i=c) = \frac{N_c}{N},\quad \Pr(x_{il}|y_i=c) = \frac{N_{x_{il},c}+1}{N_{x_{il}} + N},
\end{align}
where $N_c$ is the number of samples that belong to class $c$; $N$ is the total number of samples; $N_{x_{il},c}$ is
the number of occurrences of feature $x_{il}$ in class $c$; $N_{x_{il}}$ is the number of occurrences of feature
$x_{il}$ among all classes for the whole training set; and the additive terms in the numerator and denominator, $1$ and
$N$ respectively, are smoothing terms to avoid zero probabilities for any classes.
To further increase speed and avoid underflow when evaluating $\Pr(y_i=c)$ in Eq.~\eqref{eq:naive_Bayes_indep}, the
log-space is considered such that the product becomes a sum.
Thus, we can write the Bayes classifier in the log-space as
\begin{align}\label{eq:upd_Bayes_class}
\hat{c} = \arg\!\max_{c\in\calC} \quad \log \Pr(y_i=c) + \sum_{l=1}^{d} \log \Pr(x_{il}|y_i=c).
\end{align}
Note that the Bayes classifier is linear in the input features $x_{il}$ in the log-space, and due to this reason it is
also considered as a linear classifier.

\subsection{Support Vector Machine}\label{subsub:svm}
Using a linear prediction functions with a bias $b$ as some methods above, the \ac{SVM} is used to
provide nonlinear decision boundaries when the two classes (in the binary case) are non-separable.
The \ac{SVM} function in Table~\ref{tab:loss_func} refers to a binary classification, and the multiclass classification
can be done in a similar manner to the logistic regression.
The objective function of \ac{SVM} is the hinge loss, which is convex and nonsmooth thus suggesting the use of
subgradient method to obtain its minimizer.
In addition, the prediction function may be different than the linear $\vtW^T\vtX$ over $\vtW$, such as polynomial
with $d$-degree with $h(\vtX;\vtW)=(1+\vtW^T\vtX)^d$, and radial (also known as Gaussian) with $h(\vtX;\vtW) = \exp
(-\gamma\Norm{\vtW-\vtX}^2)$, where $\gamma$ is a parameter that controls the variance of the model and the smoothness
of the decision boundary.
Such examples are part of the polynomial kernel \ac{SVM}, whereas the first \ac{SVM} mentioned in
Table~\ref{tab:pred_func} is the linear kernel \ac{SVM}.

\subsection{Decision Trees and Random Forest}\label{subsub:random_forest}
Another popular classification and regression method is random forest~\citep{Hastie2009}, which builds on two other
methods: decision trees and bagging.
Hence, we describe briefly decision trees and bagging before introducing random forest.
Decision trees partition the feature space into a set of regions and fit a simple model, such as a constant, to each
region, in which the partitions are usually binary because it does not fragment the data too quickly.
Due to this, the method can be represented as a binary tree such that samples belonging to a region are assigned to the
left branch, and the others to the right branch.
For regression problems, we can define the prediction function for a sample $\vtX_i$
\begin{equation}
h(\vtX_i) = \sum_{m=1}^{M} c_m I(\vtX_i\in \calR_m),
\end{equation}
where $M$ is the number of regions that will be partitioned, $c_m$ is the constant that will model region $m$,
$\calR_m$ is the set that represents region $R_m$, and $I(\vtX_i)$ is the indicator function, i.e., $I(\vtX_i)=1$ if
$\vtX_i\in\calR_m$ and $0$ otherwise.
If the objective function is the \ac{MSE}, the optimal $\hat{c}_m$ is the average of the output
$y_i,\;i=1,\ldots,N$ in the region $\calR_m$: $\hat{c}_m = \text{ave}(y_i|\vtX_i\in\calR_m)$.
The tree size $M$ is usually a tuning parameter that controls the complexity of the model, and the optimal tree size
can be obtained adaptively from the data.
To obtain the regions $\calR_m$, the \ac{MSE} is generally computationally infeasible and often a greedy algorithm is
used.
For further information about the optimization of the regions and tree size, we refer the interested
reader to~\cite[Section 9.2.2]{Hastie2009}.
For classification problems, the modifications needed in the method are related to the criteria for splitting
nodes and optimizing the tree size $M$, which are discussed with more details in~\cite[Section 9.2.2]{Hastie2009}.
Hence, it is necessary to use a different loss function.
Let us denote by $N_m$ the number of samples that belong to region $\calR_m$, and by $m$ the terminal node in region
$\calR_m$.
Then, we define the proportion of class $k$ samples in node $m$ as
\begin{equation}
\hat{p}_{mk}=\dfrac{1}{N_m}\sum_{\vtX_i\in\calR_m} I(y_i=k).
\end{equation}
We classify the samples in node $m$ to class $k(m)$ as $k(m)=\ArgMax{k}{\hat{p}_{mk}}$, thus being the prediction
function for a sample $x_i$.
Based on this, the different loss functions (shown in Table~\ref{tab:loss_func}) are the misclassification error, the
Gini index, and the cross-entropy.
These functions are similar, but the Gini index and  cross-entropy are differentiable, which is desirable in numerical
optimization.

Overall, decision trees have a high variance and a small change in the data can result in very different data splits.
To alleviate this problem, a popular option is to use bagging to average many trees and reduce the overall variance of
the model~\citep[Section 8.7]{Hastie2009}.
For the training dataset as $\calD = \{(\vtX_i,y_i)\}_{i=1}^N$, let us randomly draw $B$ datasets, each of size $N_B$,
with replacement from the training dataset $\calD$, in which we refer to the $b$-th dataset as $\calD_b,\;b=1,\ldots,B$.
Specifically, each dataset $\calD_b$ is also referred to as a bootstrap dataset in the literature.
Then, we fit our model for each dataset $\calD_b$, providing predictions $h^{\star b}(\vtX),\; b=1,\ldots,B$.
With this, we define the bagging estimate as
\begin{align}
h_{\text{bag}}(\vtX_i) = \frac{1}{B} \sum_{b=1}^{B} h^{\star b}(\vtX_i).
\end{align}
Note that bagging averages the prediction of a collection of sample (bootstrap) datasets, thereby reducing the variance
of the estimated prediction function for a particular model.
Moreover, bagging can be applied to several types of model, and one of the popular uses is exactly with decision trees
for regression and classification.
For regression, the bagging is the average of a regression tree using the bootstrap sampled versions of the training
data; whereas for classification it is predicted class that received more votes in a committee of trees.

Towards the reduction of the variance in decision tree methods, random forest was proposed as a substantial
modification of bagging that averages a collection of decorrelated trees~\citep[Section 15.1]{Hastie2009}.
The objective of random forest is to further reduce the variance from bagging trees by reducing the correlation between
the trees, but without increasing the variance too much.
This can be achieved by random selection of the input variables used when growing the trees.
Specifically, select $d_m$ input variables at random from the $d_x$ input variables for each bagging prediction $b$ and
create a tree prediction for this bootstrap dataset $\calD_b$.
After growing $B$ trees, the random forest predictor for regression and classification problems are:
\begin{align}
h^B_{\text{rf}}(\vtX)&=\frac{1}{B} \sum_{b=1}^{B}  h^{\star b}_{\text{rf}}(\vtX_i),\\
h^B_{\text{rf}}(\vtX)&=\texttt{maj}\{ h^{\star b}_{\text{crf}}(\vtX_i)\}_1^B,
\end{align}
where $h^{\star b}_{\text{rf}}(\vtX_i)$ is the predictor function for the tree using the random forest sampling
mentioned above for regression problems, $h^{\star b}_{\text{crf}}(\vtX_i)$ is the class prediction of the $b$-th
random forest tree using the sampling mentioned above for classification problems, and
$\texttt{maj}\{\cdot\}_b^B$ is the majority vote for a class considering all the bootstrap datasets B.
Note that the parameter $d_m$ is a design choice, and typical values for regression problems are $\Floor{\sqrt{d_x}}$
or as low as $1$; and for classification problems are $\Floor{d_x/3}$ or as low as $5$.
Nowadays, random forest methods are popular due to its computational simplicity, variance reduction compared to normal
decision tree methods, and the low tuning parameters required.

\subsection{Hidden Markov Models}\label{subsub:hidden_markov}
Another interesting modelling and learning technique is the \ac{HMM}, which can be used for supervised and unsupervised
learning techniques.
In the following, we discuss the three main categories of \ac{HMM} that can be used for supervised and
unsupervised learning.
To understand \ac{HMM}, first we need to explain Markov chains. A Markov Chain is a stochastic model that considers
stochastic processes in which future events depend on and can be predicted, to some extent, by the most recent past
event~\citep{Bertsekas2002,Jurafsky2020}.
Formally, a Markov chain is defined by: a set of $N$ states $\calS=\{S_1,\;S_2,\;\cdots, S_N\}$; a transition
probability matrix $\mtA\in\bbR^{N\times N}$, where each $a_{ij}$ represents the probability of transitioning from
state $i$ to state $j$ and with $a_{ij} \geq 0,\;\forall i,j\in\calS$ and $\sum_{j=1}^{N} a_{ij}=1,\;\forall
i\in\calS$; an initial probability distribution over states, $\pi_i$, which states the probability that the Markov
chain will start in state $i$ with $\pi_i\geq 0,\;\forall i,$ and $\sum_{i=1}^{N} \pi_i=1$; and the Markov property,
which states that a sequence of events $S_0,S_1,S_2,\ldots$ that take values in $\calS$ satisfy
$\Pr(S_{n+1}=j|S_n=i,S_{n-1}=i_{n-1},\ldots,S_0=i_0)= \Pr(S_{n+1}=j|S_n=i) = a_{ij}$ for all times $n$, states
$i,j\in\calS$,
and possible sequences of earlier states $i_0,\ldots,i_{n-1}$.
The use of Markov chains is useful when dealing with a sequence of observable events, but in many applications we must
deal with \emph{hidden} events (states), i.e., not observable directly.
For such hidden events, we need to use \ac{HMM} to quantify the hidden events (states) in the probabilistic model.

With respect to \ac{HMM}~\citep{Rabiner1989,Jurafsky2020}, they have the same properties of a Markov chain but
with the inclusion of the following properties~\citep[Appendix A]{Jurafsky2020}: a set $\calQ=\{Q_1,Q_2,\ldots,Q_N\}$ of
$N$ states, which are used to represent the hidden states; a sequence of $T$ observations
$\calO=\{=O_1,O_2,\ldots,O_T\}$, each one drawn from a vocabulary $\calV=\{V_1,V_2,\ldots,V_V\}$; a sequence of
observation likelihoods, also called emission probabilities $B=B_i(O_t)$, each expressing the probability of an
observation $O_t$ being generated from a state $i$; the output independence property, which states that the probability
of an output observation $O_i$ depends only on the state that produced the observation $S_i$ and not on any other
states or any other observations, $\Pr(O_i|Q_1,\ldots,Q_T,O_1,\ldots,O_T)= \Pr(O_i|Q_i)$.
From the seminal tutorial in~\cite{Rabiner1989}, \ac{HMM} can be characterized by three fundamental problems: 
likelihood,
decoding, and learning.

In a likelihood problem, given an \ac{HMM} $\lambda=(A,B)$ and an observation sequence $\calO$, the goal is to
determine the likelihood $\Pr(O|\lambda)$ without knowing the hidden state sequence $\calQ$.
Since $\calQ$ is unknown, the probability of an observation sequence needs to be computed over all possible hidden
state sequences and weighted by their respective probability.
For an \ac{HMM} with $N$ hidden states and observation sequence of $T$ observations, there are $N^T$ possible hidden
sequences.
The likelihood calculations can be computationally prohibitive for practical scenarios with large number of
observations or number of hidden states.
The forward (or backward) algorithm is the most common solution to this problem, which uses a single forward trellis to
make the calculations in a dynamic programming fashion and with a computational complexity of $O(N^2T)$.
We refer to~\cite[Appendix A.3]{Jurafsky2020} for a detailed explanation of the forward algorithm.

In a decoding problem, given an \ac{HMM} $\lambda=(A,B)$ and an observation sequence $\calO$, the objective is to
determine the best hidden state sequence $\calQ^\star=\{Q_1,\ldots,Q_T\}$ such that $Q^\star
=\ArgMax{Q_1,\ldots,Q_t}{\Pr(Q|O,\lambda)}$.
The decoding term refers to the task of determining which sequence of variables in a model is the underlying source of
some sequence of observations.
In the \ac{HMM} case, we could run the forward algorithm and compute the likelihood of an observation sequence given
the hidden state sequence.
With this, we could simply choose the hidden state sequence with the maximum likelihood.
However, this is computationally prohibitive given the exponentially large number of state sequences.
The most common decoding algorithm in \ac{HMM} is the Viterbi algorithm, which is based on dynamic programming and
makes uses of a dynamic programming trellis with computational complexity of $O(N^2T)$.
We refer to~\cite[Appendix A.4]{Jurafsky2020} for a detailed explanation of the Viterbi algorithm.

In a learning problem, given an observation sequence $\calO$ and the set of states $\calQ$ in the \ac{HMM}, learn the
\ac{HMM} transition probabilities $A$ and the emission probabilities $B$.
The input to this problem is an unlabelled sequence of observations $\calO$ and a vocabulary of potential hidden states
$\calQ$.
The standard algorithm for this \ac{HMM} learning problem is the forward-backward (or Baum-Welch), which is discussed
with more details in~\cite[Appendix A.5]{Jurafsky2020}.

\subsection{Feedforward Neural Networks}\label{subsub:ANN}
Different from the \ac{HMM}, the central idea in \acp{ANN} is to extract combinations of the inputs as derived
features and then model the target as a nonlinear function of these features.
The result is a powerful learning method that can be applied into regression or classification models~\citep{Hastie2009}.
For regression, the common loss function is the \ac{MSE}, whereas for classification the loss function is the
cross-entropy.
An \ac{ANN} is composed of a predefined number of layers $L+2$, each with a predefined number of nodes $N_l$ with
$l=0,\ldots,L+1$, also known as \emph{neurons}, that perform an affine operation followed by a point-wise nonlinear
operation, also known as \emph{activation function}.
The zero-th layer is defined as the \emph{input layer}, whereas the last layer $L+1$ is known as the \emph{output
layer}.
The $L$ layers between both the input and output layers are defined as the \emph{hidden layers} of the
\ac{ANN}.\\
\indent The layer $l$ is defined by the weights $\mtW_l\in\bbR^{N_{l-1}\times N_l}$, a bias term
$\vtB_l\in\bbR^{N_l}$, and the activation function $\sigma_l:\bbR^{N_l}\rightarrow\bbR^{N_l}$.
When layer $l$ receives an input $\vtR_{l-1}\in\bbR^{N_{l-1}}$ from layer $l-1$, the resulting output
$\vtR_l\!\!\in\!\!\bbR^{N_l}$ is evaluated as $\vtR_l\!\!\vcentcolon=\!\! \sigma_l \left(\mtW_l\vtR_{l-1} + \vtB_l
\right)$.
The processing at each layer can be viewed as two steps: first, an affine combination of the inputs with the weights
and bias; second, a nonlinear point-wise operation by applying the activation function.
For the last hidden layer $L$, the output is defined as
\begin{align}\label{eq:predic_func_ann}
\vtR_{L} =& \sigma_{L} \big(\mtW_{L}(\sigma_{L-1}(\mtW_{L-1}(\ldots(\sigma_1(\mtW_1\vtR_0 +
\vtB_0)\ldots) + \nonumber\\
&\hspace*{1.4cm} \vtB_{L-1})))\allowbreak \! +\! \vtB_{L}\! \big),
\end{align}
where $\vtR_0$ is the input $\vtX_i$ for a sample $i$.
Since $\vtR_0=\vtX_i$, it is commonly assumed the weight matrix $\mtW_0$ is an identity matrix.
The \ac{ANN} defined above is the classical feedforward neural network with fully-connected layers, also known as
multilayer perceptron, which is the most common \ac{ANN} model used in literature.
If the \ac{ANN} has $L=1$, i.e., a single hidden layer, it is referred to as a \emph{shallow network}; whereas if
$L>1$, it is referred to as a \emph{\acl{DNN}} (\acs{DNN})\acused{DNN}.\\
\indent
From Eq.~\eqref{eq:predic_func_ann}, the prediction function  $h(\vtX;\mtW)$ can be understood as $\vtR_{L}$, given
that it is the final output of the model.
It represents a series of linear product operations between the weight of current layer, $\mtW_L$, and the input to the
current layer from previous layer, $\vtR_{L-1}$.
Note that the prediction function $h(\vtX;\mtW)$ is nonlinear in the vast majority of the cases due to the activation
function.
There are many choices of activation function nowadays and the most popular are the sigmoidal, hyperbolic
tangent, softmax, and \ac{ReLU} functions~\citep{Zappone2019}.
Their applicability depend on whether layer $l$ is an input, hidden, or output layer, as well as on the structure of
the neural network.

The nonlinear operations are important because they represent the universal approximation characteristic, which was
formally established in~\cite{Hornik1989}.
This is known as the \emph{universal approximation theorem} of \ac{ANN}, which states that an \ac{ANN} can approximate
arbitrarily well any deterministic continuous function over a compact set by using even a single layer with enough
neurons and different activation functions~\citep{Hornik1989,Zappone2019}.
Moreover, several empirical results have shown that deep neural networks seem to require a lower number of
neurons and provide lower test errors than shallow architectures~\citep[Sec. 6.4.1]{goodfellow2016book}, which explain
the vast applicability of \ac{DNN} nowadays in several scientific fields.
In practice, the weights $\mtW_l$ and bias $\vtB_l$ of layer $l$ are obtained using \ac{SGD} together with
back-propagation~\citep[Sec. 6.5]{goodfellow2016book}.

\subsection{Convolutional Neural Networks}\label{subsub:CNN}
Different from feedforward neural networks, the other most popular architectures are \acp{CNN} and
\acp{RNN}. 
The \ac{CNN} architecture was introduced as a means to deal with inputs that have a spatial or grid structure, such as
images and time series, and to handle classification tasks.
Hence, the input $\vtX_i$ is usually represented as a matrix $\mtX_i\in\bbR^{d\times d\times N_c}$, where $N_c$ is the
number of channels and is typically equal to $1$ or $3$ if black-and-white or color images are processed, respectively.
Similarly, the weight $\vtW$ is represented as a matrix $\mtW\in\bbR^{F\times F\times N_c}$, where $F\leq d$ 
represents the dimension of the sliding window of a convolution operation.
Instead of using matrix multiplication between the weight $\mtW$ and the input $\mtX_i$, \acp{CNN} perform a
discrete convolution operation, between the input and weights, resulting in an output matrix
$\mtY_i\in\bbR^{d-F+1\times d-F+1}$ whose $(l,m)$-th element is defined as~\citep{Zappone2019}
\begin{align}\label{eq:cnn_conv_oper}
\mtY_i(l,m) = \sum_{j=1}^{F} \sum_{k=1}^{F} \sum_{n=1}^{N_c} \mtW(j,k,n) \mtX(j+l-1,k+m-1,n).
\end{align}
As can be seen from Eq.~\eqref{eq:cnn_conv_oper}, \acp{CNN} perform a cross-correlation instead of a pure convolution.
Nevertheless, the name \emph{convolutional} remains because the cross-correlation simply represents a partial 
convolution, i.e., no flipping of the matrix $\mtW$ , and because it does not typically affect efficacy while bringing 
practical benefits~\citep{goodfellow2016book}.
Together with the convolution operation, which is an affine transformation, the nonlinear activation function
$\sigma(\cdot)$ is applied point-wise, in which the most commonly used is the \ac{ReLU}.
Then, the output of the activation goes through a \emph{pooling} operation to make the representation become
approximately invariant to small translations of the input.
The most common pooling operations used are the \emph{max pooling}, resulting in the maximum over a rectangular
neighbourhood, and the \emph{average pooling}, resulting in the average over a rectangular neighbourhood.
A typical \ac{CNN} $l$-th layer is represented by the cascade of these three operations: convolution ($\ast$),
activation layer, and pooling ($\verb|pool|$); whose mathematical definition is
\begin{align}\label{eq:cnn_layer}
\mtR_{L} &= \texttt{pool}\big(\sigma_{L}\big(\mtW_{L} \ast \mtV_{L-1}\big) \big).
\end{align}
In practice, it is common to use several \ac{CNN} layers in the first layers, decreasing the size of the data, and
employ fully-connected layers at the second to last layer, benefiting from a more manageable dimension of the data.
Finally, a softmax activation function at the end is commonly used to convert the predictions into probabilities.
Thus, the prediction function for a \ac{CNN} with $L$ layers can be understood as the aforementioned cascade of
nonlinear and affine operations, which is represented as
\begin{align}\label{eq:predic_func_cnn}
\vtR_{L+2} &= \texttt{softmax}\big(\vtW_{L+1}^{\text{T}}\vect{\mtV_{L}}\big),
\end{align}
where $\texttt{softmax}(\cdot)$ represents the softmax operation, $\vtW_{L+1}$ is the weight vector of the
fully-connected layer $L+1$, and $\vect{\mtV_{L}}$ is the vectorized output of layer $L$.

\subsection{Recurrent Neural Networks}\label{subsub:RNN}
Similarly to \acp{CNN}, the \ac{RNN} architecture was introduced as a means to deal with sequential data, usually present
in sequences with input values $\vtX^{(1)},\ldots,\vtX^{(\tau)}$ with $\tau$ steps, which could correspond to a
time series $\vtX^{(t)}$ or sequence of words~\citep{goodfellow2016book,Zappone2019,Jagannath2019}.
Different from feedforward neural networks, in which the information propagates forward, in \ac{RNN} the information is
also allowed to flow in loops.
The term \emph{recurrent} is due to the use of recurrent connections between hidden units defined as
\begin{equation}\label{eq:hidden_rnn}
\vtH^{(t)} = f\left(\vtH^{(t-1)},\vtX^{(t)};\vtTheta \right),
\end{equation}
where $f(\cdot)$ is the activation function of a single unit, $\vtH^{(t)}$ is the state of the hidden unit at time $t$,
$\vtX^{(t)}$ is the input sequence at time $t$, and $\vtTheta$ represents the weight parameters of the network that
remain the same for all time indices.
We can represent the operations between the hidden units, inputs, and weights as
\begin{align}\label{eq:predic_func_rnn}
\vtA^{(t)} &= \mtU\vtX^{(t)} + \mtW\vtH^{(t-1)} + \vtB,\\
\vtH^{(t)} &= \tanh\left(\vtA^{(t)} \right),\\
\vtO^{(t)} &= \mtV\vtH^{(t)} + \vtC,\\
\hat{\vtY}^{(t)} &= \texttt{softmax}\left(\vtO^{(t)} \right),
\end{align}
where the parameters $\mtU,\mtW,\mtV$ are the weight matrices for the input-to-hidden, hidden-to-hidden, and
hidden-to-output connections, the parameters $\vtB, \vtC$ are the bias vectors.
Note that the activation function used for the hidden unit is the hyperbolic tangent, and the output activation
function is the softmax.
Hence, the prediction function for \acp{RNN} can be represented by Eqs.~\eqref{eq:predic_func_rnn}.

A problem that may occur with \acp{RNN} is the appearance of exploding or vanishing gradients when dealing with
long-term dependencies~\citep{goodfellow2016book}.
This is a general problem that may occur with general \acp{DNN} but it is more pronounced in \acp{RNN} due to the
repeated application of the same parameters at each time step in a long temporal sequence.
A popular solution to overcome this problem is the \ac{LSTM} architecture, in which the \ac{RNN} is extended to
consider a self loop that includes an input, output, and forget gates, which are referred to as \emph{LSTM
cells}~\citep{Jagannath2019}.
The gates in the \ac{LSTM} cell are responsible to control the storage, when to allow read, write, and forget into the
state of the cell.
These modifications allow the network to keep or forget information over a long period of time.
Specifically, we can represent the operation of the input, output, and forget gates in an \ac{LSTM} cell as
follows~\citep{goodfellow2016book}:
\begin{align}\label{eq:predic_func_lstm}
\vtI^{(t)} &= \sigma\left( \mtW_i \vtX_t + \mtU_i \vtH_{t-1} + \vtB_i \right),\\
\vtO^{(t)} &= \sigma\left( \mtW_o \vtX_t + \mtU_o \vtH_{t-1} + \vtB_o \right),\\
\vtF^{(t)} &= \sigma\left( \mtW_f \vtX_t + \mtU_f \vtH_{t-1} + \vtB_f \right),\\
\vtC^{(t)} &= \vtF^{(t)}\odot \vtC^{(t-1)} \!+\! \vtI^{(t)}\!\odot\!\tanh\left( \mtW_c \vtX_t + \mtU_c \vtH_{t-1} +
\vtB_c \right) ,\\
\vtH^{(t)} &= \vtO^{(t)}\odot \tanh\left(\vtC^{(t)}\right),
\end{align}
where $\vtI^{(t)},\vtO^{(t)},\vtF^{(t)}$ are the input, output, and forget vector gates, respectively; $\vtC^{(t)},
\vtH^{(t)}$ are the cell state vector and the hidden state vector, respectively; $(\mtW_i, \mtU_i, \vtB_i), (\mtW_o,
\mtU_o, \vtB_o), (\mtW_f, \mtU_f, \vtB_f)$ are the weight, transition, and bias matrices and vector for the input,
output and forget gates, respectively; and $\odot$ represents the Hadamard product.
Similar to the \ac{RNN} architecture, the prediction function of the \ac{LSTM} architecture can be represented by
Eqs.~\eqref{eq:predic_func_lstm}.

\section{Unsupervised Learning}\label{sub:unsup_learning}
In unsupervised learning, the goal is to learn some properties of the generation of the data and help to explain the
structure of the data~\citep{Simeone2018,Hastie2009}.
Let us define a set of $N$ observations $(\vtX_1,\vtX_2,\ldots,\vtX_N)$ of a random $d_x$-vector $\vtX\in\bbR^{d_x}$
with joint probability density $\Pr(\vtX)$.
Then, the goal is to directly infer the properties of this probability density without the help of a supervisor (or
teacher) providing correct answers or notion of error for each observation.
The correct answers can be understood as the labels $\{\vtY_i\}_{i=1}^N$, which may not be available in unsupervised
learning.
Specifically, one of the key reasons for using unsupervised learning is to provide labels for unlabelled data.

The input dimension $d_x$ is usually much higher than in supervised learning problems, which make it difficult to 
represent the density $\Pr(\vtX)$ graphically and infer some of its properties.
Usually, we evaluate rough estimates of the global model using various simple descriptive statistics that characterize
$\Pr(\vtX)$.
These descriptive statistics attempt to characterize $\vtX$-values, or collections of such values, where
$\Pr(\vtX)$ is relatively large.
This provides information about the associations among the variables and whether or not they can be considered as
functions of a smaller set of variables, also called \emph{hidden} or \emph{latent} variables.
Some important tasks that analyse such associations include density estimation, clustering, dimensionality reduction,
and feature extraction.
For example, clustering attempts to find multiple convex regions of the $\vtX$-space that contain modes of
$\Pr(\vtX)$.
This indicates whether or not $\Pr(\vtX)$ can be represented by a mixture of simpler densities representing
distinct types or classes of observations.
For density estimation, mixture models are popular because their association rules attempt to construct simple
descriptions of high density regions when using very high dimensional binary-valued data.

Different from supervised learning that has a clear measure of success through the expected loss of the joint
distribution $\Pr(\vtX,\vtY)$, unsupervised learning has no such direct measure of success.
It is difficult to validate the output drawn from most unsupervised learning methods, which is the reason that most
unsupervised learning methods resort to heuristic arguments to motivate their solutions.
Hence, the effectiveness of many methods is subjective and cannot be verified directly.
In the following, we present some unsupervised learning methods that are commonly used in the spectrum sharing
literature, such as the K-means clustering, mixture models and autoencoders for density estimation.

\subsection{K-Means}\label{subsub:K_means}
K-means is a clustering method used to form descriptive statistics mentioned that determine distinct subgroups of the
data, such that each subgroup, termed cluster, represents different properties of the data~\citep{Hastie2009}.
To provide such distinction between groups, clustering methods use a measure to determine the degree of dissimilarity
between each individual object within the subgroups.
This measure varies depending on the type of variable, e.g., quantitative or categorical, and has a role similar to the
role of the loss function in supervised learning.

In general, clustering methods assign each sample of data to a cluster without regard to a probability model describing
the data.
Given a predetermined number of clusters $K<N$, each sample $i=1,\ldots,N$ is uniquely assigned to a cluster, which can
be represented mathematically by $C(i)=k$ , which states that sample $i$ belongs to cluster $k$.
One approach is to directly specify a mathematical loss function and attempt to minimize it through some combinatorial
optimization algorithm.
Since the goal of clustering is to group the data that have similar descriptive statistics, we can model this by
minimizing a function that measures the dissimilarity between points that belong to the same cluster.
Hence, this loss function is termed as the \emph{within-cluster} point scatter and defined as~\citep{Hastie2009}
\begin{equation}
W(C)=\frac{1}{2} \sum_{k=1}^{K} \sum_{C(i)=k} \sum_{C(i')=k} d(\vtX_i,\vtX_{i'}),
\end{equation}\label{eq:within_cluster_loss}
where $d(\vtX_i,\vtX_{i'})$ is the dissimilarity measure.
Another possible loss function is the \emph{between-cluster} point scatter, which measures the distances between
clusters and is maximized in some clustering methods.
The between-cluster loss is defined as
\begin{equation}\label{eq:between_cluster_loss}
B(C)=\frac{1}{2} \sum_{k=1}^{K} \sum_{C(i)=k} \sum_{C(i')\neq k} d(\vtX_i,\vtX_{i'}).
\end{equation}
Note that minimizing (or maximizing) the loss function $W(C)$ (or $B(C)$) requires solving a combinatorial problem
over all possible assignments of $N$ samples into $K$ clusters.
The exhaustive search solution is prohibitive for such problems, so the practical solution is to use greedy solutions
to minimize (or maximize) the loss function.
However, these greedy solutions converge to a local optimal solution that may be very far from the global optimum
solution.

Specifically, the K-means method is one of such greedy solutions and among the most popular clustering methods for
quantitative variables.
K-means considers the Euclidean distance as the dissimilarity measure and has the within-cluster loss function defined
as~\citep{Hastie2009}
\begin{equation}
W(C)= \sum_{k=1}^{K} N_k \sum_{C(i)=k} \Norm{\vtX_i-\bar{\vtX}_{k}}^2, \label{eq:within_cluster_kmeans_loss}
\end{equation}
where $\bar{\vtX}_{k}$ is the mean vector (centroid) of cluster $k$ , and $N_k=\sum_{i=1}^{N} I(C(i)=k)$ with $I(x)$
being the indicator function, i.e., $I(x)=1$ if $x$ is true and $I(x)=0$ otherwise.
Hence, K-means minimizes the average dissimilarity of the samples from the cluster mean in a greedy solution approach.
The first step is the choice on the number of clusters $K$, while the second step is the assignment of $K$ points
$\bar{\vtX}_{k}$ to act as centroids.
These $K$ centroids can be chosen randomly from the samples or randomly within the domain of the data.
The third step consists in assigning each sample to the closest centroid, meaning that $C(i)=\underset{1\leq k\leq K}
{\arg\min} \Norm{\vtX_i- \bar{\vtX}_{k}}$.
In the fourth step, the centroids are updated using the samples assigned to the newly formed clusters.
Then, the third and fourth steps are iteratively updated until either the centroid of the newly formed clusters do not
change, or the samples remain in the same cluster, or a maximum number of iterations has been reached.
As mentioned, K-means is one among many clustering methods and we refer to~\cite[Chapter 14.3]{Hastie2009} for a more
in-depth analysis of more clustering methods.

\subsection{Mixture Models}\label{subsub:mixture_models}
Different from clustering, density estimation aims to provide a good approximation of the distribution
$\Pr(\vtX)$ that can be used for estimation problems, dimensionality reduction, and detection of
outliers~\citep{Simeone2018,Murphy2013}.
A common approach in density estimation, and many unsupervised learning problems, is to assume that the observed
variables are correlated because they arise from a hidden common cause (similar to the \ac{HMM}), which are termed
latent variables.
Such models are useful because they have usually fewer parameters than models with direct correlations that are not
hidden, and because they represent a compressed representation of the data.

First, let us define as $\vtTheta\in\bbR^d$ the vector of parameters to be optimized, such that we want to
characterize the distribution $\Pr(\vtX|\vtTheta)$.
Then, let us define $\{\vtZ\}_{i=1}^N\in\bbR^{d_z}$ as the $N$ latent variables with dimension $d_z\ll d_x$.
We define a generative model as a model in which the distribution $\Pr(\vtX|\vtTheta)$ is
defined by a parameterized prior $\Pr(\vtZ|\vtTheta)$ of the latent variable $\vtZ$ and by parameterized
conditional distributions, such as $\Pr(\vtX|\vtZ,\vtTheta)$ or $\Pr(\vtX,\vtZ|\vtTheta)$, which defines
the relationship between latent and observed variable $\vtX$.
For discrete latent variables $\vtZ\in\{1,\ldots,K\}^{d_z}$, $\Pr(\vtX|\vtTheta)=\sum_{\vtZ}
\Pr(\vtZ|\vtTheta) \Pr(\vtX|\vtZ,\vtTheta)$ and $\Pr(\vtX|\vtTheta)=\sum_{\vtZ}
\Pr(\vtX,\vtZ|\vtTheta)$ (the continuous case is treated similarly, replacing the sums by integrals).

Using this modelling, \acp{GMM} can be described by the generative model with the following
distributions~\citep{Simeone2018,Murphy2013}:
\begin{equation}
	\vtZ_i\sim\text{Cat}(\mtPi), \quad \vtX_i|\vtZ_i = k\sim \calN(\vtMu_k,\mtSigma_k),
\end{equation}
where the latent variable $\vtZ_i$ follows a multinomial (categorical) distribution and the parameter vector $\vtTheta$
with parameter vector is composed of $\vtTheta = [\mtPi, \{\vtMu_k\}_{k=1}^K, \{\mtSigma_k\}_{k=1}^K]$.
Thus, the conditional $\Pr(\vtX|\vtTheta)$ is modelled as:
\begin{equation}
\Pr(\vtX|\vtTheta) = \sum_{k} \pi_k \calN(\vtMu_k,\mtSigma_k),
\end{equation}
where $\sum_k \pi_k =1$ with $0\leq \pi_k\leq 1$, and $\calN(\vtMu_k,\mtSigma_k)$ represents a multivariate Gaussian
distribution with mean $\vtMu_k$ and covariance matrix $\mtSigma_k$.
Note that the conditional probability $\Pr(\vtX|\vtTheta)$ is modelled as a convex combination of $K$ Gaussian
distributions, and hence the name of Gaussian mixture model.
In general, mixture models can use any density distribution instead of the Gaussian, such as the Dirichlet
distribution that gives rise to the Dirichlet mixture model, but the Gaussian mixture model is the most popular.
The parameters are usually fit by maximum likelihood using the \ac{EM} algorithm, which has an expectation step
followed by a maximization step.

The basic idea of the \ac{EM} algorithm is to maximize the log-likelihood of the observed data~\citep[Section
11.4.2]{Murphy2013}:
\begin{equation}
l(\vtTheta)\! = \!\sum_{i=1}^N\! \log (\Pr(\vtX_i|\vtTheta)) \!=\! \sum_{i=1}^N \!\log \left( \sum_{\vtZ_i}
\Pr(\vtX_i,\vtZ_i|\vtTheta)\right).
\end{equation}
This is a complicated problem due to the logarithm of the sums, so for the sake of simplicity let us define the
complete data log-likelihood $l_cl(\vtTheta)=\sum_{i=1}^{N} \log (\Pr(\vtX_i,\vtZ_i|\vtTheta))$.
The log-likelihood $l_{cl}(\vtTheta)$ cannot be computed because $\vtZ_i$ is unknown.
However, we can compute the expected complete data log-likelihood $l_{cl}(\vtTheta)=\sum_{i=1}^{N} \log
(\Pr(\vtX_i,\vtZ_i|\vtTheta))$ as $Q(\vtTheta,\vtTheta^{t-1})=\bbE[l_{cl}(\vtTheta)|\vtX_i,\vtTheta^{t-1}]$,
where $t$ is the current iteration number of the algorithm and $Q$ is defined as the auxiliary function.
After some manipulation, the expected complete data log-likelihood can be written as
\begin{equation}
Q(\vtTheta,\vtTheta^{t-1}) \!=\! \sum_i\!\sum_k\! r_{ik}\log \pi_k + \sum_i\!\sum_k\! r_{ik}\log
\Pr(\vtX_i|\vtTheta_k),
\end{equation}
where $r_{ik}=\Pr(z_{ik}=k|\vtX_i,\vtTheta^{t-1})$ is the responsibility that component $k$ takes for sample $i$.
Then, the expectation step computes the expected complete data log-likelihood by setting $r_{ik}$ as
\begin{equation}
r_{ik} = \frac{\pi_k \Pr(\vtX_i|\vtTheta_k^{t-1}) }{\sum_{k'=1}^K \pi_{k'} \Pr(\vtX_i|\vtTheta_{k'}^{t-1})
}.
\end{equation}
Then, the maximization step optimizes $Q(\vtTheta,\vtTheta^{t-1})$ with respect to $\vtPi$ and $\vtTheta$.
For $\vtPi$, the updates are $\pi_k=\frac{1}{N}\sum_i r_{ik}$.
For $\vtTheta$, the updates are related to the distribution chosen for the mixture model, and specifically for
\ac{GMM}, the updates are
\begin{align}
\vtMu_k = \frac{\sum_i r_{ik}\vtX_i }{\sum_i r_{ik}},\quad \mtSigma_k = \frac{\sum_i r_{ik}\vtX_i \Transp{\vtX}_i}
{\sum_i r_{ik}} -\vtMu_k\Transp{\vtMu}_k.
\end{align}
Then, the expectation and maximization steps, for all $k=1,\ldots,K$, continue iteratively until convergence of the
expected complete data log-likelihood is achieved.
We refer the interested reader to~\cite[Section 11.4.2]{Murphy2013} for detailed information about the \ac{EM}
algorithm for other distributions.

\subsection{Autoencoders}\label{subsub:autoencoders}
Differently from generative models, autoencoders are part of a parameterized discriminative model
$\Pr(\vtZ|\vtX,\vtTheta)$ that produces the hidden variables $\vtZ$ from the data $\vtX$ using a parameterized
generative model $\Pr(\vtX|\vtZ,\vtTheta)$~\citep{Simeone2018}.
The parameterized model $\Pr(\vtZ|\vtX,\vtTheta)$ is known as encoder, while the parameterized model
$\Pr(\vtX|\vtZ,\vtTheta)$ as decoder.
Accordingly, the latent variables are also referred to as the code.
In practice, autoencoders are trained to reproduce the data $\vtX$ at the output, thus turning the unsupervised problem
into a supervised one with labels given by the data point $\vtX$ itself.
Due to this reason, autoencoders are also part of the class of semi-supervised learning problems, which generalizes
both supervised and unsupervised learning problems to situations in which not all the data is unlabelled.

The most typical implementation, named deterministic autoencoder, uses parameterized deterministic functions $\vtZ =
F_{\vtTheta}(\vtX)$ and $x = G_{\vtTheta}(\vtZ)$ instead of the probabilistic models $\Pr(\vtZ|\vtX,\vtTheta)$
and $\Pr(\vtX|\vtZ,\vtTheta)$, which are then termed variational autoencoders~\citep{Simeone2018}.
Then, the encoder $\vtZ = F_{\vtTheta}(\vtX)$ is concatenated with a decoder $x = G_{\vtTheta}(\vtZ)$, which gives the
input-output $\vtT=G_{\vtTheta}(F_{\vtTheta}(\vtX))$.
We can formalize this as the minimization of a loss function (such as Euclidean distance) over the parameters
$\vtTheta$ as
\begin{equation}
\underset{\vtTheta}{\min} \sum_{i=1}^{N} l\big(\vtX_i,G_{\vtTheta}(F_{\vtTheta}(\vtX_i))\big).
\end{equation}
In fact, this problem is trivially solved by the identify function $G_{\vtTheta}(F_{\vtTheta}(\vtX_i))=\vtX_i$.
Thus, it is more practical to consider constraints on the encoder-decoder such that the latent variables $\vtZ$ have
lower dimensionality or sparsity.

Moreover, the associated functions $F_{\vtTheta}(\vtX)$ and $G_{\vtTheta}(\vtZ)$ could be either linear or nonlinear.
If linear, an example is to set the encoder as $F_{\vtTheta}(\vtX)=\Transp{\mtW}\vtX$, with $\mtW\in\bbR^{d_z\times
d_x}$ and decoder $G_{\vtTheta}(\vtZ)=\mtW\vtZ$ together with a quadratic function, thus aiming to minimize the
optimization problem $\underset{\mtW}{\min} \sum_{i=1}^{N} \NormTwo{\vtX_i-\mtW\Transp{\mtW}\vtX}^2 $.
This example can be solved in closed form~\citep{Simeone2018}, whose solution is given by the sample covariance matrix
$\frac{1}{N}\sum_i \vtX_i \Transp{\vtX}_i$.
However, the most popular use of autoencoders are when the encoder and decoder are nonlinear and modelled by an
\ac{ANN}, which are far more powerful than linear autoencoders.
In the case of one hidden layer \ac{ANN}, the encoder can be written as $F_{\vtTheta}(\vtX_i)=\vtZ_i=\sigma_1(\mtW_1
\vtX_i + \vtB_1)$ while the decoder can be written as $G_{\vtTheta}(\vtZ_i)=\sigma_2(\mtW_2 \vtZ_i + \vtB_2)$.
Then, the aim is to solve the optimization problem
\begin{equation}
\underset{\substack{\mtW_1,\mtW_2,\\ \vtB_1,\vtB_2}}{\textup{minimize}} \sum_{i=1}^{N} \NormTwo{\vtX_i- \sigma_2(\mtW_2
\sigma_1(\mtW_1 \vtX_i + \vtB_1) + \vtB_2) }^2.
\end{equation}
As expected, the number of layers for the encoder and decoder may be higher than one to increase the representative
power of the autoencoder model.

\section{Reinforcement Learning}\label{sub:reinfo_learning}
In reinforcement learning, the goal is to maximize a numeric reward value that is accumulated through several actions
done by an agent (decision maker), that reward being positive or negative, occurring in a dynamical environment~\citep{Sutton2018}.
The learning algorithm needs to discover which actions to take in order to maximize the accumulated reward, which
implies that actions may impact the current and subsequent rewards.
Note the word \emph{discover}, which is important because the learning algorithm may not take the action that
maximizes the current reward (exploitation).
Instead, it can decide to \emph{explore} a suboptimal action that may provide a better reward in the future
(exploration).
The trade-off between exploitation and exploration, together with the dynamic decision making scenario, are important
characteristics of reinforcement learning.

Different from supervised learning, reinforcement learning does not have access to examples (labels) of a desired
action that is both correct and representative of all the situations in which the agent has to act~\citep{Sutton2018}.
For this reason, the agent must use its own experience to maximize the reward.
With respect to unsupervised learning, both reinforcement and unsupervised learning reinforcement learning do not have
access to labels.
However, both learning paradigms are remarkably different because reinforcement learning is not interested in
understanding a hidden structure of the data.
Instead, reinforcement learning is interested in maximizing a numeric reward through several actions in a dynamic
system.

The decision maker is called the \emph{agent} and everything the agent interacts with is called the
\emph{environment}~\citep{Sutton2018}.
The agent interacts with the environment at discrete time steps by selecting actions, in which the environment responds
with a numerical reward to these actions and further presenting new situations to the agents.
Ultimately, the agents seeks to maximize the numerical reward over time through a series of actions.
Due to the dynamical environment in which the agent is performing actions, the modelling of this system is done in a
probabilistic manner, which builds on the Markov Chain framework, presented in Section~\ref{sub:sup_learning}, and is
called a \ac{MDP}.
We define herein the mathematical notation commonly used for \ac{MDP} following the notation from~\cite{Sutton2018},
and then extend it to reinforcement learning problems.

At each discrete time step $t=0,1,2,\ldots$, the agent receives a representation of the environment's state
$s_{t}\in\calS$, where $\calS$ is the set of all possible states.
Based on state $s_t$, the agent selects action $a_t\in\calA(s)$, where $\calA(s)$ is the finite set of actions given
that the agent is at state $s_t$.
As a consequence of its action, the agent receives a numerical reward $r_{t+1}\in\calR\succ\bbR$ and moves to state
$s_{t+1}$, where $\calR$ is the set of numerical rewards.
Note that it is common to refer to the reward for the state $s_t$ and action $a_t$ as $r_{t+1}$, instead of $r_t$, to
emphasize that the reward and new state $s_{t+1}$ are jointly determined~\citep{Sutton2018}.
When the sets of states $\calS$, actions $\calA$, and rewards $\calR$ are finite, we have a finite \ac{MDP}.
In this case, the random variables $r_t$ and $s_t$ have well defined discrete probability distribution that depend only
on the preceding state and action, i.e., the probability distribution follows the Markov property.
With this, we define the probability of state $s'\in\calS$ with reward $r\in\calR$ at time $t$ as
$p:\calS\times\calR\times\calS\times\calA\rightarrow[0,1]$ with
$p(s',r|s,a)=\Pr(s_t=s',r_t=r|s_{t-1}=s,a_{t-1}=a)$,
for all $s',s\in\calS$, $r\in\calR$, and $a\in\calA(s)$.
The probability distribution $p$ characterizes the dynamics of the \ac{MDP}, which allows us to compute the state
transition probabilities as $p(s'|s,a)=\Pr(s_t=s'|s_{t-1}=s,a_{t-1}=a)=\sum_{r\in\calR} p(s',r|s,a)$; the
expected reward for state-action pairs as $r(s,a)=\bbE[r_t|s_{t-1}=s,a_{t-1}=a]=\sum_{r\in\calR} r\sum_{s'\in\calS}
p(s',r|s,a)$; and the expected reward for the state-action-next state triples as:
\begin{align}
r(s,a,s')=&\bbE[r_t|s_{t-1}=s,a_{t-1}=a,s_t=s'] = \sum_{r\in\calR} r\; \frac{p(s',r|s,a)}{p(s'|s,a)}.
\end{align}

As mentioned before, the goal of reinforcement learning is to maximize the reward accumulated over time.
Using the definitions before, the goal at time step $t$ is to maximize the expected return, defined as:
\begin{align}\label{eq:g_t_return}
g_t=&r_{t+1}+\gamma r_{t+2}+\gamma^2 r_{t+3}+\cdots =  r_{t+1}+\gamma g_{t+1} = \sum_{k=0}^\infty \gamma^k r_{t+k+1},
\end{align}
where $0\leq \gamma\leq 1$ is the discount rate.
Note that the discount rate determines how much impact future rewards will have on the current goal.
For $\gamma=0$, the agent is \emph{myopic} and only interested in maximizing the immediate reward $r_{t+1}$; whereas
when $\gamma$ approaches $1$ the return objective takes future rewards into account and the agent becomes
\emph{farsighted}.
Usually, $\gamma$ is lower than 1 when the reward is nonzero and constant to ensure that the infinite sum converges.
In case of a finite sum in $g_t$, i.e., we have a finite number of time steps $T$, $\gamma$ can be equal to $1$, the
final state $s_T$ is called the \emph{terminal state}, and the final number of steps $T$ itself is a random variable.
Overall, tasks with $T=\infty$ are termed \emph{continuous tasks}, usually with $\gamma<1$, while tasks with finite $T$
are termed \emph{episodic tasks}, usually with $\gamma=1$, and its expected return $g_t$ is termed discounted return.

Using the expected return $g_t$, we need two more definitions: a \emph{policy} and \emph{value function}.
A policy $\pi:\calS\times\calA\rightarrow[0,1]$ is a mapping function from states to probabilities of selecting
possible actions.
We denote $\pi(a|s)$ as the probability that the agent is following policy $\pi$ at time $t$ given that $a_t=a$ and
$s_t=s$.
Then, we define the value function $v_{\pi}(s)$ of state $s$ under a policy $\pi$ as the expected return when starting
at state $s$ and following policy $\pi$ afterwards by the following expression:
\begin{equation}\label{eq:state_value_func}
\hspace*{-0.3cm}v_{\pi}(s)\!=\!\bbE_{\pi}[g_t|s_t\!=\!s]\!=\!\bbE_{\pi}\Big[\sum_{k=0}^\infty \gamma^k
r_{t+k+1}|s_t\!=\!s\Big],\!\! \forall s\in\calS.
\end{equation}
Similarly, we define the value of taking action $a$ in state $s$ under a policy $\pi$, denoted by $q_{\pi}(s,a)$, as
the expected return when starting from state $s$, taking action $a$, and following policy $\pi$ afterwards by the
following expression
\begin{align}
q_{\pi}(s,a)=&\bbE_{\pi}[g_t|s_t=s,a_t=a],\nonumber\\
=&\bbE_{\pi}\Big[\sum_{k=0}^\infty \gamma^k r_{t+k+1}|s_t=s,a_t=a\Big].
\end{align}
We call the function $v_{\pi}(s)$ as the \emph{state-value function} for policy $\pi$, and the function $q_{\pi}(s,a)$
as the \emph{action-value function} for policy $\pi$.

A fundamental property of values function is a recursive relation that stems from Eqs.~\eqref{eq:g_t_return}
and~\eqref{eq:state_value_func}:
\begin{align}\label{eq:value_bellman}
v_{\pi}(s) =& \bbE_{\pi}[g_t|s_t=s] = \bbE_{\pi}[r_{t+1} + \gamma g_{t+1}|s_t=s],\nonumber\\
\!=\!& \sum_a \pi(a|s)  \sum_{s'} \sum_r p(s',r|s,a) [r + \gamma v_{\pi} (s')],
\end{align}
where this recursive equation holds for all $s,s'\in\calS$, $a\in\calA(s)$, and $r\in\calR$.
This fundamental equation is called the \emph{Bellman equation} for the value function $v_{\pi}$, and can also be
understood as an expected value over the variables $a, s',$ and $r$ using the probability $\pi(a|s) p(s',r|s,a)$.
Using the state-value function, and its formulation as the Bellman equation, we can now decide which policy to follow
in order to maximize the reward in the long term.
Thus, we define that a policy $\pi$ is better than or equal to policy $\pi$ if its expected return is greater than or
equal to that of $\pi'$ for all states, i.e., $\pi\geq \pi'$ if and only if $v_{\pi}(s)\geq v_{\pi'}(s)$ for all
$s\in\calS$.
With this partial ordering over policies, we define an optimal policy $\pi_{\star}$ as the policy that is better than
or equal to all other policies.
Moreover, one can prove that for \acp{MDP} there exists at least one optimal policy $\pi_{\star}$~\citep{Sutton2018}.
Following an optimal policy, the optimal state-value function $v_{\star}(s)$ and the optimal
action-value function $q_{\star}(s,a)$ are defined as
\begin{align}
v_{\star}(s) =&\underset{\pi}{\max}\; v_{\pi}(s), \forall s\in\calS,\\
q_{\star}(s,a) =& \underset{\pi}{\max}\; q_{\pi}(s,a),\forall s\in\calS,a\in\calA(s).
\end{align}

Note that $v_{\star}(s)$ is a value function for a policy, so it must satisfy the Bellman
equation~\eqref{eq:value_bellman} for state values.
Since it is the optimal value function, it can be written without reference to any specific policy by noting that the
value of a state under an optimal policy must be equal to the expected return of the best action from that state.
Specifically, we can write $v_{\star}(s)$ as:
\begin{align}\label{eq:opt_value_bellman}
v_{\star}(s) =& \underset{a\in\calA(s)}{\max}\;q_{\pi_{\star}}= \underset{a}{\max}\; \bbE_{\pi_{\star}}
[g_t|s_t=s,a_t=a],\nonumber \\
=& \underset{a}{\max}\; \sum_{s'} \sum_r p(s',r|s,a) [r + \gamma v_{\star}(s')].
\end{align}
Similarly, we can write $q_{\star}(s,a)$ as:
\begin{align}\label{eq:opt_act_value_bellman}
q_{\star}(s,a) =& \bbE[r_{t+1} + \gamma\; \underset{a'}{\max}\; q_{\star}(s_{t+1},a')|s_t=s,a_t=a],\nonumber\\
=&\sum_{s'}\sum_r p(s',r|s,a) [r + \gamma\; \underset{a'}{\max}\; q_{\star}(s',a')].
\end{align}
These specific equations for the optimal state-value and action-value functions are called the \emph{Bellman optimality
equation} for $v_{\star}(s)$ and $q_{\star}(s,a)$, respectively.
Solving the Bellman optimality equations is one route to obtain the optimal policy $\pi_{\star}$, and thus solving a
\ac{MDP}.

The solution to \ac{MDP} through the Bellman optimality equations use dynamic programming algorithms~\citep{Sutton2018}.
These algorithms are composed of two fundamental parts: \emph{policy evaluation} and \emph{policy improvement}.
Policy evaluation, also referred to as the \emph{prediction problem}, is related to the iterative computation of the
value functions for a given policy and all states $s\in\calS$ (see Eq.~\eqref{eq:value_bellman}).
Policy improvement, also referred to as the \emph{control problem}, is related to the computation of an improved
policy, usually a greedy policy calculated as
$\pi'(s)=\underset{a}{\arg\max}\; q_{\pi}(s,a)$, given the value function for that policy.
Thus, a policy $\pi$ is evaluated using $v_{\pi}(s)$, we use its value to improve policy $\pi$ towards policy $\pi'$,
and this iterative process is repeated until the optimal policy and optimal value function are obtained.
The most popular dynamic programming algorithms are the value iteration and policy iteration, which use these two
fundamental parts in different manners.
For an in-depth analysis of these algorithms, we refer the interested reader to~\cite[Chapter 4]{Sutton2018}.

However, solving the Bellman optimality equations is rarely available in practice because of at least three
assumptions.
The first assumption is that we accurately know the dynamics of the environment, which are in the form of the
probability distribution $p(s',r|s,a)$ with its four arguments.
The second assumption is that we have enough computational resources to complete the computation of the solution, which
is similar to an exhaustive search due to its need to look ahead at all possibilities for the states/actions.
Finally, the third assumption is the Markov property, which states that the next state and action pair
$(s_{t+1},a_{t+1})$ depend only on the preceding state and action $(s_{t},a_{t})$ instead of the whole history until
time $t$ as $(s_{1},a_{1},s_{2},a_{2}\cdots s_{t},a_{t})$.
For \ac{MDP}, the first and third assumptions hold but the second assumption depends on the finite number of states and
actions.

In reinforcement learning, the first and third assumptions do not hold, which makes the distinction with respect to
\acp{MDP} clear.
Hence, reinforcement learning does not have full knowledge about the dynamics of the environment and the Markov
property does not hold.
Since there is no full knowledge of the environment, reinforcement learning methods rely on the estimation of the
environment.
Depending on the estimation used, if the estimation of the state-value and action-value functions, or the policy,
reinforcement learning methods can be subdivided into three categories: temporal-difference learning, policy gradients,
or actor-critic methods.
In the following, we discuss the solution methods for each category.

Temporal-difference learning methods use raw experience without a model of the environment's dynamics.
Since the model is not available, state values alone are not sufficient to determine a policy.
It is necessary to explicitly estimate the value of each action in order for the values to be useful in suggesting a
policy.
Thus, one of our primary goals is to estimate action-value $q_{\pi}(s,a)$ instead of state-values $v_{\pi}(s)$.
To make this estimation, the agent needs to learn from the state and actions that have been visited, which implies that
we need to ensure that all the state-action pairs are visited.
There are two approaches to ensuring this, resulting in what we call on-policy methods and off-policy methods.
On-policy methods attempt to evaluate or improve the policy that is used to make decisions, whereas off-policy methods
evaluate or improve a policy different from that used to generate the data.
In off-policies, we consider that every pair has a nonzero probability $\epsilon$ of being visited.
We formally define this by using a stochastic policy $\pi(s,a)$ defined as
\begin{equation}\label{eq:stoch_policy_epsilon}
\pi(s,a)=\begin{cases}
\frac{\epsilon}{|\calA|} + (1-\epsilon), &\text{if } a = \underset{a'}{\arg\max}\; q_{\pi}(s,a'),\\
\frac{\epsilon}{|\calA|}, &\text{otherwise}.
\end{cases}
\end{equation}
Hence, this policy \emph{explores} a suboptimal action with probability $\epsilon$ and \emph{exploits} the gained
knowledge in a greedy manner with probability $1-\epsilon$.
Due to this greedy approach using a probability, this policy is called an $\epsilon$-greedy policy.
In on-policies, two types of policies are used: one that is learned about and that becomes the optimal policy, and
one that is more exploratory and is used to generate behavior.
The policy being learned about is called the target policy, and the policy used to generate behavior is called the
behavior policy.
The target policy is usually the deterministic greedy policy with respect to the current estimate of the action-value
function $q_{\pi}(s,a)$.
This policy becomes a deterministic optimal policy while the behavior policy remains stochastic and more exploratory,
usually an $\epsilon$-greedy policy.

\subsection{TD(O) Methods}\label{subsub:td-zero}
For both on- and off-policies, we need to provide the estimates for the action-value function $q_{\pi}(s,a)$ that will
be used.
First, we provide an estimate of the state-value function $v_{\pi}(s)$ using an stochastic approximation method,
named Robbins-Monro, in the form of:
\begin{equation}\label{eq:v_func_gen_approx}
v(s_t) = v(s_t) + \alpha [g_t -v(s_t)],
\end{equation}
where $\alpha$ is a constant step-size parameter with the property of $\sum_k \alpha_k > \infty$ and $\sum_k \alpha_k^2
< \infty$, $g_t$ is the actual return following time $t$, and we omitted the policy $\pi$ for simplicity.
Note that Eq.~\eqref{eq:v_func_gen_approx} requires the return $g_t$ after following time $t$, i.e., it is not
available until the end of this episodic task because it depends on the obtained reward and the next state.
The Eq.~\eqref{eq:v_func_gen_approx} with the general reward $g_t$ is the basis for Monte Carlo methods in
reinforcement learning, and the state-value function $v(s)$ can be updated using this equation.
In temporal-difference learning, the simplest algorithm is called TD(0) and extends Eq.~\eqref{eq:v_func_gen_approx} by
using an approximate version of the return $g_t$ as
\begin{equation}\label{eq:v_func_gen_approx_td0}
v(s_t) = v(s_t) + \alpha [r_{t+1} + \gamma v(s_{t+1}) -v(s_t)],
\end{equation}
where the term $r_{t+1} + \gamma v(s_{t+1})$ estimates the return $g_t$ using only the next time step $t+1$.
The TD(0) algorithm can be seen as the algorithm to solve the prediction problem, i.e., the policy evaluation step that
was also part of dynamic programming algorithms.
Using this update on the state-value function together with any fixed policy, the algorithm TD(0) provides an online
and
incremental solution approach that converges to $v_{\pi}$ while \emph{only} requiring us to wait for a single time step.
However, this need to wait for one time step is crucial in several practical applications that cannot wait for such
time step.
Due to this reason, we resort to the traditional methods that estimate the action-value function $q_{\pi}(s,a)$: 
\ac{SARSA}
and Q-learning.
Moreover, \ac{SARSA} and Q-learning can be seen as the algorithms to solve the control problem, i.e., the policy 
improvement
step from dynamic programming.

\subsection{SARSA and Q-Learning}\label{subsub:sarsa_q_learning}
To overcome the need to wait for one time step in TD(0), we need to estimate the action-value function $q_{\pi}(s,a)$.
As mentioned before, we can do this via an on-policy or off-policy, where the temporal-difference method with an
on-policy is named \ac{SARSA}, and the method with an off-policy is named Q-learning.
In \ac{SARSA}, the $\epsilon$-greedy policy is used together with the following estimate for the action-value function
$q_{\pi}(s,a)$~\citep{Sutton2018}:
\begin{equation}\label{eq:q_func_gen_approx_sarsa}
q(s_t,a_t) = q(s_t,a_t) + \alpha [r_{t+1} + \gamma q(s_{t+1},a_{t+1}) -q(s_t,a_t)],
\end{equation}
where this update is done after every transition from a non-terminal state-action pair.
Note that Eq.~\eqref{eq:q_func_gen_approx_sarsa} follows the structure of Eq.~\eqref{eq:v_func_gen_approx_td0}, but now
for the action-value function $q_{\pi}(s,a)$.
This update rule uses the quintuple of events $(s_t,a_t,r_{t+1},s_{t+1},a_{t+1})$, which gives the name \ac{SARSA} to 
the
algorithm.
Overall, the \ac{SARSA} algorithm converges to the optimal policy and action-value function if all state-action
pairs are visited and the policy converges to the greedy policy, which can be ensured by using an $\epsilon$-greedy
policy such as in Eq.~\eqref{eq:stoch_policy_epsilon}.
In Q-learning, the goal is not only to estimate the action-value function $q_{\pi}(s,a)$ but to estimate the optimal
action-value function $q_{\star}(s,a)$, independent of the policy followed.
Specifically, the the action-value function $q_{\pi}(s,a)$ is calculated as:
\begin{align}
\label{eq:q_func_gen_approx_qlearn}
q(s_t,a_t) = q(s_t,a_t) \!+\! \alpha \big[r_{t+1} \!+\! \gamma\; \underset{a}{\arg\max}\; q_{\pi}(s_{t+1},a)  \!-\!q(s_t,a_t) \big].
\end{align}
The reason that Q-learning is an off-policy method is due to the use of a target policy $\pi(s)$ that is greedy with
respect to $q(s,a)$ in Eq.~\eqref{eq:q_func_gen_approx_qlearn}, as $\pi(s_{t+1})=\underset{a}{\arg\max}\;
q_{\pi}(s_{t+1},a)$, and the use of a behavior policy $\mu(s,a)$ that is $\epsilon$-greedy with respect to $q(s_t,a_t)$.
Overall, the Q-learning algorithm converges to the optimal policy and action-value function under the same assumptions
as \ac{SARSA}.

\subsection{Approximation Methods}\label{subsub:approx_techniques}
Note that \ac{SARSA} and Q-learning need to update a table of values that scales with the number of states $\|\calS\|$ 
and
number of actions $\|\calA\|$~\citep{Sutton2018}.
However, many applications have a combinatorial number of possible states and actions, which makes it impossible to 
obtain an optimal policy or optimal value function.
Using \ac{SARSA} or Q-learning in such scenarios is prohibitive due to the memory needed to store such large tables 
and
the time and data needed to fill every state accurately, even if it has not been seen before.
To make decisions in such unforeseen states, it becomes necessary to generalize from previous visits to similar states.
For such scenarios, the goal is to obtain a good approximate solution for the optimal policy, or optimal value
function, that is able to generalize.
Creating approximate solutions that generalize well belongs to the subject of \emph{function approximation}
in supervised learning.
Hence, we can use function approximation techniques from supervised learning, such as \ac{ANN}, to approximate the
optimal policy or optimal value function in reinforcement learning.

We represent the value function $v_{\pi}(s)$ as a parameterized approximate value $\hat{v}(s,\vtW)$ of a state $s$
given a weight parameter $\vtW\in\bbR^d$~\citep{Sutton2018}.
For example, $\hat{v}(s,\vtW)$ may be a linear function in the feature of the states $s$ using the vector $\vtW$ of
feature weights or a multi-layer \ac{ANN} where the weights $\vtW$ represent the connection weights in all layers.
Note that the number of weights $d$ is much less than the number of states $\|\calS\|$, which means that we must
specify which states we care most about.
Hence, let us denote by $\mu(s)\geq 0, \sum_s \mu(s)=1$ a state distribution representing how much we care about the
error in state $s$.
This error represents the error between the approximate value function $\hat{v}(s,\vtW)$ and the true value function
$v_{\pi}(s)$, which can be represented by the mean squared value error $\overline{\text{VE}}$ as:
\begin{align}\label{eq:mse_value}
\overline{\text{VE}} (\vtW) = \sum_{s\in\calS} \mu(s)[\hat{v}(s,\vtW) -v_{\pi}(s) ]^2.
\end{align}
The minimization of $\overline{\text{VE}} (\vtW)$ can be used as the performance objective in reinforcement learning and
$\mu(s)$ is usually selected as a fraction of the time spent in $s$, which can be selected differently if an on- or
off-policy is used.

Since we aim at minimizing the value error $\overline{\text{VE}} (\vtW)$, we can use the \ac{SGD} method to adjust
the weight vector $\vtW$ towards the direction that minimizes the objective function $\overline{\text{VE}} (\vtW)$.
Let us assume that $\hat{v}(s,\vtW)$ is a differentiable function of $\vtW$ for all $s\in\calS$, and that at each
discrete time step $t=0,1,\ldots$ we will be updating the weights $\vtW_t$.
Moreover, let us assume that on each time step $t$ we observe a new state $s_t$ and its true value $v_{\pi}(s)$ under
the selected policy $\pi$.
Hence, we write the weight updates as~\citep{Sutton2018}:
\begin{align}\label{eq:weight_upd_value_error}
\vtW_{t+1} = \vtW_t + \alpha[v_{\pi}(s_t) - \overline{\text{VE}} (s_t,\vtW_t)]\nabla \overline{\text{VE}} (s_t,\vtW_t),
\end{align}
where $\alpha$ is a positive step-size parameter, and $\nabla \overline{\text{VE}} (s_t,\vtW_t)$ denotes the gradient
of
$\hat{v}(s,\vtW)$ with respect to $\vtW_t$.
However, Eq.~\eqref{eq:weight_upd_value_error} assumes that the value function $v_{\pi}(s_t)$ is independent of
$\vtW_t$, which may not be the case if the value function is unknown and a noisy approximation $u_t$ is used (such as
TD(0)).
For this reason, we make the assumption that $v_{\pi}(s_t)$ is independent of the weight parameter $\vtW$ and we
call this a \emph{semi-gradient} method.
For instance, we can derive a semi-gradient TD(0) using the approximation $u_t=r_{t+1}+\gamma \overline{\text{VE}}
(s_{t+1},\vtW)$

The drawback of the semi-gradient methods is that the convergence is not as robust as gradient methods.
Nevertheless, these methods do converge reliably in some cases such as linear approximation functions.
Specifically, let us define the vector $\vtX(s)\in\bbR^d$ as the feature vector representing state $s$, in which each
component $x_i(s)$ is the value of a function $x_i:\calS\rightarrow\bbR$ called feature of $s$.
Using this feature vector, linear methods approximate the state-value function as~\citep{Sutton2018}:
\begin{align}\label{eq:apprx_value_func_linear}
\overline{\text{VE}} (s,\vtW) = \Transp{\vtW} \vtX(s) = \sum_{i=1}^d w_i x_i (s).
\end{align}
For linear methods, the features are the basis functions that span a set of approximate functions.
The \ac{SGD} update in a semi-gradient TD(0) method is simply $\vtW_{t+1} = \vtW_t + \alpha[u_t - \overline{\text{VE}}
(s_t,\vtW_t)] x(s_t)$.
For the semi-gradient TD(0), linear methods are proved to converge to a bounded expansion of the local optimal,
specifically to a $\overline{\text{VE}}(\vtW_{TD(0)})\leq
\frac{1}{1-\gamma}\underset{\vtW}{\min}\;\overline{\text{VE}}(\vtW)$.
Other approximation functions can be used, which can be polynomials, Fourier series using sums of sine and cosine
basis, \ac{ANN}, or radial Gaussian functions.
For an in-depth analysis of other approximation functions, we refer the interested reader to~\cite[Chapter
9]{Sutton2018}.

Recently, \acp{ANN} are commonly used as function approximations, which give rise to the deep reinforcement
learning methods~\citep{Sutton2018}.
Among these methods, one of the most popular is the deep Q learning method is the that uses function approximation 
in
Q-learning with an \ac{ANN} as the approximation function~\citep{Mnih2013}.
The action-value function $q(s_t,a_t)$ is approximated as $q(s_t,a_t,\vtW_t)$, referred as a deep Q-network, using an
\ac{ANN} that is trained using the \ac{SGD} method.
Specifically, the \ac{SGD} updates for deep Q learning is denoted as:
\begin{align}\label{eq:weight_upd_deep_q}
\vtW_{t+1} =& \vtW_t + \alpha\big[(r_{t+1}+\gamma q(s_{t+1},a_{t+1},\vtW_t)) -\nonumber\\
&\hspace*{1.4cm}q(s_{t},a_{t},\vtW_t))\big]\nabla q(s_{t},a_{t},\vtW_t)),
\end{align}
where the method remains an off-policy method with a mix between target policy and behavior policy.

\subsection{Policy Gradient and Actor-Critic Methods}\label{subsub:policy_actor_critic}
As mentioned, reinforcement learning methods can be further classified as temporal-difference, policy gradients,
or actor-critic methods~\citep{Sutton2018}.
We focus on temporal-difference methods because they are among the most used in the wireless spectrum sharing
literature, with the most popular methods being Q-learning and deep Q-learning.
Nevertheless, we discuss now briefly policy gradients and actor-critic methods.
In policy gradient methods, the objective is to learn a parameterized policy that can select actions without consulting
a value function.
The value function may still be used to learn the policy parameter, but it is not required for selecting an action.
Specifically, we denote by $\vtTheta\in\bbR^{d'}$ the policy parameter's vector and define the approximated policy as
$\pi(a|s,\theta)=\Pr\big(a_t=a|s_t=s,\vtTheta_t=\vtTheta\big)$.
Note that $\pi(a|s,\theta)$ denotes the probability that action $a$ is taken at time $t$ considering that the state at
time $t$ is $s$ with parameter $\vtTheta$.
Finally, actor-critic methods are methods that learn approximations to both policy and value functions, where the
\emph{actor} is a reference to the learned policy and \emph{critic} refers to the learned value function.
For an in-depth analysis of both policy gradients and actor-critic methods, we refer the interested reader
to~\cite[Chapter 13]{Sutton2018}.

\section{Summary}\label{sub:summary}
In this chapter, we revised the fundamentals of the \ac{ML} methods that are often used in 
spectrum sharing.
Specifically, we discussed the most often used methods in supervised, unsupervised, and reinforcement learning.
In the following chapters, we will discuss the state-of-the art in spectrum sharing methods that use \ac{ML}
methods to solve spectrum sharing problems.
For each type of spectrum sharing, we present the most often used \ac{ML} methods, and describe how they are 
adopted to the specific spectrum sharing problem. 

\chapter{Spectrum Sensing}\label{Sec:Sensing}
\section{Introduction}

\begin{figure*}[!t]
	\centering
	\includegraphics[width=1\textwidth]{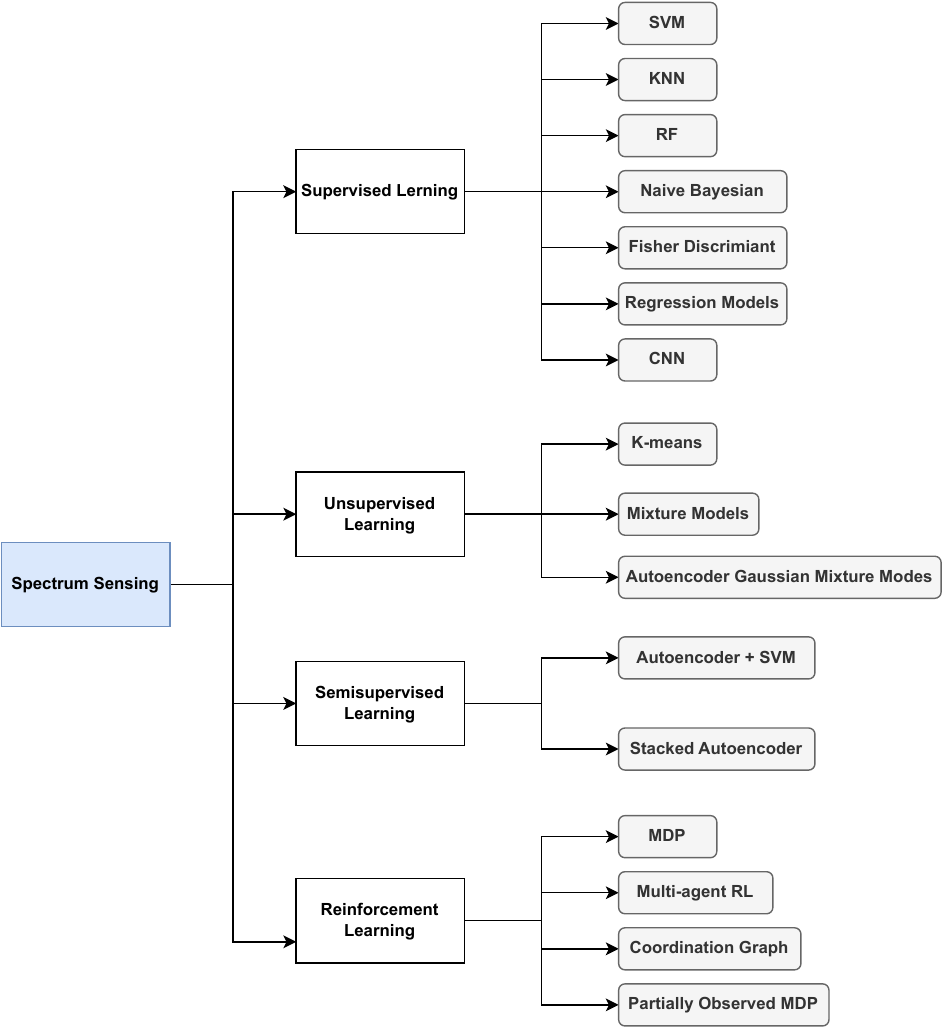}
	\caption{\ac{ML} approaches for spectrum sensing covered in this survey.}\label{Fig:Spectrum_Sensing}
\end{figure*}

In the previous chapter, we have surveyed the most prominent \ac{ML} methods that are used for spectrum sharing. In this section, we see in the detail how these methods are used for spectrum sensing. A summary of the \ac{ML} methods most frequently used in spectrum sensing is showed in  Fig. \ref{Fig:Spectrum_Sensing}. 

In the classical non-cooperative spectrum sensing unlicensed users sense radio frequencies bands in a multi-dimensional space (such as 
time, spatial, and frequency domain) to obtain information about channel usage and quality \footnote{In some database assisted 
spectrum sharing schemes such as TV white spaces and licensed shared access, for example, spectrum sensing does not exist because 
the sensing information is provided by spectrum databases.}.  
These users with sensing capabilities 
aim to detect whether a \ac{PU}, i.e., the one who owns the spectrum, is using the spectrum to make access decisions and/or parameters configuration. These capabilities improve network efficiency since power/interference constraints can be satisfied via some optimization task. 

The coexistence of multiple networks combined with the uncertainty of the wireless environment lead to various challenges on applying non-cooperative spectrum sensing \cite{Letaief2009}. The main challenge is related to the severe shadow situations faced by the \ac{SU} transmitters due to multipath fading or building obstruction leading to a hidden terminal problem. Another issue is related to different \ac{PU} devices on each \ac{CRN} using various transmission power levels, modulation schemes and/or data rates. To deal with these problems, cooperation among \acp{SU} was proposed \cite{Sendonaris2003}.
In  \ac{CSS} (\acl{CSS}), different closely located \ac{SU} nodes or wireless sensors can sense the \ac{PU} signals and share the obtained information among \acp{SU}.

Both non-cooperative and \ac{CSS} can be divided into two main categories according with the frequency range of sensing \cite{Arjoune2019}:
\begin{enumerate}
	\item \textit{Narrowband}: it refers to the case when the sensing is done in the frequency channel of interest. Different methods are proposed in the literature for narrowband spectrum sensing such as energy detection, cyclostationary feature detection, matched filter detection and eigenvalue detection \cite{Cabric2004,Zeng2009}. Although these schemes are simple and easy to implement, they are not a practical solution for high data rate requirements where a large bandwidth will be needed for wireless systems. Narrowband schemes only check the availability of particular channel and high bandwidth solutions are often achieved by carrier aggregation. 
	\item \textit{Wideband}: the sensing is done over a frequency range wider than the channel bandwidth to overcome narrowband low rate issues. Common schemes such Nyquist-based and sub-Nyquist-based,  known as \textit{compressive sensing}, are examples of wideband approaches \cite{Sun2013}. However, these schemes have complex implementation and they may also lead to high latency sensing time which leads to an impractical solution for real-time.
\end{enumerate}

The usage of \ac{ML} for spectrum sensing problem arose as an alternative to overcome those disadvantages for both narrowband and wideband schemes. Albeit \ac{ML} algorithms are more complex to apply, they outperform the classical state-of-the-art sensing algorithms in terms of detection accuracy and latency due their capacity to learning from the environment and surroundings as well their ability to adapt to environment changes \cite{Thilina2013}.

Regardless of which bandwidth or cooperative paradigm is used, the main goal of spectrum sensing is to detect the presence or absence of a transmitting signal in a given time instance $t$. Therefore, we have two hypothesis, $\mathcal{H}_1$ and $\mathcal{H}_0$ , where the former represents the presence of a \ac{PU} transmission and the latter indicates absence of transmissions in such channel. Thus, we can model the receiving signal $y_i$ at the cognitive node $i$ as:
\begin{equation}
	\begin{aligned}
&\mathcal{H}_0: y_i = n_i, \\
&\mathcal{H}_1: y_i = x_i + n_i,
	\end{aligned}\label{eq:spectrum_sensing}
\end{equation}
where $n_i$ is the additive white Gaussian noise and $x_i$ is the signal to be detected for user $i$.

\begin{figure}[!t]
	\centering
	{\includegraphics[width=0.75\textwidth]{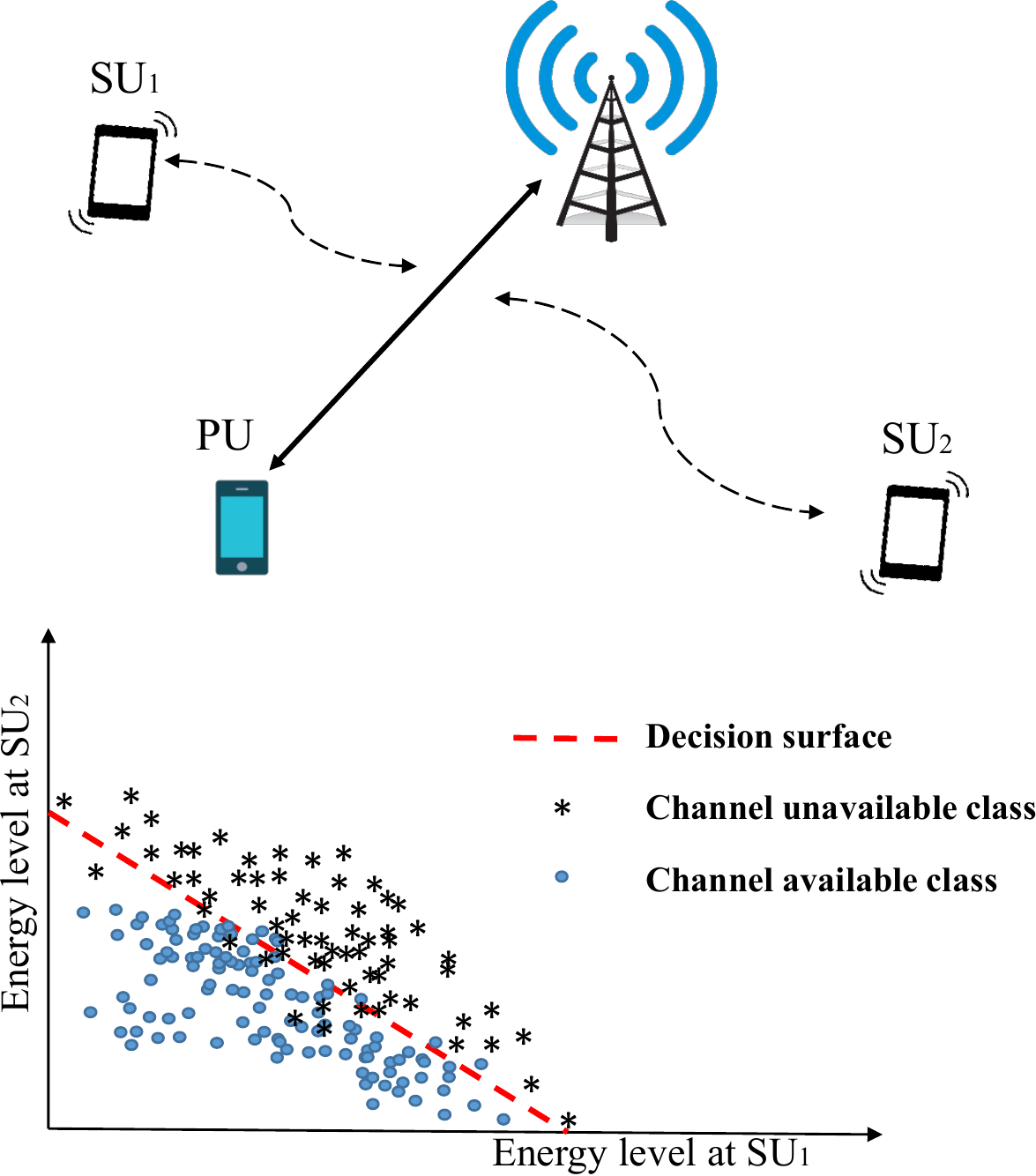}}
	\caption{Illustration of \ac{ML} usage for cooperative spectrum sensing. The \acp{SU} sense the desired channel and the classifier decides which class the collected signal belong to. }
	\label{fig:spectrumsensing}
\end{figure}

Then, spectrum sensing can be viewed as a classification problem where the  classifier has two possible classes: channel free class ($\mathcal{H}_0$) or channel occupied class ($\mathcal{H}_1$). The availability check is done based on energy statistic, occupancy over time, probability vectors or other features obtained from signal statistics which are feed into a classifier. Fig. \ref{fig:spectrumsensing} illustrates the usage of binary classification for a \ac{CSS} scenario based on energy levels. The spectrum is considered occupied when both $SU_1$ and $SU_2$ measures are above the decision surface, otherwise, the channel is available for transmission.

\ac{ML}  can deal with the spectrum sensing classification problem using three different paradigms: supervised learning only, 
unsupervised learning only
or semisupervised learning 
frameworks. Another possibility is to solve the classification problem by \ac{RL} approaches.

\section{Supervised Spectrum Sensing}

As we saw in Section \ref{sub:sup_learning}, supervised schemes focus on solutions where labeled training data is needed. In this scheme, each feature vector is labeled with its corresponding channel availability class ($\mathcal{H}_0$ or $\mathcal{H}_1$). Despite this requirement, supervised spectrum sensing tends to exhibit a good performance due the channel availability information obtained at the corresponded labels.  Many supervised classifiers have been considered in the literature for spectrum sensing problem.

Motivated by the results found in \cite{Thilina2013} and \cite{Zhang2011}, many works explored \ac{SVM} classification 
for the spectrum sensing problem due to its easier implementation, capacity to handle non-linear data, good accuracy 
detection when binary classes are used and high performance in terms of \ac{ROC} 
\cite{Subekti2018,Saber2019,Li2018a,Awe2018,Coluccia2019,Awe2013,Sabra2023,Tang2023,Jan2018,Lu2016}. 

\ac{k-NN} is one of the simplest \ac{ML} approaches (see Section \ref{subsub:k-nn}). The classification is based on neighbor votes, which are the 
training points of the feature vectors. 
For $k = 1$, the classification problem relies on calculate the Euclidean distance between $y_i$ and the feature vector. The chosen label is the same label as the point with minimum distance. 
This algorithm is very suitable for low complex requirements and it is also a stable \ac{ML} algorithm 
\cite{Kiang2003}.

Differently from \ac{k-NN}, \ac{SVM} (see Section \ref{subsub:svm}) tries to find a linearly separating hyperplane, with 
the help of support vectors, by satisfying the following condition
	\begin{equation}
		\begin{aligned}
			h(\vtX_i;\vtW) &\geq \mu_0, \quad  \text{if channel is free},\\
			h(\vtX_i;\vtW) &\leq \mu_1, \quad  \text{if channel is busy},
		\end{aligned}
	\end{equation}
where $\mu_0$ and $\mu_1$ are the decision surface for class $\mathcal{H}_0$ and $\mathcal{H}_0$ respectively\footnote{Usually $\mu_0$ = 1 and $\mu_1$ = -1 for binary classification problems.} and $h(\vtX_i;\vtW)$ is one of the prediction function in Table \ref{tab:pred_func} of Section \ref{sub:sup_learning}.

References \cite{Thilina2013} and \cite{Lu2016} showed that linear kernel \ac{SVM} has a lower training computational 
complexity with a better \ac{ROC} performance than polynomial kernel \ac{SVM}, however, the later presents a lower 
classification delay with better probability detection. 

Although, in general, \ac{SVM} algorithms outperform \ac{k-NN} 
in terms of \ac{ROC} performance \cite{Thilina2013}, \ac{k-NN} can exhibit a better classifier accuracy depending on 
which \ac{k-NN} and \ac{SVM} classifiers are used and the which feature vector is used for classification 
\cite{Mikaeil2014,Saber2019}.

Authors in \cite{Li2018a} proposed a group optimization algorithm for \ac{CSS} to reduce the cooperation overhead. The energy vectors are collected from the receiving signal by each group of users and then fed into the \ac{SVM} classifier, improving the sensing efficiency. Using energy vectors results in low performance at low \ac{SNR}, other works considered  different strategies to overcome this issue. Reference \cite{Awe2018} explored a beamforming aided energy vector strategy  to increase the \ac{SNR} for \ac{SVM} training. The proposed algorithm provides high performance for a multi \ac{PU} temporal and spatial detection problem where the main task was to find the number of active \acp{PU} as well as their location. A combination of energy values and eigenvectors as feature vectors and their high-order statistics was proposed by \cite{Coluccia2019}. The authors showed that the chosen combination is able to separate the two available classes and identify the \ac{PU} presence even  under low \ac{SNR}.  Instead of using energy vectors, references \cite{Awe2013} and \cite{Sabra2023} explore covariance matrix features to sense spectrum. In the former, the authors performed an eigenvalue estimation with \ac{SVM} in a \ac{PU}-\ac{SU} scenario, leading to a  good performance in detection rate accuracy with minimum sensors aid. On the other hand, in the latter, the authors showed the impact of realistic correlated noise on \ac{SVM}-assisted \ac{CSS}-based technique.  In \cite{Tang2023}, the authors used \ac{SVM} to classify the  maximum value of the spatial spectrum and the angle of arrival difference. The proposed approach provided robustness against noise uncertainty outperforming energy detection and covariance absolute value approaches at low \ac{SNR}. The channel free class was divided into various classes in \cite{Jan2018} so the \ac{SU} could decide a better transmission power and optimize the network performance. Using heuristic statistical features for the multi-class \ac{SVM} sensing problem, the proposed method increases the \ac{SU} capacity outperforming the conventional state-of-the-art algorithms.

Other supervised learning algorithms were used in literature for spectrum sensing problem such as:  Fisher discriminant 
\cite{Ding2013,Thilina2013}, naive Bayesian \cite{Xu2018}, 
 regression methods \cite{Khalfi2017,Reddy2023,Wang2016,Valadao2021,Wang2023}, and \acl{PSO} \cite{Ghazizadeh2016}, as we describe in the detail in the following.

As shown in Section \ref{subsub:fisher_analysis}, Fisher discriminant aims to solve Eq.(\ref{eq:fisher_problem_form}) to find the best feature set to separate $\mathcal{H}_0$ from $\mathcal{H}_1$. In this approach the dataset $x_i$ is mapped into a feature space via linear or non-linear function.  Reference \cite{Ding2013} explored  both kernel solutions for  \ac{CSS} \ac{PU} detection. The results showed that non-linear kernel outperforms  linear approach \ac{SVM} in terms of lower average error probability at the expense of higher computational complexity. Although they did not achieve a better performance when compared to the likelihood-ratio test detector, the proposed solutions are more practical in terms of implementation which justifies their application in the studied scenario. On the other hand, as can be observed in Eq. (\ref{eq:gen_Bayes}), the Bayesian approach (see Section \ref{subsub:bayes_class})  returns the class which has the maximum posterior probability in the sample. This method is used jointly with \ac{HMM} in \cite{Xu2018} to capture the vector features using a naive Bayesian learning algorithm. The \ac{HMM} (see Section \ref{subsub:hidden_markov}) computes a probability distribution over possible sequences of labels, and chooses the best label sequence. This approach, however, can lead to redundant states. To deal with this issue, the authors included a self-transition bias weight in the algorithm to enhance self-transition probability. This strategy is know as sticky \ac{HMM} \cite{Fox2011}. The results showed that the  \ac{CSS} performance can be increased and better results are achieve by using spatio-temporal correlation to improve the sensing capability using the minimum of cooperating \acp{SU}.

In Table \ref{tab:pred_func} we presented \ac{ML} regression models and their loss and prediction functions that often are used in spectrum sharing problems. Linear regression models (see Section \ref{subsub:lin_log_regre}) were used in \cite{Khalfi2017} to address wideband \ac{CSS} problems. By using past spectrum occupancy information and current spectrum measurement as classifier features, the authors used gradient descent and \ac{SVR} as linear regression methods. The results showed that both methods present a better performance than nonlinear \ac{SVR} and conventional wideband sensing schemes. In linear regression models, the probable value of an unknown variable is estimated by a linear mixture of values. Therefore, the expected value of the response variable (component of at least one variable) is considered a constant. To overcome this issue, reference \cite{Reddy2023} applied a generalized linear model. Instead of using a linear function to estimate the expected value of the response variable, this model uses a linear mixture of inputs to a nonlinear function. This results in a response variable with arbitrary distributions, increasing the prediction performance. Another regression method used in the literature for spectrum sensing is the random forest algorithm (see Section \ref{subsub:random_forest} for details). One of its advantages is the low training time and acceptable accuracy for small number of decision classes. These characteristics motivated authors in \cite{Wang2016} to apply random forest in a three class spectrum sensing problem. The classifier receives the energy data feature and predicts which of the three available classes the signal belongs to: idle (no \ac{PU}), busy (a nearby \ac{PU} is transmitting) or middle (the transmitting \ac{PU} is relatively far from the \ac{SU}). By allowing the \ac{SU} to transmit on both idle and middle classes, the \ac{SU} throughput was increased improving the overall network performance in a non-cooperative spectrum sensing. On the other hand, authors in \cite{Valadao2021} proposed a \ac{CSS} framework where each unlicensed user applies a random forest algorithm to detect whether a licensed user is present. The sensing results are combined in a fusion center which determines each sensing band status by applying a residual neural network model. Another spectrum sharing framework was presented in \cite{Wang2023}. The authors explored classification and regression tree algorithms to determine the \ac{RSS} classification of the training data.

Differently from the aforementioned works, reference \cite{Ghazizadeh2016} combined  \ac{SVM}, \ac{k-NN} and naive 
Bayes algorithms by using \acl{PSO}. The authors trained the three classifiers with the energy vectors given a 
coefficient between 1 and 0 for each label. The \acl{PSO}, then, adjusts the classifier weights in a optimal way. 
Although it requires more time for training and needing to build three classifiers, the proposed method reduced both 
error rate and the error of channel available class.

Adaptive neural networks can adjust their learning parameters when the  environment changes, such as the addition of more \acp{PU} or power profile changing, presenting a stable performance if the number of training samples are sufficient. This feature was explored in  \cite{Tang2010,Vyas2017,Wang2022a,RajeshBabu2022,Zheng2020,Lee2019,Liu2019c,Liu2019d,Xie2019,Lees2019,He2019,,Yuan2023,Han2017,OShea2018} to classify the sensing data and determine the channel availability for \ac{SU} transmission. Hybrid detection schemes  where the energy detection was combined with cyclostationary feature \cite{Tang2010} and likelihood ratio test statistic \cite{Vyas2017} were proposed to improve the energy detection performance for low \ac{SNR} values, proving the neural network effectiveness for spectrum sensing problem. An adversarial learning strategy combining three \acp{NN} (a feature extractor, a predictor, and a discriminator) was proposed by \cite{Wang2022a}. This approach reduces the \ac{SNR} dependence of the extracted features and improves the sensing detection. Extreme learning  technique can also be applied to increase detection rate \cite{RajeshBabu2022}. 

In recent years, \ac{DL} has called attention due to its performance gain when applied to large amount of data \cite{Zhang2022}. Due to the possible coexistence of multiple \acp{CRN}, the demand for \acp{SU} with capability to process and classify lots of sensing data motivated the usage of \ac{DL} architecture for spectrum sensing.  This approach is also explored in the literature to build spectrum sharing architectures for \ac{LTE} and nextG networks \cite{Baldesi2022}. Among the \ac{DL} approaches, \ac{CNN} (see Section \ref{subsub:CNN}) has been widely applied due to its excellent performance as a classifier and less parameter training needed when compared to a fully connected neural network. This efficiency was demonstrated by \cite{Lees2019} in radar detection at 3.5 GHz band for energy based methods and \cite{Han2017} for cyclostationary scheme at different \acp{SNR}  using a combination of energy vectors and cyclostationary features. The \ac{CNN} schemes offered a better accuracy-complexity tradeoff than the compared methods. Moreover, reference \cite{Zheng2020} showed that \acp{CNN}  can also be applied in general sensing problems, where the signal to be sensed is unknown, being robust to noise uncertainty even in the presence of colored noise.

\ac{CNN} has a strong capability in extracting features of matrix-shaped data which motivates the use of the covariance matrix as the input in \cite{Liu2019d,Liu2019c,Xie2019}. This approach relies on the fact that priori \ac{PU} information is not available in practice so, to deal with this issue, the covariance matrix from both signal and noise is taken into account. Then, Equation (\ref{eq:spectrum_sensing}) turns into:
\begin{equation}
	\begin{aligned}
		&\mathcal{H}_0: \mathbf{R_{y_i}} = \sigma^2_{n_i}\mathbf{I_m}, \\
		&\mathcal{H}_1: \mathbf{R_{y_i}} = \mathbf{R_{x_i}} + \sigma^2_{n_i}\mathbf{I_m},
	\end{aligned}\label{eq:hipothesis_cov_matrix}
\end{equation}
where $\sigma^2_{n_i}$ is the noise variance, $\mathbf{I_m}$ stands for the identity matrix of order $m$ and $\mathbf{R_{y_i}} = E[aa^H]$ is the covariance of a random variable $a$, where $E[\cdot]$ is the expectation operator and $H$ denotes the conjugate.

Reference \cite{Liu2019d} proposed a solution based on two stages. A offline training using a \ac{CNN} is first done to generate a test statistic. Then, an online decision is performed by comparing the generated statistic with a predefined threshold designed to satisfy a desired probability of false alarm. This work was extended by \cite{Liu2019c} where a theoretical analysis of the \ac{CNN} performance was provided. On the other hand, authors in \cite{Xie2019} took into consideration both present and historical sensing data to train the \ac{CNN}. The active pattern based scheme improved the \ac{PU} detection outperforming both estimator-correlator detector and the \ac{HMM}-based detector. \acp{CNN} can also be applied in \ac{CSS} problems. Based on the \ac{CNN} characteristic to extract spatial features from the input data, reference \cite{Lee2019} proposed an architecture that explored the spatial and spectral energy vectors correlation from the cooperative \acp{SU}. This scheme outperformed \ac{SVM} for \ac{PU} detection with a small sized \ac{CNN} structure. \ac{CNN} can be used in cooperative spectrum sensing with \ac{LSTM} (see Section \ref{subsub:RNN}) to explore its ability to learn	temporal features from sequential data \cite{Li2023,Janu2023}.  model based on \ac{CNN} to extract spatial features and  to extract time features is proposed. This approach was shown to be effective for low \ac{SNR} values and capable to improve detection probability and classification accuracy.

Traditional \ac{CNN} and \ac{LSTM} approaches, however, are not good at capturing the temporal correlation features from spectrum sensing data \cite{Xiao2021,Schuster1997}. To overcome this limitation, authors in \cite{Xing2022} proposed a three-layer \ac{DNN} combined by a 1-dimensional \ac{CNN}, responsible to analyze and extract local signal features, a bidirectional  \ac{LSTM} which captures the global correlation features in both directions, and a self-attention network  to help the classifier on relating temporal features in the time series. This strategy leads to low probability of miss detection and false alarm, especially at low \ac{SNR} values.

Other neural networks were also used in literature for spectrum sensing problems. Reference \cite{OShea2018} explored the residual networks for signal classification. Differently from \cite{Valadao2021}, the authors considered channel impairments such as carrier frequency offset and multipath fading. The work showed that residual networks are more effective than a traditional \ac{CNN} approach to build deep networks. Another relevant work is presented in \cite{He2019} where authors have developed a graph neural network (GNN) architecture to improve the overall system energy efficiency for \ac{CSS} in a sensor-aided \ac{CRN}. The network weights were trained using Q-learning so the best parameters could be found. The GNN is also explored in \cite{Yuan2023} to allow multiple satellites to fuse their signals for spectrum sensing.

Table \ref{tab:spectrum_sensing_supervised} presents a summary of supervised learning approaches for spectrum sensing problem.

\begin{table*} 
	\centering
	\caption{Summary of \ac{ML} works for spectrum sensing using supervised learning approach.}
	{\scriptsize  \begin{tabularx}{\linewidth}{
				|>{\hsize=0.55\hsize}X|
				>{\hsize=1.35\hsize}X|
				>{\hsize=1.1\hsize}X|
			}
			\hline
			\centering \textbf{Supervised \\ Learning Approach} &
			\vspace*{0.10cm} \centering \textbf{Comments}  & 
            \vspace*{0.10cm}  {\hspace*{0.9cm}\textbf{Related Works}} 
			\\ \hline		
			\vspace*{0.7cm} \centering Support Vector Machine  &   
            \begin{tabitemize}
                \item High performance on \ac{PU} detection;  
                \item The increment of the number of training samples increases the  classification delay;
                \item Classification time depends on the dimension of  feature vectors, the 
                number of support vectors and the Kernel function.
            \end{tabitemize} &  
		\vspace*{0.7cm} \parbox[c]{4cm}{\centering	\cite{Li2018a,Awe2018,Tang2023,Awe2013,Zhang2011,Coluccia2019,Sabra2023,Jan2018,Lu2016,Ghazizadeh2016,HaozhouXue2015,Saber2019}}
			     \\ 
			\hline
			\vspace*{0.6cm} \centering K-nearest-neighbor   & 
            \begin{tabitemize}
                \item Low training duration;
                \item High classification delay ;
                \item Good performance on \ac{PU} detection; 
                \item Works well for low \acp{SNR} if trained at these conditions; 
                \item Excellent accuracy for classifying new frames.
            \end{tabitemize}  
            & \vspace*{0.3cm} \parbox[c]{4cm}{\centering \cite{Thilina2013,Tiwari2018,Mikaeil2014,Ghazizadeh2016,Saber2019} }   \\  
			\hline
			\vspace*{0.6cm} \centering Fisher \mbox{Discriminant}    & 
            \begin{tabitemize}
                \item Low training duration  and classification delay; 
                \item Performance of non-linear detector increases with the number of training samples; 
                \item Non-linear kernel has high computational complexity.  
            \end{tabitemize} 
            &   \vspace*{0.6cm} \parbox[c]{4cm}{\centering\cite{Ding2013,Thilina2013} }   \\
			\hline		
			\vspace*{0.6cm} \centering Naive Bayesian   & 
            \begin{tabitemize}
                \item Low classification duration; 
                \item High performance if predictors independence holds; 
                \item Can be used jointly with \ac{HMM} to improve classification performance;
                \item It does not work well for a category not present in the training set.  
            \end{tabitemize}
            &   \vspace*{0.5cm} \parbox[c]{4cm}{\centering\cite{Xu2018,Ghazizadeh2016,Mikaeil2014}  }   \\
			\hline
			\vspace*{0.6cm} \centering Random Forest    &  
            \begin{tabitemize}
                \item Low training duration; 
                \item Good performance for small number of classes; 
                \item Overfitting can be reduced by tunning hyper parameters; 
                \item Large datasets can lead to a slow classification  performance.
            \end{tabitemize}  
            & \vspace*{0.5cm} \parbox[c]{4cm}{\centering \cite{Wang2016,Valadao2021,Wang2023}  }    \\
			\hline
			\vspace*{0.8cm} \centering Regression Models    &  
            \begin{tabitemize}
                \item Reduces the sensing measurements;  
                \item Works well to predict spectrum occupancy;  
                \item Linear regression has a better false alarm and  miss-detection performance; 
                \item Non linear Support Vector Regression has a better performance  on spectrum occupancy prediction. 
            \end{tabitemize}
             &   \vspace*{0.8cm} \parbox[c]{4cm}{\centering \cite{Khalfi2017,Reddy2023} }   \\
			\hline
		\vspace*{1cm} \centering Neural Network  & 
            \begin{tabitemize}
                \item Dynamic learning from the signal features; 
                \item Ability to detect untrained signals; 
                \item Capability to classify huge amount of sensing data; 
                \item Some models are vulnerable to modeling uncertainties;  
                \item Some models are complex to achieve high performance, increasing offline training duration. 
            \end{tabitemize}
             & 
             \vspace*{1cm} \parbox[c]{4cm}{\centering \cite{Tang2010,Vyas2017,Wang2022a,RajeshBabu2022,Zheng2020,Lee2019,Liu2019c,Liu2019d,Xie2019,Zhang2022,Baldesi2022,Lees2019,He2019,Yuan2023,Han2017,OShea2018,Hayashida2019,Subekti2018,Cheng2019,Li2023,Janu2023,Xing2022}}
                 \\
			\hline
	\end{tabularx}}
	\label{tab:spectrum_sensing_supervised}
\end{table*}

\section{Unsupervised Spectrum Sensing}

For unsupervised learning, no information about channel availability or \ac{PU} knowledge is required (see Section \ref{sub:unsup_learning}). The signal features needed are collected from sensing measurements which are naturally unlabeled and do not require any explicit training. The unsupervised algorithms learn hidden patterns from sensing data via clustering, association or dimensional reduction techniques.

The K-means algorithm (see Section \ref{subsub:K_means}) was explored in some works in the literature for different aspects of spectrum sensing problems. Considering the problem defined in Eq. (\ref{eq:spectrum_sensing}), the number of clusters K is, in general, defined as two, one for each class ($\mathcal{H}_0$ and $\mathcal{H}_1$). Then, two centroids are defined and Eq.(\ref{eq:within_cluster_kmeans_loss}) is used to define the classes.  References \cite{Thilina2013} and \cite{Lu2016} analyzed the performance of K-means comparing with other \ac{ML} algorithms for \ac{CSS} scenarios. While the former used energy vector as signal feature, the latter considered a probability vector since it can lower the energy vector dimension to a bi-dimensional vector, decreasing the training and classification time at the expense of a worse detection probability. The K-means algorithm determines from which class the feature vector belongs to by clustering the received data into one of the available classes. The works showed that the K-means algorithm presents  good performance in terms of \ac{PU} detection probability with low delay on training and classification stages being a suitable approach for \ac{CSS}.

A different use case of the K-means algorithm was explored in \cite{Hayashida2019,Subekti2018,HaozhouXue2015}. In these works, the sensing process is divided into two phases: the first one using an unsupervised learning K-means and the second phased using a supervised learning approach. Section \ref{subsec:Unsup+Sup} discusses in detail the case when semisupervised learning paradigm is used in the spectrum sensing problem.

Differently from K-means, \acp{MM} are probabilistic methods that obtain a classification under uncertainty of which group the collected data belong to (see Section \ref{subsub:mixture_models}). They work well on identifying a \ac{PU} presence problem where the collected signal sample needs to be classified with best accuracy as possible even in the presence of outliers due to fading and/or shadowing. \ac{MM} algorithms were applied in the literature for both narrowband \cite{Zhou2023,Zhang2019,Tiwari2018,Xie2020,Thilina2013} and wideband \cite{Qi2018a} spectrum sharing problems. 

The authors in \cite{Thilina2013} showed that \acp{GMM} can adjust their decision surface adaptively for different \ac{CR} scenarios. Moreover, this adjustment can be close to optimal if  energy vectors feature are considered, improving the \acp{PU} detection.  Although \acp{GMM} requires a high training time,  reference \cite{Zhou2023} combined \ac{GMM} with a \ac{PSO} estimator as an alternative to maximum likelihood approach for channel coefficient estimation due its lower computational complexity. In \cite{Zhang2019}, it was used jointly with a conjugate Dirichlet process in  to identify the current \ac{PU} power transmission for a multi-level spectrum sensing. While the \ac{GMM} clustered the \acp{PU} signals into different classes, the  conjugate Dirichlet process \ac{GMM} was used to determine the current \ac{PU} power level based on previous \ac{GMM} classification. To overcome the training delay from \ac{GMM}, authors in \cite{Tiwari2018} used a log-Rayleigh \ac{MM} to capture signal distribution information. Since in the log scale the magnitude of the complex noise follows a log-Rayleigh distribution, the proposed \ac{ML} solution has a faster training response when compared to TxMiner \cite{Zheleva2015}, which uses \ac{GMM}, presenting a high accuracy performance for multi-transmitter detection. Reference \cite{Xie2020} used a variational \ac{AE} combined with the \ac{GMM} creating a \ac{DL} structure using features extracted from the covariance matrix. The proposed model only requires a small amount of samples collected in absence of the \ac{PU}'s signal since both variational \ac{AE} (see Section \ref{subsub:autoencoders}) and \ac{GMM} are robust with random initialization. The proposed model approached supervised learning performance with less training requirements and no prior knowledge about noise power or signal's statistics.

For wideband compressive sensing, \cite{Qi2018a} observed that if energy statistic is considered as features for sensing, the channel free class follows a central chi-square distribution while the channel occupied class follows a noncentral chi-square distribution. Based on this observation, the authors proposed an adaptive sensing scheme based on chi-square-\ac{MM} to maintain constant false alarm rates in channel energy detection validated by simulation results.

Table \ref{tab:spectrum_sensing_unsupervised} presents a summary of unsupervised learning approaches for spectrum sensing problem.

\begin{table*}[!t] 
    \centering
    \caption{Summary of \ac{ML} works for spectrum sensing using unsupervised learning approach.}
    {\footnotesize  \begin{tabularx}{\linewidth}{
                |>{\hsize=0.65\hsize}X|
                >{\hsize=1.35\hsize}X|
                >{\hsize=1.00\hsize}X|
            }
            \hline
            \centering \textbf{Unsupervised Learning Approach}  & 
            \vspace*{0.10cm} \centering \textbf{Comments}  & 
            \vspace*{0.10cm}  {\hspace*{0.6cm}\textbf{Related Works}} 
            \\ \hline	
           \vspace*{1.2cm} \centering K-means   &  
            \begin{tabitemize}
                \item Number of samples does not affect the classification delay; 
                \item Good performance on \ac{PU} detection;  
                \item Classification time depends only on the dimension of the testing feature vector.
            \end{tabitemize}
            &  \vspace*{0.9cm} \parbox[c]{3cm}{\centering \cite{Thilina2013,Lu2016,HaozhouXue2015,Hayashida2019}    }     \\
            \hline
            \vspace*{1.3cm} \centering Mixture Models  &
            \begin{tabitemize}
                 \item Good performance on multi-level classification; 
                 \item Good performance on multi-transmitter detection; 
                 \item Requires proper parameter initialization; 
                 \item The increment of the number of training samples increases the classification delay.  
            \end{tabitemize}
            & \vspace*{0.8cm} \parbox[c]{3cm}{\centering \cite{Zhou2023,Zhang2019,Tiwari2018,Xie2020,Thilina2013,Qi2018a} }     \\
            \hline
           \vspace*{0.4cm}  \centering Autoencoder Gaussian Mixture Model       & 
            \begin{tabitemize}
                \item Suitable as a test statistic in spectrum sensing; 
                \item Low susceptibility to overfitting; 
                \item Low accuracy without training.
            \end{tabitemize}
            & \vspace*{0.7cm}  \parbox[c]{3cm}{\centering\cite{Xie2020}  }       \\
            \hline		
    \end{tabularx}}
    \label{tab:spectrum_sensing_unsupervised}
\end{table*}

\section{Semisupervised Spectrum Sensing}\label{subsec:Unsup+Sup}
In the previous sections just one classification paradigm was used to perform spectrum sensing. Although unsupervised algorithms converge faster, are simpler to implement and do not need any prior knowledge of the sensing data, they provide a worse detection accuracy when compared with supervised learning algorithms. Some works explored the advantages of both learning tasks and developed a hybrid sensing scheme composed of two stages. The first stage is done by a supervised learning algorithm whose output is used at the second stage to train the supervised learning algorithm. The main objective is to reduce the training delay observed at supervised learning algorithms while keeping a good sensing performance.

Reference \cite{HaozhouXue2015} combined K-means and \ac{SVM} algorithms. Firstly, K-means is used to discover \ac{PU}'s pattern based on covariance matrix features. Then, the \ac{SU} uses the labels provided by K-means to train the \ac{SVM} algorithm to perform the \ac{PU} detection. This method is blind since it does not need any a prior information about \ac{PU}'s signal, channel or the noise power. Another strategy using K-means was explored in \cite{Hayashida2019} benefiting from its fast convergence and high performance in the absence of outliers. This feature fits well in sensing-location problems where the main goal is to find the geographic location of \acp{PU} via clustering sensing data. K-means is used to generate a location estimation cluster for the sensing data and then a \ac{DNN} was to establish a relationship between the clustered data and the measured delay profile.

Other works explored \acp{AE} as unsupervised learning method. In \cite{Subekti2018} a spectrum monitoring scheme using an \ac{AE} neural network and \ac{SVM} was proposed to determine what user is accessing the spectrum. Using an image based classifier, the \ac{SVM} receives the \ac{AE} \acl{NN} output and it classifies the signal as \ac{PU} or \ac{SU} with high  accuracy. On the other hand, the authors in \cite{Cheng2019} used a stacked \ac{AE} network, which combines an \ac{AE} with logistic regression, to sense orthogonal frequency division multiplexing signals. This strategy requires less connections than \ac{CNN} to retain essential information about signal features.

Table \ref{tab:spectrum_sensing_semisupervised} presents a summary of semisupervised learning approaches for spectrum sensing problem.

\begin{table*} [!t]
	\centering
	\caption{Summary of \ac{ML} works for spectrum sensing using semi-supervised learning approach.}
	{\footnotesize  \begin{tabularx}{\linewidth}{
				|>{\hsize=0.9\hsize}X|
				>{\hsize=1.35\hsize}X|
				>{\hsize=0.75\hsize}X|
			}
			\hline
			\parbox[c]{2.5cm}{\centering\vspace*{0.05cm}\textbf{\mbox{Semi-supervised} \mbox{Learning Approach}}}  & 
			\parbox[c]{4cm}{\centering\vspace*{0.05cm}\textbf{Comments}} & \parbox[c]{2.5cm}{\centering\vspace*{0.05cm}\textbf{Related 
			Works}}
			\\ \hline	
			
			\vspace*{0.07cm}\centering Autoencoder and \\ Support Vector \\ Machine   & 
            \begin{tabitemize}\vspace*{-0.2cm}
                \item Lowers the number of labels for \ac{SVM} classification; 
                \item Performance depends on \ac{SVM} Kernel.  
            \end{tabitemize}
             & \vspace*{0.3cm} \parbox[c]{2.5cm}{\centering\cite{Subekti2018} }    \\
			\hline
			\vspace*{0.9cm} \centering Stacked \\ Autoencoder      &  
            \begin{tabitemize}
                \item Architecture suitable for practical scenarios with a limited amount of labeled data;  
                \item Retain essential signal features; 
                \item Simple and easy training; 
                \item Intermediate to high online computational complexity; 
                \item Large amount of parameters.
            \end{tabitemize}
			& \vspace*{1.2cm} \parbox[c]{2.5cm}{\centering\cite{Cheng2019} }        \\
			\hline
	\end{tabularx}}
	\label{tab:spectrum_sensing_semisupervised}
\end{table*}

\section{Reinforcement Learning Spectrum Sensing}\label{subsec:RL_Sensing}
Different from the tradition spectrum sensing classification using unsupervised, supervised or semisupervised learning, most recent 
works considered \ac{RL} and \ac{DRL} to solve the sensing problem (see Section \ref{sub:reinfo_learning} for reinforcement learning). 
These approaches are most effective when optimizing a certain objective (throughput, energy consumption and spectrum usage, e.g.) or 
imbalanced data distributions \cite{Lin2020}. This is true because \ac{RL} enables the \acp{SU} to provide a good performance by 
choosing a set of actions that conserve time and energy while providing a good performance \cite{Pham2023}. 

The most common \ac{RL} models for spectrum sensing include Markov Decision Process  \cite{Xu2020a,Zhang2019b,Li2020a,Sarikhani2020,Liu2021,Bokobza2023,Sun2022,Ngo2023}, multi-agent \ac{RL} \cite{Jiang2021,Kim2023,Li2023a,Zhang2023}  and coordination graph \cite{Cai2020}.

In Markov decision process, a simple approach is to model each channel with two possible status (free or occupied) and define a transition matrix of the Markov	chain and the channel state transition. The \ac{RL}, then, is used to implement the best sensing strategy based on action and reward to optimize a given objective. This strategy is often used jointly with a \ac{DQN} to enable the agent to deal with complex states and actions and also to reduce Q-learning complexity.

In \cite{Xu2020a}, the authors combined the \ac{RL} concept with spectrum
sensing to propose an architecture to optimize \acp{CRN} performance on different \ac{5G} scenarios. The \ac{RL} learning engine was designed to strengthen sensing accuracy and false alarm probability indicators. In \cite{Zhang2019b}, the \ac{DRL} was used to perform cooperative spectrum sensing in \acp{CRN}. The main idea is to capture sensing results in a distributive way so \acp{SU} can learn the best sensing strategy. The algorithm was combined with to upper confidence bounds with Hoeffding-style to improve the exploration efficiency.  Reference \cite{Li2020a} applied a \ac{DQN} to reduce computational complexity and to address the uncertainties of wireless networks dynamics on spectrum sensing scenarios. The \ac{DQN} framework can successfully sense and select a channel segment according to the user bandwidth demand by interacting with the environment with no training data set. On the other hand, \cite{Sarikhani2020} used \ac{DQN} to reduce the \acp{SU} signaling overhead. The energy samples are collected from neighbor nodes in different bands to feed the \ac{CNN} in order to obtain the global sensing result. Applying \ac{RL} in a spectrum sensing scenario yields an unknown environment with a large state space.  To efficiently improve learning in a such scenario, a combination of \ac{DQN} and \ac{ANN}, called \ac{DDQN} \cite{VanHasselt2015}, can be used. In this strategy, each user can  learn the channel correlations by individual training to detect if the acknowledgment signal is received correctly \cite{Liu2021}. This approach results in decreasing the sensing bandwidth with only a minor decrease in the throughput \cite{Bokobza2023}. It can also be used to perform a skip sensing \cite{Sun2022}, where an idle time slot can be skipped from sensing to avoid unnecessary spectrum sensing process. Another strategy to increase spectrum sensing performance, is to consider both current and past sensing data \cite{Ngo2023}. In this case, a \ac{RNN} can be used to capture the \ac{PU} spectrum occupancy  historical data and a post-decision state learning accelerates the learning speed.

Another \ac{RL} strategy is \ac{MARL}. In this approach, the spectrum sensing scenario is modeled the same way as Markov decision process, however, the actions from all agents have influence in the state and in the reward of each individual agency. In \cite{Jiang2021}, the authors modeled a \ac{UAV} network spectrum sensing problem as a \ac{MARL} where the real-time reward was composed by a sensing-transmission cost and utility. Q-learning and deep Q-learning algorithms are proposed based on independent learner to perform a dynamic cooperative spectrum sensing. Reference \cite{Kim2023} combined \ac{MARL} with negotiated aspirations bargaining solution to share the limited spectrum resource by explore a dynamic cooperation game model to efficiently control the spectrum sharing process in various system scenarios. Another application explored in the literature was done in \cite{Li2023a} for cellular vehicle-to-everything networks. The authors combined an Indian buffet process to predict the channel selection probability with a \ac{DQN} to select the best sensing channel based on this probability. A \ac{DDQN} solution with \ac{MARL} is addressed in \cite{Zhang2023} where spectrum sensing is done jointly with spectrum access. In this work, each \ac{SU} learns the channel characteristics to adjust its the sensing window and perform a power allocation strategy based on \ac{DQN} algorithm. This strategy enables \acp{SU} to achieve high data rate with low mutual interference.

A different spectrum sensing strategy is to use a coordination graph to reduce the problem sparsity \cite{Cai2020}. This is done by decomposing the global Q-learning reward into a max-plus problem which can be solved by message passing. The authors also applied a \ac{DQN} to increase the algorithm convergence speed. 

Table \ref{tab:spectrum_sensing_semisupervised} presents a summary of \ac{RL}  approaches for spectrum sensing problems.

\begin{table*}[!t]
	\centering
	\caption{Summary of \ac{ML} works for spectrum sensing using \ac{RL} approach.}
	{\footnotesize \begin{tabularx}{\linewidth}{
			|>{\hsize=1\hsize}X|
			>{\hsize=1.5\hsize}X|
			>{\hsize=0.7\hsize}X|
			>{\hsize=0.8\hsize}X|
		}
		\hline
		\parbox[c]{2.5cm}{\centering{\textbf{\ac{RL} model}}} & \parbox[c]{4cm}{\centering{\textbf{Algorithm}}} & {\textbf{Spectrum Sensing}}    &  {\textbf{Reference}} \\ \hline
		
		\multirow{5}{*}{\parbox[c]{2.5cm}{\centering\vspace*{0.05cm}Markov Decision Process}} &   \parbox[c]{2.5cm}{\centering\vspace*{0.05cm}  Trials in Markov Decision Process}   &   Non \ac{CSS} & \parbox[c]{2cm}{\centering \cite{Xu2020a}  }       \\\cline{2-4}
		&   \makecell{Deep Q-Network \\ \parbox[c]{3.5cm}{\centering\vspace*{0.05cm}Upper confidence bounds with Hoeffding-style}}    & \ac{CSS} & \parbox[c]{2cm}{\centering \cite{Zhang2019b} }    \\\cline{2-4}
		&   Deep Q-Network       & Non \ac{CSS}  & \parbox[c]{2cm}{\centering\cite{Li2020a} }        \\\cline{2-4}
		& Deep Q-Network & \ac{CSS}         & \parbox[c]{2cm}{\centering  \cite{Sarikhani2020}} \\\cline{2-4}
		& Dueling Deep Q network &  \ac{CSS} &\parbox[c]{2cm}{\centering  \cite{Liu2021}} \\\cline{2-4} 		
		& Double Deep Q network &  \ac{CSS} & \parbox[c]{2cm}{\centering \cite{Bokobza2023}} \\\cline{2-4} 	
		& Double Deep Q network &  \ac{CSS} &\parbox[c]{2cm}{\centering  \cite{Sun2022}} \\\cline{2-4} 
		& Deep Q-Network & \ac{CSS}       & \parbox[c]{2cm}{\centering  \cite{Ngo2023}} \\\hline
		{\parbox[c]{2.5cm}{\centering\vspace*{0.05cm}Multi-agent reinforcement learning}} &     \makecell{\parbox[c]{3.5cm}{\centering\vspace*{0.05cm}Independent learner Q-Learning} \\ \parbox[c]{3.5cm}{\centering\vspace*{0.05cm}Independent learner Deep Q-Learning}}   &   \ac{CSS}  & \parbox[c]{2cm}{\centering \cite{Jiang2021}  } \\\cline{2-4} 
		& Multi-agent Q-learning & \ac{CSS}   & \parbox[c]{2cm}{\centering  \cite{Kim2023}} \\\cline{2-4}
		& Deep Q-Network & \ac{CSS}      &  \parbox[c]{2cm}{\centering \cite{Li2023a}} \\\cline{2-4}		
		& Double Deep Q-Network & \ac{CSS}  & \parbox[c]{2cm}{\centering  \cite{Zhang2023}} \\\cline{2-4}
		\hline
		\parbox[c]{2.5cm}{\centering\vspace*{0.05cm}Coordination Graph} & Deep Q-Network & \ac{CSS}  & \parbox[c]{2cm}{\centering \cite{Cai2020} }\\ \hline
	\end{tabularx}}
	\label{tab:spectrum_sensing_RL}
\end{table*}

\section{Summary}\label{sub:Chap_Sensing:summary}
In this chapter, we addressed spectrum sensing mechanisms. We modeled the spectrum sensing problem as a classification problem and showed how \ac{ML} can be used to solve it. 

Supervised, unsupervised and semisupervised learning strategies were used in literature to solve the spectrum sensing classification problem. We provided a comprehensive survey of recent works that applied these techniques and we summarized them showing their strong points and main contributions.

We also addressed \ac{RL} techniques for spectrum sensing. Instead of solving a classification problem, in \ac{RL} approaches the main goal is to optimize some objective. We surveyed recent works and we presented a summary of the \ac{RL} model and algorithms used in these works to solve the spectrum sensing problem.

\chapter{Spectrum Allocation}\label{Sec:Allocation}
\section{Introduction}

In the previous chapter we have discussed \ac{ML} methods of spectrum sensing which aim to determine the channel occupancy, i.e., to decide if there is a transmission over the sensed channel or it is idle. The next major task is the spectrum allocation where we have the following objectives: (1) ensure that all users (licensed and unlicensed) have access to the available channel frequency respecting the allocation and access policies; and (2) optimize the spectrum efficiency by reducing interference among network users. For \acp{CRN} the information obtained after the sensing process is used  to map the cognitive users into the available licensed channels. On the other hand, in non cognitive equipment, the channel assignment can be done  by either a \ac{BS} or a central scheduling unity which uses  channel state information from the users to find the optimal channel allocation solution.

Channel allocation methods can be roughly classified into fixed and dynamic schemes \cite{MacDonald1979,Tekinay1991}. In fixed assignment, a frequency channel is allocated to each cell or user based on a frequency planning process. This process is not optimal and it results in sub-utilization of the channel resources reducing the spectral efficiency.  On the other hand, in dynamic allocation, all frequency channels are available and they are assigned by users' request. This on demand strategy is proposed to maximize the spectrum utilization and to allow the coexistence of multiple networks sharing the same resource.

The dynamic channel allocation is done according to some optimization task where conditions and constraints should be satisfied. However, depending on the channel assignment objectives, solving the optimization problem can be a hard task due to complex cost functions or non-convex constraints. To overcome this issue, a trial-error approach (\ac{RL}) can approximate the optimization problem with robustness. As discussed in Section \ref{sub:reinfo_learning}, in \ac{RL} the learners implement actions to explore the environment without any prior knowledge to learn the optimal control policy which can be implemented in both centralized or distributed fashion.

\begin{figure}[!t]
	\centering
	{\includegraphics[width=0.75\textwidth]{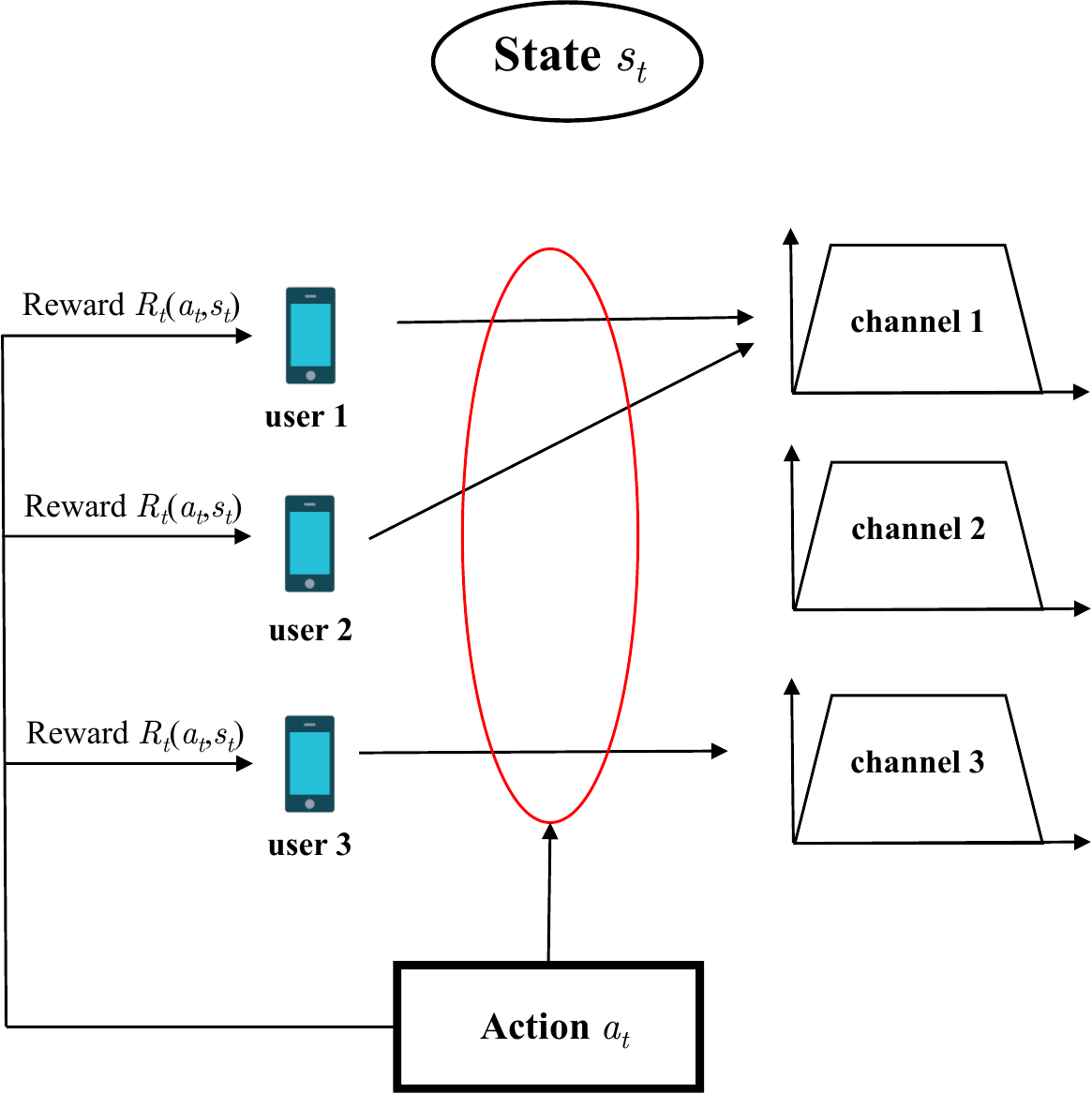}}
	\caption{Illustration of spectrum allocation. In this example $K = M = 3$ and each user is allocated to a single channel.}
	\label{fig:spectrumallocation}
\end{figure}

Channel assignment is a particular case of resource allocation problem. To employ \ac{RL}, we have to determine the system state $s_t$, its behavior, the allocation cost and reward $R_t(a_t,s_t)$ for a given action $a_t$, and the next state $s_{t+1}$. For an easier explanation, let us consider, without loss of generality, a system with $K$ users and $M$ available channels \footnote{A similar approach is found in \cite{Wang2018}.}. The usage matrix $\mathbf{U}$ represents the channel status at a given instant of time $t$, where:
\begin{equation}
u_{k,m} =
\begin{cases}
1, & \text{if channel $m$ is in use by user k},\\
0, & \text{otherwise}.
\end{cases}       
\end{equation} 
Figure \ref{fig:spectrumallocation} illustrates an example of three users channel allocation and three available channels. The \textit{state} $s_t = \mathbf{U}_t$ represents the channel mapping at a time instant $t$. Initially $\mathbf{U}_t = \mathbf{0}$, then an \textit{action} $a_t$ which is defined by an assignment or removal of a user $k$ to/from a channel $m$ is performed. After $a_t$, the usage matrix is:
 \begin{equation}
	\mathbf{U}=\left(\begin{array}{ccc}
		1 & 0 & 0 \\
		1 & 0 & 0 \\
		0 & 0 & 1 \\
	\end{array}\right).
\end{equation}

To each action is associated a \textit{reward},  $R_t$, which is related with an action-state pair $(a_t,s_t)$. It captures the benefit for each user to be allocated or removed  to/from a specified channel. For example, if a user is allocated in a channel with other users or suffers from deep fading, this action receives a low reward. On the other hand, if the assigned channel increases the user performance, the reward is high. In the example, users 1 and 2 will receive a low reward because they are allocated to the same channel, while user 3 will get a higher reward.  A \textit{cost} is used to represent the overall impact of an action $a_t$ into a state $s_t$ taking into account all individuals rewards. Many functions or methods can be used to determine the action cost and reward, depending on system parameters, characteristics or optimization objective.  Finally, the \textit{next state}, $s_{t+1}$, represents a transition from the previous state, $s_t$, after an action $a_t$.

Applying \ac{RL} to channel allocation problem brings some advantages like \cite{Wang2016a,Wang2018}: (1) adaptation to environment 
changes due to continuous learning; (2) real-time environment monitoring; (3)  capacity of modeling and solving hard-to-optimize 
problem with non-convex constraints; and (4) capacity of learning from historical data. Therefore, many works in literature made use of 
\ac{RL} strategies to address channel allocation problem by Q-learning based methods,
deep Q-learning based methods
or other \ac{RL} methods.

\section{Q-learning Based Methods}

Q-learning (see Section \ref{subsub:sarsa_q_learning}) is an effective solution when there is no fixed allocation between channel and users \cite{JunhongNie1999}. In a shared environment, the available channels need to be frequently updated since they depend on users activity which may change dynamically. Then, spectrum allocation can learn the best assignment strategy, i.e., the best available frequency channel by optimizing users or network long-term reward in an incremental way instead of exchanging sensing information about the channel availability and negotiating the channel allocation based on selfish requirements \cite{Li2009}. Therefore, the network can adapt the channel selection strategy until reaching convergence allowing coexistence of different technologies \cite{Bennis2010}.

The potential of Q-learning solutions for channel allocation of \ac{LTE-U} in the unlicensed 5 GHz band and its coexistence with WiFi system was investigated in \cite{Liu2017a,Hu2017,Sallent2015}. Q-learning can dynamically allocate blank subframes improving the spectrum utilization for both WiFi and \ac{LTE-U} networks while it reduces \acs{LTE} delay \cite{Liu2017a}. Work in \cite{Hu2017} proposed a Q-learning scheme with downlink-uplink decoupling technique. The authors showed that the network could achieve high efficiency in terms of rate and load balance with limited information on its state. Also, a distributed Q-learning scheme can be used at different \ac{MNO} coordination levels to achieve the best channel selection strategy design for \ac{LTE-U} networks \cite{Sallent2015}. The authors in \cite{Srinivasan2016} explored this idea and they proposed an inter-operator proximal spectrum sharing scheme for \acp{MNO} based on a Q-learning framework. The algorithm learns the spectral needs for the \acp{BS} and it applies the best channel sharing policy resulting into a self-organized network. This approach guarantees that users with high \ac{QoE} requirements could be served by a \ac{BS}.

Q-learning based schemes were also investigated for \ac{D2D} technology \cite{Luo2014,AlQerm2016,Zia2019}. A single tier network resource allocation  Q-learning scheme was proposed in \cite{Luo2014} to maximize the system throughput while reference \cite{AlQerm2016} proposed a two-tier cooperative learning strategy where devices learn the best strategy by sharing their information to meet their \ac{QoS} requirements. On the other hand, reference \cite{Zia2019} considered a multi-tier heterogeneous network where a multi-agent Q-learning algorithm was used for channel resource allocation. Each \ac{D2D} user assumed an agent role by taking actions based on \ac{SINR} and data rates to maximize its own throughput while keeping the cellular users requirements.

The usage of \ac{UAV} networks has been explored recently for different kind of applications. On these networks there is a high data demand due large volume data applications such as aerial imaging, surveillance, communication relays, and real-time monitoring. Also the high mobility users yield to a dynamically environment that requires robust communication and fast adaptability to channel variations. Therefore, these networks often require more spectrum to meet users' requirements, which are not guaranteed by traditional spectrum sharing mechanisms. To overcome this limitations, Q-learning was used in \cite{Kawamoto2019} to properly allocate channels for those kind of networks. The authors used the total transmitted data size of a physical channel period as an action reward. This scheme allows the users to learn the distribution of time slots and the allocation of the modulation and codification scheme for each pair.

Q-learning decisions take into account cumulative rewards related to an action that cause a state transition. Some works modified the Q-learning algorithm to improve system performance \cite{Faganello2013,Wilhelmi2017,Zhao2019}. Authors in \cite{Faganello2013} proposed three modifications in the Q-learning algorithm. The first one, Q-learning+, limits the number of epochs to be considered for reward calculation. The second one, called Q-Noise, takes into account the channel conditions by evaluating the \ac{SINR} of different channels. The third one, named Q-Noise+, is a joint implementation of the two previous algorithms. In an decentralized scenario, no information of other users are available and the system can be described only by the set of actions and rewards. Based on this idea, reference \cite{Wilhelmi2017} proposed a stateless Q-learning where each user chooses an action in an ordered way decided in the beginning of each iteration. The designed scheme was shown to achieve close-to-optimal solutions. To reduce the resource consumption due to the periodical learning process, the work in \cite{Zhao2019} proposed an event-driven Q-learning mechanism. This mechanism established a condition based on the degree of change in state value Q by defining a changing tolerance threshold for the degree of network disturbance. The agent will only update its strategy and action if the condition is above the threshold, otherwise it performs the last action. 

The Q-Learning based spectrum allocation works are summarized in Table \ref{tab:spectrum_allocation_QL}.

\begin{table*}[!t]
	\centering
	\caption{Description summary of \ac{ML} works for spectrum allocation using Q-Learning strategy.}
	{\footnotesize \begin{tabularx}{\linewidth}{
				|>{\hsize=0.6\hsize}X|
				>{\hsize=0.4\hsize}X|
				>{\hsize=2\hsize}X|
			}
			\hline 
			\parbox[c]{2cm}{\centering\textbf{Reference}}   & \parbox[c]{1.5cm}{\centering\vspace*{0.05cm}\textbf{Network type}}  & \parbox[c]{7cm}{\centering\textbf{Comment}}  \\ \hline
			 \parbox[c]{2cm}{\centering \vspace*{0.2cm} \cite{JunhongNie1999}} &   \parbox[c]{1.3cm}{\centering \vspace*{0.2cm} Cellular system }      &  \vspace*{-0.4cm}     Demonstrated the viability of Q-learning to solve dynamic channel assignment problem efficiently.  
			\\ \hline
			\parbox[c]{2cm}{\centering \vspace*{0.2cm} \cite{Liu2017a,Hu2017,Sallent2015}} &  \parbox[c]{1.3cm}{\centering \vspace*{0.2cm} \ac{LTE-U} and WiFi}    & \vspace*{-0.5cm} Investigated Q-learning mechanisms for channel allocation allowing coexistence in the unlicensed spectrum.     \\ \hline
			\parbox[c]{2cm}{\centering \vspace*{0.4cm}\cite{Srinivasan2016} } &   \parbox[c]{1.3cm}{\centering\vspace*{0.4cm} \ac{LTE}  }   &    Proposed an inter operator scheme that allows mobile network operators to achieve high user quality of experience and spectral efficiency.     \\ \hline
			\parbox[c]{2cm}{\centering \vspace*{0.2cm}\cite{Luo2014,AlQerm2016,Zia2019}}  &  \parbox[c]{1.3cm}{\centering\vspace*{0.05cm} \ac{D2D}  }       &  Reduced co-tier and/or cross-tier interference by applying channel and power allocation scheme.     \\ \hline
			\parbox[c]{2cm}{\centering \vspace*{0.5cm}\cite{Kawamoto2019}}& \parbox[c]{1.3cm}{\centering\vspace*{0.5cm}   \ac{UAV}    }  &   Reduced transmission requests  by allowing users to learn the distribution of time slots and the allocation of modulation and codification schemes for each users pair.    \\  \hline
			\parbox[c]{2cm}{\centering \vspace*{0.2cm}\cite{Faganello2013}} &  \parbox[c]{1.3cm}{\centering\vspace*{0.05cm} Cognitive sensors networks  }    & \vspace*{-0.5cm}   Proposed three improvements on Q-learning algorithm: Q-learning+, Q-Noise and Q-Noise+.  \\\hline
			\parbox[c]{2cm}{\centering \vspace*{0.5cm}\cite{Wilhelmi2017}} & \parbox[c]{1.3cm}{\centering\vspace*{0.5cm}  Wireless networks   }   &  \vspace*{-0.5cm} Proposed a decentralized stateless Q-learning scheme where each user chooses an action in an ordered way decided in the beginning of each iteration.    \\\hline
			\parbox[c]{2cm}{\centering \vspace*{0.2cm}\cite{Zhao2019}} & \parbox[c]{1.3cm}{\centering\vspace*{0.05cm} Wireless local network }   &  \vspace*{-0.5cm} Provided an event-drive Q-learning algorithm that establishes a threshold to determine whether or not an action should be taken.       \\ 
			\hline
	\end{tabularx}}
	\label{tab:spectrum_allocation_QL}
\end{table*}

\section{Deep Q-learning Based Methods}

High dimensional problems are intractable in traditional Q-learning methods due to the high number of states and actions to be stored in the Q-table. Moreover, as Q-learning is a gradual optimization process, this situation leads to a slow convergence to find the optimum action. To deal with this problem, the \ac{DRL} combines the traditional Q-learning with \ac{DNN} (see Section \ref{subsub:ANN} and Section \ref{subsub:approx_techniques} for \ac{DNN} and \ac{DRL} details) to compensate its limitations by approximating the action-value function.

The \ac{DQN} approximates the optimal action-value from the Q-function. Often this approach is used with experience replay to learn on/off policies to reduce the divergence in the parameters. The \ac{DQN} framework outperforms Q-learning in terms of energy efficiency and achieving a better network control \cite{Li2018b} and can  can effectively allocate resources to network slices \cite{Shi2023}. The work in \cite{Zhou2019} incorporated a \ac{LSTM} (see Section \ref{subsub:RNN}) into the \ac{DQN} framework to improve multi-\ac{UAV} system performance. This modification preserves the historical observation data when a recursive \ac{DNN} is used helping the network to learn from the past actions and to adapt to the environment changes.

The traditional \ac{DRL}, however, is unstable when a nonlinear function is used to represent the action-value function \cite{Sutton2018}. To cope with this issue, the experience replay and target network can used to improve stability \cite{Mnih2015,Fedus2020}. The experience replay uses a buffer to store the most recent transitions determined by the control policy breaking the temporal correlation in the training data. It allows data to be reused in training instead of being discarded after collection. On the other hand, a target network is a copy of the Q-function that is fixed as a stable target for a certain number of steps. 

Based on these ideas, several works \cite{Wang2020b,Ye2019,Liu2018a,Liu2020,Li2022,Yu2023,Guo2022,Qi2023,Guan2023,Xiang2022,Nakashima2020} made use of these tools on their \ac{DRL} algorithms  for different kind of networks and optimization tasks. The work in \cite{Wang2020b} studied the channel assignment for \ac{NOMA} system to solve the energy efficiency maximization problem. Due to the non convexity of this problem, the authors proposed three \ac{DQN}-based frameworks that were jointly trained to find the best channel and power allocation policy. Combined with a event trigger strategy, the proposed method was shown to improve the uplink energy efficiency performance while reduces the computational time. In \cite{Ye2019}, the authors showed that the \ac{DQN} can efficiently allocate resources form both unicast and broadcast scenarios in a \ac{V2V} network. References \cite{Liu2018a} and \cite{Liu2020} addressed the channel allocation problem for satellites systems. While the former aimed to minimize the service blocking probability in a  multibeam allocation fashion, the latter proposed a centralized \ac{DRL} algorithm to reduce the data transmission delay.

Pure \ac{RL} approaches are impracticable for high dimensional environments due to the difficult to achieve the optimal solution in low probability of occurrence states \cite{Qi2021}. To overcome this issue, federated learning has arisen as a technique where agents can work in a cooperative manner by sharing their local parameters while exploring independently the state space. Each user link  can be considered as an individual agent who trains local models which are aggregated in the \ac{BS} to provide a global information of the system to perform the best channel allocation \cite{Li2022,Yu2023}. The authors in \cite{Guo2022} introduced three optimization criteria to allocate spectrum for multiuser mobile edge computing systems. Federated learning is, then, used to optimize these criteria and to allocate more bandwidth to the user with a higher task priority.  

Additional methods can also be used to improve \ac{DQN} performance for channel allocation.  The \ac{DDQN} approach is used in various work due its capacity on handling continuous state spaces efficiently. Reference \cite{Qi2023} proposed an algorithm for \ac{D2D} spectrum allocation with energy harvesting.  \ac{DDQN} is jointly applied with the spectrum slice degree by \cite{Guan2023}. This strategy determines the spectrum segmentation due an arrival of new services. Based on this information, a \ac{RL} algorithm is applied to determine the best resource allocation, decreasing the blocking probability. Reference \cite{Xiang2022} applied a \ac{DDQN} with priority sampling algorithm to enhance achievable rate for \ac{D2D} communications. The priority is introduced during the experience replay and it is defined based on temporal difference error between the estimated and actual Q-value. This strategy allows the agent to learn more dominant features. In \cite{Nakashima2020}, the authors combined experience replay with  \ac{DDQN}  and dueling network \cite{Wang2016b} to avoid overestimation when different networks are employed and to learn the state values without taking the action effects, respectively. This strategy was combined with a graph convolutional network to efficiently assign channels in a short period of time with maximized cumulative reward. The use of experience replay, however, is problematic due to the possible issues with  non-stationarity  when applied with a multi-agent \ac{DQN}. To overcome this issue, a fingerprint method \cite{Foerster2017} can be used to track the trajectory of the policy change of other agents. This approach uses the policy changes of other agents as the input of the action-valued function in the Q-table instead of all other agent parameters. The fingerprint is correlated with the training iteration number so it can capture the agents policies with a smooth variation over time.   Reference \cite{Liang2019} proposed a combination of multi-agent \ac{RL} with \ac{DL} with a fingerprint method. By tracking the polices changes, each agent received a common reward preventing instability and improving the network sum capacity. The fingerprint method was also explored in \cite{Ji2023}. The authors proposed a channel allocation under harsh conditions and high data transmission requirements by combining dueling \ac{DDQN} with low-dimensional fingerprints and soft-update	architecture. 

Differently from previous works, authors in \cite{Bhattacharya2022} integrated \ac{DQN} with blockchain. In this scheme, a DeepBlocks strategy is proposed where there is a probability $\epsilon$ to perform an action outside the Q-table, increasing exploitation. To deal with the time varying nature of the user mobility in vehicular networks, reference \cite{Kumar2022} combined \ac{LSTM} with \ac{DQN} and \ac{A2C}. This strategy was proven to allocate spectrum efficiently in environments with user mobility and high demand variations. \ac{MARL} was proven to be efficient for spectrum allocation tasks due the improvement of state space and action space in \ac{RL} \cite{Busoniu2010}. The work in \cite{Yang2022} adopted a \ac{MARL} dueling \ac{DQN}. This algorithm allows the users to learn the action-value distribution by estimating both the state-value and action advantage functions, increasing learning efficiency, network data rate, and \ac{QoS} satisfaction probability. An actor-critic strategy was also adopted in \cite{Urmonov2023}, where the authors performed a centralized off-line training, allowing	all agents to share their observations over critic networks and mapping the Q-value for all feasible actions in the given state. This strategy resulted in a higher packet reception ratio when compared with \ac{DDQN}. 

The Deep Q-Learning based spectrum allocation works are summarized in Table \ref{tab:spectrum_allocation_DQL}.
\begin{table*}[!pt]
	\centering
	\caption{Description summary of \ac{ML} works for spectrum allocation using deep Q-Learning strategy.}
	{\scriptsize 	\begin{tabularx}{\linewidth}{
				|>{\hsize=0.8\hsize}X|
				>{\hsize=0.6\hsize}X|
				>{\hsize=1.6\hsize}X|
			}
			\hline 
			\parbox[c]{3cm}{\centering \textbf{Reference}}   & \parbox[c]{2cm}{\centering\vspace*{0.05cm}\textbf{Network type} } & \parbox[c]{5cm}{\centering \textbf{Comment}}  \\ \hline
			
			\parbox[c]{3cm}{\centering \vspace*{0.3cm}\cite{Li2018b}} &  \parbox[c]{2cm}{\centering\vspace*{0.2cm} Ultra dense network}   &    Determined the best channel allocation for \acp{BS} through an on/off policy to enhance energy efficiency.    \\  
			\hline
			\parbox[c]{3cm}{\centering \vspace*{0.3cm}\cite{Shi2023}}  &  \parbox[c]{2cm}{\centering\vspace*{0.05cm}\ac{5G}}   &  Proposed a robust algorithm to admit requests and allocate resources as a low-complexity solution.     \\ 
			\hline
			\parbox[c]{3cm}{\centering \vspace*{0.3cm}\cite{Zhou2019}} &   \parbox[c]{2cm}{\centering\vspace*{0.05cm}\ac{UAV}}    &  Incorporated a \acl{LSTM} into the \acl{DQN} framework for channel allocation.    \\ 
			\hline
			\parbox[c]{3cm}{\centering \vspace*{0.2cm}\cite{Wang2020b,Ye2019,Liu2018a,Liu2020}} &   \parbox[c]{2cm}{\centering\vspace*{0.05cm}Satellite \ac{IoT}}    &  Decreased \acl{DQN} learning instability  by using experience replay and target network.        \\  
			\hline		
			\parbox[c]{3cm}{\centering \vspace*{0.3cm}\cite{Li2022,Yu2023}}  & \parbox[c]{2cm}{\centering\vspace*{0.15cm} Vehicular networks }   &  Applied federated learning multi-agent deep reinforcement learning for power and channel allocation.      \\		\hline	
			\parbox[c]{3cm}{\centering \vspace*{0.3cm}\cite{Guo2022}} & \parbox[c]{2cm}{\centering\vspace*{0.15cm} Wireless mobile edge computing}     & Proposed a federated learning approach to solve a three criteria optimization problem to allocate bandwidth based on users' priority.       \\
			\hline
			\parbox[c]{3cm}{\centering \vspace*{0.2cm}\cite{Qi2023}}  &  \parbox[c]{2cm}{\centering\vspace*{0.05cm} \ac{D2D} } &    Adopted a \ac{DDQN} algorithm for channel allocation combined with energy harvesting.     \\
			\hline
			\parbox[c]{3cm}{\centering \vspace*{0.2cm}\cite{Guan2023}} & \parbox[c]{2cm}{\centering\vspace*{0.05cm}  Elastic optical network }    &  Combined \ac{DDQN} with spectrum slice degree to reduce the blocking probability.     \\
			\hline 
			\parbox[c]{3cm}{\centering \vspace*{0.3cm}\cite{Xiang2022}} &  \parbox[c]{2cm}{\centering\vspace*{0.15cm}\ac{D2D} and cellular network  }   &  Proposed a \ac{DDQN} with  priority sampling distributed algorithm to enhance both \ac{D2D} and cellular.     \\
			\hline
			\parbox[c]{3cm}{\centering \vspace*{0.35cm}\cite{Nakashima2020}} & \parbox[c]{2cm}{\centering\vspace*{0.35cm} Wireless local network }    &  Employed experience replay with \acl{DDQN} to avoid overestimations and dueling network to learn state values without taking an action.     \\
			\hline 
			\parbox[c]{3cm}{\centering \vspace*{0.35cm}\cite{Liang2019}} & \parbox[c]{2cm}{\centering\vspace*{0.35cm}Vehicular network } &   Developed a multi-agent \acl{DQN} framework incorporating experience replay and a fingerprint-based method to address non-stationary issues.         \\ \hline		
			\parbox[c]{3cm}{\centering \vspace*{0.3cm}\cite{Ji2023}} & \parbox[c]{2cm}{\centering\vspace*{0.15cm}Vehicular network}  &   Proposed a dueling \ac{DDQN} \ac{RL} algorith with low-dimensional fingerprints and soft-update architecture.        \\ \hline
			\parbox[c]{3cm}{\centering \vspace*{0.4cm}\cite{Bhattacharya2022}} &\parbox[c]{2cm}{\centering\vspace*{0.5cm} \ac{6G}}   &  Proposed a \ac{DQN} scheme integrated with blockchain. The Q-table is updated by a DeepBlock strategy via $\epsilon$-greedy method, where there is a probability of $\epsilon$ to the agent take an action outside the Q-table.    \\ \hline
			\parbox[c]{3cm}{\centering \vspace*{0.3cm}\cite{Kumar2022}} & \parbox[c]{2cm}{\centering\vspace*{0.1cm}Vehicular network }  &   Developed \ac{LSTM}-\ac{DQN} and \ac{LSTM}-\ac{A2C} algorithms for efficiently allocate spectrum in vehicular networks.     \\  \hline 	
			\parbox[c]{3cm}{\centering \vspace*{0.3cm}\cite{Yang2022}} & \parbox[c]{2cm}{\centering\vspace*{0.1cm} Two-tier heterogeneous networks }  &  Proposed a \ac{MARL} with dueling \ac{DQN} scheme for device association, spectrum and power allocation strategy.     \\  \hline
			\parbox[c]{3cm}{\centering \vspace*{0.3cm}\cite{Urmonov2023}}  & \parbox[c]{2cm}{\centering\vspace*{0.1cm}Vehicular network }  &  Applied a \ac{MARL} deep learning method combined with critic and actor \acp{DNN} to efficiently allocate transport blocks.      \\ 
			\hline
	\end{tabularx} }
	\label{tab:spectrum_allocation_DQL}
\end{table*} 

\section{Other Methods}
Some works in the literature applied \ac{RL} strategies based on other algorithms than Q-learning.
Such methods achieved better performance than Q-learning and Deep Q-learning as shown in comparative results in some of these works into the considered scenario. Another reason for using non Q-learning solutions is that these have a better adjustment for some particular problem or optimization task. For example, in \cite{Miao2016} the authors explored the channel allocation problem for coexistence between Wi-Fi and \ac{LTE-U} networks.  By considering a central decision maker, a stateless \ac{RL} was implemented where a static spectrum allocation was proposed and combined with inter and intra system spectrum allocation algorithms.

Q-learning based strategies are one-step reinforcement learning approaches, thus they are not capable of predicting a sequence of future actions. Instead, \ac{LSTM}  can store information allowing a proactive network approach. This method allows the network to predict traffic patterns and properly allocate frequency resources among users \cite{Challita2018}. Another prediction strategy was used in  \cite{Tubachi2017} to predict the throughput of a \ac{CRN}. The authors used a modified actor critic learning automaton which receives \ac{RSS} indicator values and returns the label of the channel that achieves  maximum throughput.

Even \ac{DQN} methods performance degrades for large-scale networks since the number of actions and states increases with number of network users. When confronted with a high-dimensional action space, a value-based \ac{RL} algorithm cannot explore every possible action. In this case, a \ac{DDPG} strategy can be used to find the best resource allocation. This is an actor-critic solution that approximates the policy and Q-value functions  \cite{Du2023}. The authors in \cite{Sun2023} combined \ac{DDPG}  with experience replay where both channel gain and interference information were used to train the \ac{DNN} {DDPG} for optimal channel allocation. In \cite{Xu2020b}, the \ac{V2X} resource allocation problem was formulated as a Discrete-time and Finite-state Markov Decision Process where \ac{DDPG} was used to deal with continuous action space.

The work in \cite{Zafaruddin2019} proposed an online auction learning algorithm for distributed channel allocation in ad hoc networks. The problem is modeled as a multi-player \ac{MAB} where the highest bidder is the user who first access the channel. Therefore, users can learn the channel statistics in real-time without any communication among them. 

A combination of message passing and \ac{DRL} was explored in \cite{Liu2019e}. The work aimed to perform resource allocation to maximize the energy and spectrum efficiency tradeoff. Since such problems are NP-hard, the authors proposed a damped three dimensional message passing algorithm to solve the problem in a sub-optimal way with lower learning cost. A feed-forward neural network was built and the authors developed an analogous back propagation algorithm to determine the optimal parameters of the proposed message passing algorithm. Reference \cite{Ghaderibaneh2024} proposed an allocation architecture called DeepAlloc. The framework is consisted by two steps. In the first step, a \ac{CNN} model is used  to represent the spectrum allocation function input as an image. In the second step, the \ac{CNN} output is used to feed a \ac{RNN} to allocate spectrum  simultaneously to multiple \acp{SU}.

Other literature works explored non Q-learning solutions. Authors in \cite{He2023} investigated \ac{MARL} for power energy communications. In \cite{Wang2019} a multi-agent \ac{DRL} algorithm is proposed where each agent (cognitive user) interacted with the environment via a \ac{BS} control mechanism which was used to adjust the proportion of each channel resource. On the other hand, in \cite{Li2020} a centralized training where users' actions, policies and historical states were shared was performed to prevent non-stationarity issues. The deep learning network learns from historical information and updates each agent policy towards the best individual reward. Reference \cite{Morgado2022} a three-stage \ac{ML} framework for spectrum allocation. In the first stage, a \ac{LSTM} approach is used to explore time-series prediction, then a clustering \ac{ML} algorithm is used to group users/cells with similar features/requirements, and, finally, a \ac{RL} strategy is performed to dynamically allocate spectrum.

Another strategy explored in literature is to use the \ac{PPO} algorithm with \ac{DRL} for bandwidth allocation due its suitability to continuous state and action state problems. This algorithm can obtain a good performance with low implementation complexity which makes it effective for \ac{V2V} (\cite{Hu2021}) and \ac{V2I} communications (\cite{Xu2023}). Authors in \cite{Zheng2021} combined \ac{PPO} and \ac{RNN} in an actor-critic approach for channel assignment in hybrid \ac{NOMA} systems. The \ac{RNN} is used to help the actor to extract \ac{CSI} relationships while \ac{PPO} learns the best allocation policy. \cite{Di2024} used \ac{PPO} to optimize multi-slot resource allocation. The authors developed a Transform-based actor framework composed by an offline training where \ac{PPO} algorithm is used to learn the best allocation policy and an online application where \ac{PPO} is used for real-time adaptation. 

These works are summarized in Table \ref{tab:spectrum_allocation_other}.
\begin{table*}[!t]
	\centering
	\caption{Description summary of \ac{ML} works for spectrum allocation using other \ac{RL} methods.}
	{\scriptsize \begin{tabularx}{\linewidth}{
		|>{\hsize=0.6\hsize}X|
		>{\hsize=0.6\hsize}X|
		>{\hsize=1.8\hsize}X|
	}
	\hline 
	\parbox[c]{2cm}{\centering\vspace*{0.05cm}\textbf{Reference}}   & \parbox[c]{2cm}{\centering\vspace*{0.05cm}\textbf{Network type}}  & \parbox[c]{6cm}{\centering\vspace*{0.05cm}\textbf{Comment}}  \\ \hline

	\parbox[c]{2cm}{\centering \vspace*{0.15cm}\cite{Miao2016}} &  \parbox[c]{2cm}{\centering\vspace*{0.15cm}\ac{LTE-U} and  WiFi}     &   Allowed coexistence between WiFi and \ac{LTE-U} networks using a centralized stateless \ac{RL} strategy.   \\ 
	\hline
	\parbox[c]{2cm}{\centering \vspace*{0.25cm}\cite{Challita2018}} & \parbox[c]{2cm}{\centering\vspace*{0.25cm}\ac{LTE-U} and  WiFi}   & \vspace*{-0.25cm} Used a proactive \acl{DRL}  allocation algorithm based on \acl{LSTM} to allow coexistence between WiFi and \ac{LTE-U} networks.   \\
	\hline
	\parbox[c]{2cm}{\centering \vspace*{0.35cm}\cite{Tubachi2017}} &  \parbox[c]{2cm}{\centering\vspace*{0.25cm}\ac{CRN}}   & Proposed a continuous actor critic learning algorithm to predict network throughput to optimize channel allocation.    \\
	\hline 
	\parbox[c]{2cm}{\centering \vspace*{0.15cm}\cite{Du2023}} &  \parbox[c]{2cm}{\centering\vspace*{0.05cm} WiFi}   & Proposed a \ac{DDPG} strategy for channel allocation in a multi station multi access point 802.11be network.    \\
	\hline 
	\parbox[c]{2cm}{\centering \vspace*{0.25cm}\cite{Sun2023}} &  \parbox[c]{2cm}{\centering\vspace*{0.25cm} Cellular network}   & Proposed a \ac{DDPG} algorithm to optimize channel allocation and an unsupervised learning strategy for power control scheme.    \\
	\hline 
	\parbox[c]{2cm}{\centering \vspace*{0.25cm}\cite{Xu2020b}} &  \parbox[c]{2cm}{\centering\vspace*{0.25cm} \ac{V2X} }   & Proposed a \ac{DDPG} framework to solve both  frequency spectrum allocation and		transmission power control problems for sum-rate maximization task.    \\
	\hline 
	\parbox[c]{2cm}{\centering \vspace*{0.25cm}\cite{Zafaruddin2019} }&    \parbox[c]{2cm}{\centering\vspace*{0.25cm}Ad hoc network }       &   Developed a distributed online auction algorithm for channel allocation based on multi-player \acl{MAB} formulation. \\   
	\hline
	\parbox[c]{2cm}{\centering \vspace*{0.25cm}\cite{Liu2019e}} &  \parbox[c]{2cm}{\centering\vspace*{0.25cm}\ac{CRN}  }    &  Combined damped three dimensional messaging passing algorithm with \acl{DRL}.           \\ 
	\hline
	\parbox[c]{2cm}{\centering \vspace*{0.25cm}\cite{Ghaderibaneh2024}}  &     \parbox[c]{2cm}{\centering\vspace*{0.25cm}\ac{CRN} }   &   Proposed an allocation framework composed by a \ac{CNN}-based approach whose output is fed into a \ac{RNN} to allocate spectrum simultaneously to multiple \acp{SU}.        \\ 
	\hline 
	\parbox[c]{2cm}{\centering \vspace*{0.25cm}\cite{He2023} } &   \parbox[c]{2cm}{\centering\vspace*{0.25cm}\ac{UAV} and \ac{CRN} }     &   Proposed a \ac{MARL} scheme for power emergency communications where each \ac{SU} runs a  \ac{DQN} to allocate spectrum
	independently .       \\ 
    \hline
	\parbox[c]{2cm}{\centering \vspace*{0.25cm}\cite{Wang2019}}  &   \parbox[c]{2cm}{\centering\vspace*{0.25cm}\ac{CRN}   }   &   Proposed a hierarchical multi-agent \acl{RL} for frequency allocation.        \\ 
	\hline
	\parbox[c]{2cm}{\centering \vspace*{0.15cm}\cite{Li2020}} & \parbox[c]{2cm}{\centering\vspace*{0.05cm} Satellite and \acl{IoT}  }   &  \vspace*{-0.35cm}  Developed a multi-agent \acl{DRL} channel allocation framework.      \\ \hline
	\parbox[c]{2cm}{\centering \vspace*{0.35cm}\cite{Morgado2022} } & \parbox[c]{2cm}{\centering\vspace*{0.35cm}Virtualized \ac{5G} network}     & \vspace*{-0.35cm}Proposed a three-stage \ac{ML} algorithm: a time feature predictor to predict spectrum demand, a clustering strategy to identify users with similar characterization and a \ac{RL} framework to allocate spectrum dynamically.   \\ 
	\hline
	\parbox[c]{2cm}{\centering \vspace*{0.25cm}\cite{Hu2021}}  & \parbox[c]{2cm}{\centering\vspace*{0.25cm} \ac{V2V} }    &  Proposed a \ac{PPO} strategy for resource allocation to meet latency and capacity requirements on {V2V} networks. \\ 
	\hline
	\parbox[c]{2cm}{\centering \vspace*{0.35cm}\cite{Xu2023}}  & \parbox[c]{2cm}{\centering\vspace*{0.35cm} \ac{V2V} and \ac{V2I} }    &  Divided a multi-objective channel allocation problem into a set of scalar optimization subproblems where each one is solved by contribution-based dual-clip \ac{PPO} algorithm. \\ 
	\hline 
	\parbox[c]{2cm}{\centering \vspace*{0.25cm}\cite{Zheng2021}}  & \parbox[c]{2cm}{\centering\vspace*{0.25cm} Hybrid \ac{NOMA} }    &  Combined \ac{PPO} and {RNN} in an actor-critic approach for channel allocation in real-time changing environment. \\ 
	\hline  
	\parbox[c]{2cm}{\centering \vspace*{0.25cm}\cite{Di2024}}   & \parbox[c]{2cm}{\centering\vspace*{0.25cm} Multi-user cellular }    &  Applied a Transform model based \ac{PPO} algorithm for multi-slot and multi-user resource allocation to optimize spectral efficiency. \\ 
	\hline  
		\end{tabularx}}
	\label{tab:spectrum_allocation_other}
\end{table*} 

\section{Summary}\label{sub:Chap_Allocation:summary}

In this chapter, we addressed the spectrum allocation mechanism. We modeled the spectrum allocation problem as a resource allocation problem and we showed how \ac{RL} is used in literature to solve it. 

Q-learning and deep Q-learning are the main strategies used in literature for this kind of problems.  However, other \ac{RL} methods were also considered by recent works. We provided a comprehensive survey of those works  and we summarized them showing their main contributions and the network type where they were applied.

\chapter{Spectrum Access}\label{Sec:Access}

\section{Introduction}

Controlling channel access is an important feature for spectrum sharing. More than one user can be assigned in the same frequency by the spectrum allocation algorithm, which can cause severe interference if they transmit simultaneously reducing the network throughput. An intelligent and dynamic access policy should be done to reduce the adverse effects. This is done by \ac{DSA} where the channel access is controlled in an instant-by-instant basis limiting the interference among users. 

There are three main models for \ac{DSA}: dynamic exclusive use, open sharing and hierarchical access \cite{Zhao2007}. The first approach licentiates the spectrum band for exclusive use. Access can be controlled by spatial or time statistics, it is also possible for a licensed user to sell or trade spectrum. The second mode allows free spectrum access among users within a region. This mode is generally employed by services operating in the unlicensed \ac{ISM} bands. The third approach adopts the two-layer concept: primary users and secondary users. This mode is employed by \acp{CRN} and it can be divided in three paradigms according with the access technology:  underlay, overlay and interweave \cite{Goldsmith2009,Srinivasa2006}. In the \textit{underlay} approach, the \ac{SU} can simultaneously access the \ac{PU} allocated spectrum since its power is adjusted to not exceed a predefined \ac{PU}'s threshold interference, known as interference temperature. In the \textit{overlay} paradigm, the \ac{SU} detects the primary network transmission via sensing methods and transmits its data simultaneously to the licensed user. By applying signal processing techniques, both \ac{SU} and \ac{PU} can maximize the network gains and achieve the best performance in terms of capacity among the cognitive protocols \cite{Srinivasa2006}. The \textit{interweave} approach opportunistically explores ``voids'' on the frequency spectrum used by primary users. These ``voids'' are frequency resources not used by \acp{PU} detected via spectrum sensing by the \acp{SU}.  

In \ac{DSA}, the network dynamics changes constantly over time due to the variation of users accessing the channel during a time period. In such conditions, it is hard to propose a model that can capture the environment dynamics with sufficient precision. To overcome this situation, network nodes can learn by interacting with the environment instead of following a pre-determined model. 

In general, the main goal of spectrum access is to provide means for a user to transmit its data over the channel. This is done by either sensing the (pre-)allocated channels searching for white spaces or controlling the co-channel interference of other users. In the former, there is only one transmitter per channel where a user identify an idle period to transit during a certain time $T$, on the other hand, the latter allows multiple transitions in the same channel but the interference should be controlled by power allocation or beamforming strategies for example. 

\begin{figure}[!t]
	\centering
	{\includegraphics[width=0.75\textwidth]{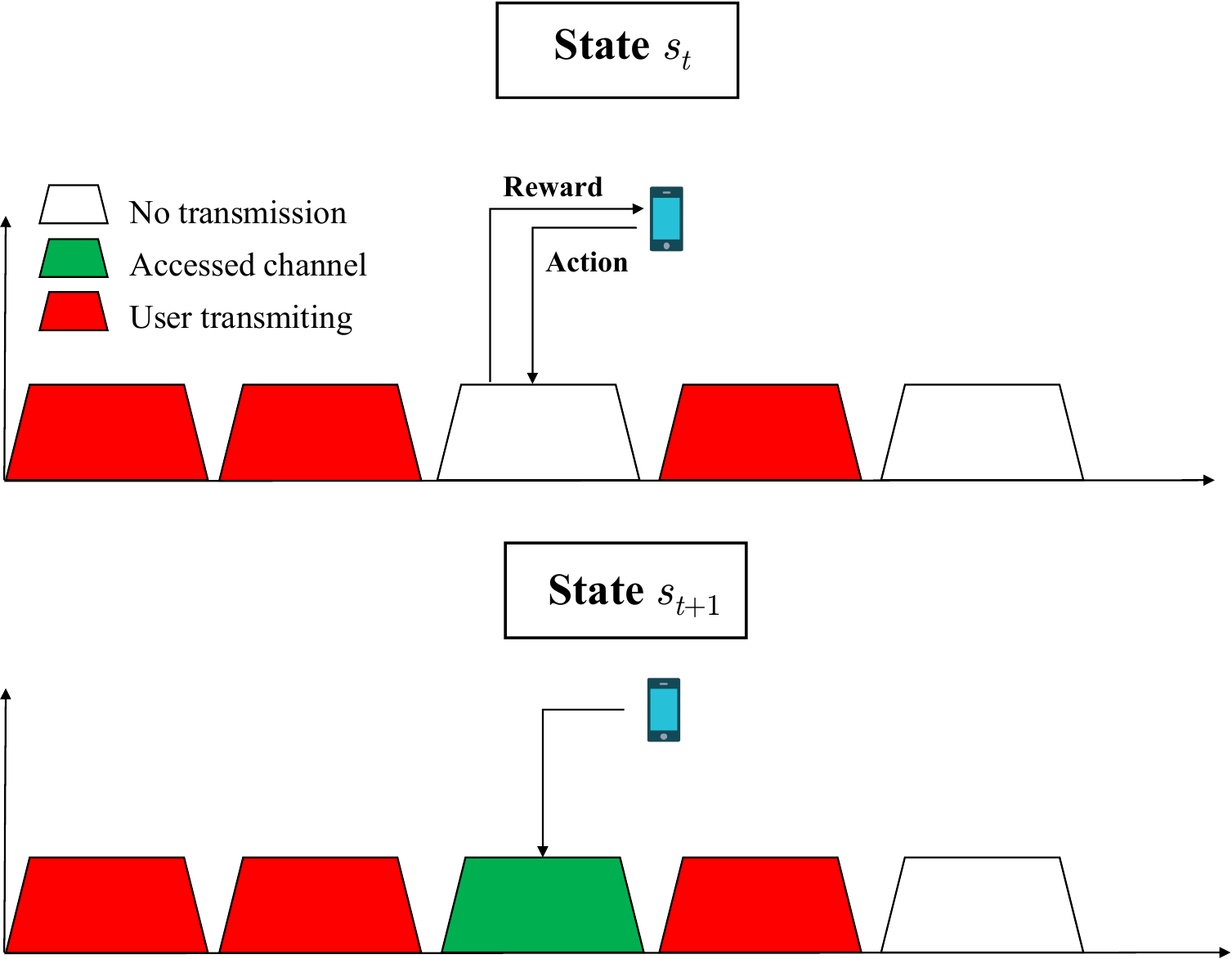}}
	\caption{Illustration of spectrum access.}
	\label{fig:spectrumaccess}
\end{figure}

This problem can be modeled as a \ac{RL} task (see Section \ref{sub:reinfo_learning} for more detailed \ac{RL} description). The \textit{state} $s_t$ can represent the channel(s) occupation at time $t$ as illustrated in Figure \ref{fig:spectrumaccess}. It can be idle, no user is transmitting over the channel,  or in use, there is a user transmitting. By observing $s_t$, the user takes an \textit{action} $a_t$ that varies according with the strategy adopted which includes, for instance, transmitting for $T$ instances of time, transmitting all the packets with a certain power, adjusting the beamforming directions or simply waiting. As a result of the observed action-state pair ($a_t$,$s_t$) the user receives an associated \textit{reward} $R_{t}$ that represents the decision impact and it changes to the \textit{next state} $s_{t+1}$ at time $t+1$. The algorithm's goal is to maximize $R_{t}$ based on some performance criterion which usually is related to packet transmission success.

Based on this general idea, many works in the literature considered \ac{RL} solutions to address \ac{DSA} problems.

\section{Q-learning Based Methods}

Q-learning based methods (see Section \ref{subsub:sarsa_q_learning}) are suitable for \ac{DSA} scenarios due to the action-reward process by trial and error procedure as demonstrated in \cite{Venkatraman2010}. Results showed that Q-learning algorithm increases the network spectral efficiency and reduce average interference power in spectrum access \cite{Liu2024}. Moreover, it can achieve a close to optimal performance for a small number of agents and actions \cite{Albinsaid2022}.

Motivated by previous results in the literature, reference \cite{Rastegardoost2018} explored the opportunistic WiFi white spaces access by \ac{LTE-U}. The agent recognizes the WiFi available channels by applying carrier sense and employs Q-learning to schedule the \ac{LTE-U} according to the idle channel duration. The proposed framework minimizing the carrier sense WiFi latency while maximizes \ac{LTE-U} access to idle channels. 

In \cite{Teng2010}, authors proposed a Q-learning based auction algorithm to control spectrum access in a \ac{CRN}. In the proposed scheme, the \acp{SU} make independent bid decisions while the \ac{PU} controls bid collisions and ensures its own access to the channel.  The non-cooperative spectrum access for \ac{CRN} was investigated in \cite{Jiang2018,Fan2018}. While \cite{Jiang2018} implemented a multi-agent Q-learning based scheme,  \cite{Fan2018} proposed a decentralized  self-adaptive Q-learning algorithm. In both cases, \acp{SU} can learn from their individual actions-reward history and update their actions to maximize their own spectral efficiency. 

Reference \cite{Liu2021a} investigated Q-learning algorithm for spectrum access in cognitive industrial \ac{IoT} for orthogonal multiple access,  underlay spectrum access and \ac{NOMA} scenarios. Authors showed that Q-learning \ac{NOMA} yielded to low outage probability, high throughput, and low	interference to the \acp{PU}. In \cite{Jiang2023}, the authors combined Q-learning with compressive random access, taking advantage of satellite access nodes sparsity. An algorithm variation was also proposed where the frame length could be adjusted to reduce collision and increase performance under a large number of access nodes.

These works are summarized in Table \ref{tab:spectrum_access_qlearning}.

\begin{table*}[!t]
	\centering
	\caption{Description summary of \ac{ML} works for spectrum access using Q-Learning strategy.}
	{\footnotesize \begin{tabularx}{\linewidth}{
				|>{\hsize=0.6\hsize}X|
				>{\hsize=0.5\hsize}X|
				>{\hsize=1.9\hsize}X|
			}
			\hline 
			\textbf{Reference}   & \parbox[c]{1.8cm}{\centering\vspace*{0.05cm}\textbf{Network type}}  & \parbox[c]{7cm}{\centering\vspace*{0.05cm}\textbf{Comment}}  \\ \hline
			
			\parbox[c]{2cm}{\centering \vspace*{0.15cm}\cite{Venkatraman2010}} & \parbox[c]{1.8cm}{\centering\vspace*{0.15cm} \ac{CRN}}   &    Q-learning efficiency for opportunistic spectrum access.   \\  
			\hline 
			\parbox[c]{2cm}{\centering \vspace*{0.25cm}\cite{Liu2024}}  &    \parbox[c]{1.8cm}{\centering\vspace*{0.05cm}Cognitive internet of vehicles }    & \vspace*{-0.4cm}Optimized spectrum selection strategies in a vehicular network for the underlay, overlay, and collaborative spectrum access modes.  \\
			\hline
			\parbox[c]{2cm}{\centering \vspace*{0.35cm}\cite{Rastegardoost2018}}  &  \parbox[c]{1.8cm}{\centering\vspace*{0.25cm}  \ac{LTE-U}  and WiFi }    & Allowed \ac{LTE-U} to access WiFi white spaces by employing carrier sensing and adjusting \ac{LTE-U} duty cycle to WiFi activity.  \\
			\hline
			\parbox[c]{2cm}{\centering \vspace*{0.25cm}\cite{Teng2010}} & \parbox[c]{1.8cm}{\centering\vspace*{0.15cm}  \ac{CRN}  }   &  Developed a Q-learning based auction algorithm for \ac{CRN} with multiple secondary users. \\
			\hline
			\parbox[c]{2cm}{\centering \vspace*{0.25cm}\cite{Jiang2018}}  & \parbox[c]{1.8cm}{\centering\vspace*{0.15cm} \ac{CRN}}  &  Investigated a multi-agent non-cooperative spectrum access with limited channel switch. \\
			\hline
			\parbox[c]{2cm}{\centering \vspace*{0.25cm}\cite{Fan2018}}  & \parbox[c]{1.8cm}{\centering\vspace*{0.15cm}Ultra dense network  }   &  Generalized temporal-spatial non-cooperative spectrum access for millimeter wave ultra dense scenarios. \\ 
			\hline
			\parbox[c]{2cm}{\centering \vspace*{0.25cm}\cite{Liu2021a}}  & \parbox[c]{1.8cm}{\centering\vspace*{0.15cm} Industrial \ac{IoT}}  &  Investigated Q-learning algorithm in three spectrum access scenarios: orthogonal multiple access, underlay spectrum access, and \ac{NOMA}. \\ 
			\hline
			\parbox[c]{2cm}{\centering \vspace*{0.45cm}\cite{Jiang2023}}  & \parbox[c]{1.8cm}{\centering\vspace*{0.35cm} Satellite \ac{IoT} } &  Proposed a diversity slotted compressive random access control scheme with Q-learning. Also applied compressive sensing to adjust frame length. \\ 
			\hline
	\end{tabularx} }
	\label{tab:spectrum_access_qlearning}
\end{table*} 

\section{Deep Q-learning Based Methods}
 
In spectrum access problems Q-learning algorithms also suffer from low performance and high computational complexity for large scale 
networks since the Q-table size increases exponentially with the number of users. Therefore, literature works 
considered \ac{DRL} solutions to overcome this issue (see Section \ref{subsub:ANN} and Section \ref{subsub:approx_techniques} for \ac{DNN} and \ac{DRL} details).

The authors in \cite{Bokobza2023} applied \ac{DQN} with a single agent to address both spectrum sensing and access policies.  Reference \cite{Li2018} addressed a \ac{CR} access for two pairs of transmitter/receiver nodes. Sensors were spatially deployed to collect the \ac{PU} signal which is used to train the \ac{DQN} such that \ac{SU} can adjust its power transmission to keep \ac{PU} \ac{QoS}. On the other hand, work in \cite{Zhu2018a} explored a \ac{DRL} framework to improve the transmission packet efficiency for cognitive \ac{IoT} networks. Authors showed that the Q-learning algorithm can efficiently help a relay to find the optimal strategy to transmit packets through multiple channels. An improved Thompson sampling algorithm for \ac{DQN} was used in \cite{He2024} to predict availability and time duration of spectral holes. The proposed algorithm took into account both the channel qualities and channel occupancy probabilities, achieving a high throughput and decreasing the number of time slots needed for spectrum sensing. Authors in \cite{Wang2018b} considered a multichannel access with correlated channels scenario. The \ac{DQN} framework was able to find near optimal solutions even for more complex scenarios. An extension was also proposed where the it periodically evaluates the reward and if the performance degrades by a certain amount, the algorithm restarts the learning process.
 
Although \ac{DQN} is an effective solution for dynamic access, it has some drawbacks such high computational complexity due to the possible large action space at user nodes and overestimation of action values at partially-observable channel conditions. To deal with \acp{DQN} poor performance in partially observable scenarios, the work in \cite{Kaur2023} integrated a gated recurrent unit over \ac{LSTM} layer to optimize spectrum access in a limited feedback scenario. To overcome learning divergence and computation overhead, the work in \cite{Pei2023} adopted the mean field technology by simplifying the agents interactions in the \ac{MARL} \ac{DQN}. The main idea is to transform a set of agent interactions in an average of these interactions, simplifying the algorithm and guaranteeing convergence.

Reservoir computing \cite{Mosleh2018}, a special type of recurrent \ac{DNN}, trains only the output weights simplifying the training process. It was used in \cite{Chang2019} to reduce the \acp{DQN} computational complexity for spectrum access. The proposed strategy allows the \acp{SU} to learn the best access polices based only on their present and past spectrum sensing outcomes without any knowledge about statistics of the system or any \acp{PU} information.

On the other hand, a \ac{DDQN} approach combined with the \ac{DQN} can be used to reduce the overestimation \cite{Xu2018a} and to perform power allocation \cite{Zhang2023}. The solution was shown to be robust when applied to complex scenarios avoiding collisions and achieving near to optimal performance.  The work in \cite{Xu2018b} addressed the partially-observable channel situation. The authors applied a deep recurrent Q network which combines a \ac{DQN} with a recurrent neural network. It can estimate the Q values with only partial information about the spectrum. The work was extended in \cite{Xu2020} by considering independent channels and partial observations on at each time step. Results showed that although the proposed scheme has a slow convergence it can achieve nearly optimal performance even with no prior knowledge being robust to fast environment changes. 

Reference \cite{Naparstek2019} addressed both Q-learning complexity and overestimation using a combination of offline training, \ac{LSTM}  (see Chapter \ref{subsub:RNN} for \ac{LSTM} details) and \ac{DDQN}. The majority of computational complexity is concentrated in the offline training which simplifies the online \acp{DQN} learning process. It can be done at a cloud or network edge, for example, taking advantage of the quasi-static environment characteristics while training still reflects the channel parameters. \ac{LSTM} and \ac{DDQN} were used to deal with the network partially observable state. The proposed framework was shown to achieve good performance for complex multi-user scenario while reducing online complexity.  In \cite{Chen2022} the authors incorporated a dueling \ac{DQN} with prioritized experience replay combined with \ac{LSTM}. The proposed strategy allowed \acp{SU} to modify their paramters to select the optimal access policy, outperforming both \ac{DQN} and dueling \ac{DQN} schemes in terms of channel throughput and convergence speed.

As discussed in Section \ref{Sec:Allocation}, federate learning allows the training data to be learned by users in a distributed way. This enhances the efficiency of training process and provides user privacy for spectrum access \cite{Liu2023}. The authors in \cite{Chang2023} incorporated \ac{MARL} to address \ac{DSA} under quantized communication. In \cite{Zhu2023}, the authors introduced weights in the learning strategy to speed up the convergence of federated \ac{DQN} algorithm. This approach took into account the delay, transmission power and the utility of \ac{SU} as the reward function improving spectrum successful access rate.

These works are summarized in Table \ref{tab:spectrum_access_dql}.

\begin{table*}[!t]
	\centering
	\caption{Description summary of \ac{ML} works for spectrum access using deep Q-Learning strategy.}
	{\scriptsize \begin{tabularx}{\linewidth}{
				|>{\hsize=0.5\hsize}X|
				>{\hsize=0.4\hsize}X|
				>{\hsize=2.1\hsize}X|
			}
			\hline 
			\parbox[c]{1.7cm}{\centering\vspace*{0.05cm}\textbf{Reference}}   & \parbox[c]{1.6cm}{\centering\vspace*{0.05cm}\textbf{Network type}}  & \parbox[c]{8cm}{\centering\vspace*{0.05cm}\textbf{Comment}}  \\ \hline
			
			\parbox[c]{1.7cm}{\centering \vspace*{0.15cm}\cite{Bokobza2023}} &  \parbox[c]{1.6cm}{\centering\vspace*{0.15cm}\ac{CRN}}      &  Proposed a The \ac{DDQN} algorithm for jointly optimizing spectrum sensing and access.     \\  
			\hline 
			\parbox[c]{1.7cm}{\centering \vspace*{0.35cm}\cite{Li2018}} &  \parbox[c]{1.6cm}{\centering\vspace*{0.25cm}\ac{CRN} }     &  Proposed a power adjustment solution for cognitive radio access for two transmitters/receivers pairs in a non-cooperative scenario.      \\  
			\hline 
			\parbox[c]{1.7cm}{\centering \vspace*{0.15cm}\cite{Zhu2018a} } &\parbox[c]{1.6cm}{\centering\vspace*{0.05cm} \ac{CRN} and \ac{IoT}}   &   Improved transmission packet efficiency for cognitive \acl{IoT} networks using a deep Q-learning scheduling.  \\
			\hline 
			\parbox[c]{1.7cm}{\centering \vspace*{0.15cm}\cite{He2024}}  & \parbox[c]{1.6cm}{\centering\vspace*{0.05cm}\ac{CRN}} &   Incorporated Thompson sampling algorithm for \ac{DQN}  to predict availability and time duration of spectral holes. \\
			\hline 
			\parbox[c]{1.7cm}{\centering \vspace*{0.15cm}\cite{Wang2018b}} & \parbox[c]{1.6cm}{\centering\vspace*{0.15cm} Wireless networks }   & Proposed an adaptive \ac{DQN} for time-varying multichannel access scenario with correlated channels. \\
			\hline	 
			\parbox[c]{1.7cm}{\centering \vspace*{0.15cm}\cite{Kaur2023} }& \parbox[c]{1.6cm}{\centering\vspace*{0.15cm} Wireless networks }   & Proposed a gated recurrent unit assisted \ac{DQN} algorithm with experience replay buffer for imperfect feedback scenario. \\
			\hline	
			\parbox[c]{1.7cm}{\centering \vspace*{0.15cm}\cite{Pei2023}} &  \parbox[c]{1.6cm}{\centering\vspace*{0.15cm}\ac{LTE-U} and WiFi }  & Applied mean field technology to reduce \ac{DQN} complexity for cooperative distributed \ac{MARL} algorithm. \\
			\hline	
			\parbox[c]{1.7cm}{\centering \vspace*{0.4cm}\cite{Chang2019}} &  \parbox[c]{1.6cm}{\centering\vspace*{0.15cm} \ac{CRN}	}    & Simplified \ac{NN} training process using resevoir computing  to investigate distributive \ac{DSA} under the presence of spectrum sensing errors. \\ 
			\hline
			\parbox[c]{1.7cm}{\centering \vspace*{0.35cm}\cite{Xu2018a}}	& \parbox[c]{1.6cm}{\centering\vspace*{0.15cm}Wireless networks } &  Proposed a \ac{DDQN} approach combined with the \acl{DQN} for discrete	channels without prior information about the system dynamics. \\
			\hline 
			\parbox[c]{1.7cm}{\centering \vspace*{0.15cm}\cite{Zhang2023}}	& \parbox[c]{1.6cm}{\centering\vspace*{0.1cm}Cellular Network}  &  Explored a multi-agent \ac{DDQN} for joint spectrum sensing and power allocation for spectrum access. \\
			\hline 
			\parbox[c]{1.7cm}{\centering \vspace*{0.15cm}\cite{Xu2018b,Xu2020}}	& \parbox[c]{1.6cm}{\centering\vspace*{0.1cm}Wireless networks}  & Proposed a deep recurrent Q network to overcome \acl{DQN} limitations in partial observations scenarios. \\
			\hline  
			\parbox[c]{1.7cm}{\centering \vspace*{0.35cm}\cite{Naparstek2019}} &  \parbox[c]{1.6cm}{\centering\vspace*{0.15cm}Wireless networks}   &  Combined offline training, \acl{LSTM} and \acl{DDQN} to increase average rate for distributed spectrum access. \\
			\hline
			\parbox[c]{1.7cm}{\centering \vspace*{0.25cm}\cite{Chen2022}} &  \parbox[c]{1.6cm}{\centering\vspace*{0.15cm}\ac{CRN}}  &  Proposed a dueling \ac{DQN} algorithm with prioritized
			experience replay combined with recurrent neural network. \\
			\hline
			\parbox[c]{1.7cm}{\centering \vspace*{0.15cm}\cite{Liu2023}} &  \parbox[c]{1.6cm}{\centering\vspace*{0.15cm}\ac{CRN} }  &  Proposed a federate \ac{DRL} algorithm for spectrum access and power allocation. \\ \hline
			\parbox[c]{1.7cm}{\centering \vspace*{0.35cm}\cite{Chang2023}} &  \parbox[c]{1.6cm}{\centering\vspace*{0.15cm} Broadband radio service } &  Proposed a federate \ac{MARL} approach for \ac{DSA} and provided theoretical analysis to evaluate communication  efficiency and system performance trade-off. \\ \hline
			\parbox[c]{1.7cm}{\centering \vspace*{0.25cm}\cite{Zhu2023}} &  \parbox[c]{1.6cm}{\centering\vspace*{0.05cm}Vehicular network}  &  Proposed an asynchronous federated
				weighted learning algorithm to increase spectrum access rate and reduce collisions. \\
			\hline 
	\end{tabularx}}
	\label{tab:spectrum_access_dql}
\end{table*} 

\section{Other Methods}

Other \ac{ML} methods were also considered for the spectrum access problem \cite{Cha2019,Foukas2019,Flandermeyer2024,Xiang2023,Jouini2009,Bonnefoi2017,Raj2018,Gui2018}. The authors in \cite{Cha2019} used a sensor-aided \ac{DSA} system based on a \ac{RL} algorithm for a secondary \ac{IoT} network. The \ac{ML} algorithm runs in a central unit after the primary network interference data are captured by the sensors controlling the spectrum access of the \ac{IoT} network. However, the \ac{RL} approach leads to a long learning period which minimizes the performance gain. As an alternative to reduce the convergence time \ac{DRL} solutions were proposed. Reference \cite{Foukas2019} proposed an indoor mobile access architecture named Iris where a \ac{DRL} strategy based on dynamic pricing mechanism was implemented. The pricing problem was modeled as a Markov decision process allowing users to request spectrum on demand. The work in \cite{Gui2018} addressed a \ac{DL}-aided \ac{NOMA} system problem. The authors proposed a pre-training structure based on restricted Boltzmann machines to train the initial input and then they used a \ac{LSTM} network to learn the channel characteristics.  The scheme was shown to achieve high performance in terms of data rate and proved to be robust and more efficient than the conventional \ac{NOMA} approaches.

In \cite{Flandermeyer2024} a \ac{RL} approach for cognitive radar was proposed. The problem is formulated as a high-dimensional continuous control task by redesigning \ac{MDP} formulation. Each agent processes both time and frequency domain observations in order to dynamically adapt the radar waveform to achieve collision	avoidance. This is achieved by balancing the radar 	performance and mutual interference mitigation at the reward function. On the other hand, a decentralized \ac{MARL} approach was proposed by \cite{Xiang2023} for vehicular networks. The agents received emergent communication via a dedicated channel to learn effective policies for channel access.
	
\ac{MAB} is an efficient formulation to keep tracking of channel occupancy for spectrum access \cite{Jouini2009}. Reference  \cite{Bonnefoi2017} showed that  \ac{MAB} algorithms can handle decentralized access for \ac{IoT} scenario. Moreover, they are robust even for a non stationary environment due to a large number of intelligent devices. Reference \cite{Raj2018} proposed a two stage spectrum access strategy through \ac{MAB} formulation and a Bayesian learning approach. In the first stage a channel is selected to be sensing via \ac{MAB} and, in the second stage, Bayesian learning is applied to determine how often the channel will be sensed. The proposed framework minimizes the sensing time while maximizing the spectrum access, improving the network throughput with a small increase on \ac{PU} interference. 

 The \ac{DDPG} approach was explored by \cite{Kassab2020} where a multi-agent \ac{RL} solution for \ac{IoT} frame access is 
proposed. This was 
done by exploring device-level correlation and time correlation of events which allowed the users to avoid collisions. The work in 
\cite{Meng2020} 
showed that \ac{DDPG} solution can deal with quantization error achieving a good performance in terms of sum-rate and robustness in a 
multi-user 
cellular network.  However, due to the instability of the actor-critic network, reaching the optimal point is challenging, especially in random 
environments like the Rayleigh fading channel \cite{Albinsaid2022}.

\ac{PPO} approach was also explored for channel access. \cite{Doshi2021} modeled the media access problem as a two state Markov decision process and applied the \ac{PPO} for intelligent spectrum access yes/no transmit decision. This work was extended by \cite{Doshi2022} where the authors generalize the previous working by allowing the users to choose the best modulation scheme. In \cite{Zhang2020}, the authors developed an asynchronous
advantage actor critic power control solution for \acp{CRN}. A distributed \ac{PPO} was used to improve the root mean square prop 
performance by using Adam optimization allowing a \ac{SU} to learn how to adjust its transmit power.

These works are summarized in Table \ref{tab:spectrum_access_other}.
\begin{table*}[!t]
	\centering
	\caption{Description summary of \ac{ML} works for spectrum access using other \ac{RL} strategies.}
	{\footnotesize \begin{tabularx}{\linewidth}{
			|>{\hsize=0.6\hsize}X|
			>{\hsize=0.4\hsize}X|
			>{\hsize=2\hsize}X|
		}
		\hline 
		\parbox[c]{1.7cm}{\centering \textbf{Reference}}  & \parbox[c]{1.5cm}{\centering\vspace*{0.05cm}\textbf{Network type}}  & 
		\parbox[c]{8cm}{\centering\vspace*{0.05cm}\textbf{Comment}}  \\ \hline
		
		\parbox[c]{1.7cm}{\centering \vspace*{0.35cm}\cite{Cha2019}}  &    \parbox[c]{1.5cm}{\centering\vspace*{0.35cm}\ac{IoT} }     &  Developed a sensor-aided \acl{DSA} with 
		\ac{RL}-based algorithm providing self-organizing feature for massive number of \ac{IoT} devices.    \\ \hline 
		\parbox[c]{1.7cm}{\centering \cite{Foukas2019}} & \parbox[c]{1.5cm}{\centering\vspace*{0.05cm}Indoor wireless networks }   &  \vspace*{-0.5cm} Developed an indoor spectrum 
		access architecture based on \ac{DRL} with dynamic pricing mechanism.  \\ \hline		
		\parbox[c]{1.7cm}{\centering \vspace*{0.25cm}\cite{Flandermeyer2024}} &\parbox[c]{1.5cm}{\centering\vspace*{0.25cm} Cognitive radar }   &  Proposed a \ac{DRL} algorithm to 
		select radar waveform to avoid collisions, to optimize bandwidth utilization ,and to mitigate distortion.  \\ \hline		
		\parbox[c]{1.7cm}{\centering \vspace*{0.25cm}\cite{Xiang2023} }& \parbox[c]{1.5cm}{\centering\vspace*{0.25cm}Vehicular network }   &  Proposed an inter-agent communication 
		mechanism into \ac{MARL} algorithm where emergent communications are traded among the agents via a dedicated channel.   \\ 
		\hline
		\parbox[c]{1.9cm}{\centering \vspace*{0.65cm}\cite{Jouini2009,Bonnefoi2017,Raj2018}} & \parbox[c]{1.5cm}{\centering\vspace*{0.65cm} \ac{CRN} } & A \ac{MAB} learning 
		frameworks to   channel usage prediction and intelligent access in a cognitive scenario to achieve higher
		throughput with lower energy consumption. \\ \hline
		\parbox[c]{1.7cm}{\centering \vspace*{0.25cm}\cite{Gui2018}} &  \parbox[c]{1.5cm}{\centering\vspace*{0.25cm} One-cell cellular network } & Combined \ac{LSTM}  with \ac{DL} for 
		\ac{NOMA} to allow spectrum access for users with random deployment served by one
		base station. \\ 
		\hline  
	\parbox[c]{1.7cm}{\centering \vspace*{0.15cm}	\cite{Kassab2020}} &  \parbox[c]{1.5cm}{\centering\vspace*{0.15cm}  \ac{IoT} } & Proposed a multi agent \ac{DDPG} approach for 
		spectrum access in \ac{IoT} networks. \\
		\hline 
		\parbox[c]{1.7cm}{\centering \vspace*{0.15cm}\cite{Doshi2021,Doshi2022}} &  \parbox[c]{1.5cm}{\centering\vspace*{0.15cm}  Cellular network } & Proposed a decentralized 
		contention-based medium access based on \ac{PPO} algorithm. \\
		\hline 
		\parbox[c]{1.7cm}{\centering \vspace*{0.25cm}\cite{Zhang2020}} &  \parbox[c]{1.5cm}{\centering\vspace*{0.25cm}  \ac{CRN} } & Proposed a power control scheme for underlay 
		spectrum access applying a distributed \ac{PPO} algorithm. \\
		\hline 
	\end{tabularx} } 
	\label{tab:spectrum_access_other}
\end{table*} 

\section{Summary}\label{sub:Chap_Access:summary}
In this chapter, we addressed the spectrum access mechanism. Specifically, we modeled this problem as a \ac{RL} task, which can be solved by Q-learning, deep Q-learning or other \ac{RL} methods. 

We also surveyed and summarized recent literature works presenting their main contributions and the network type where the \ac{RL} algorithms were applied.

\chapter{Further Aspects on Spectrum Sharing}\label{Sec:Aspects}
\section{Spectrum Handoff}  

\begin{figure}[!t]
	\centering
	{\includegraphics[width=0.75\textwidth]{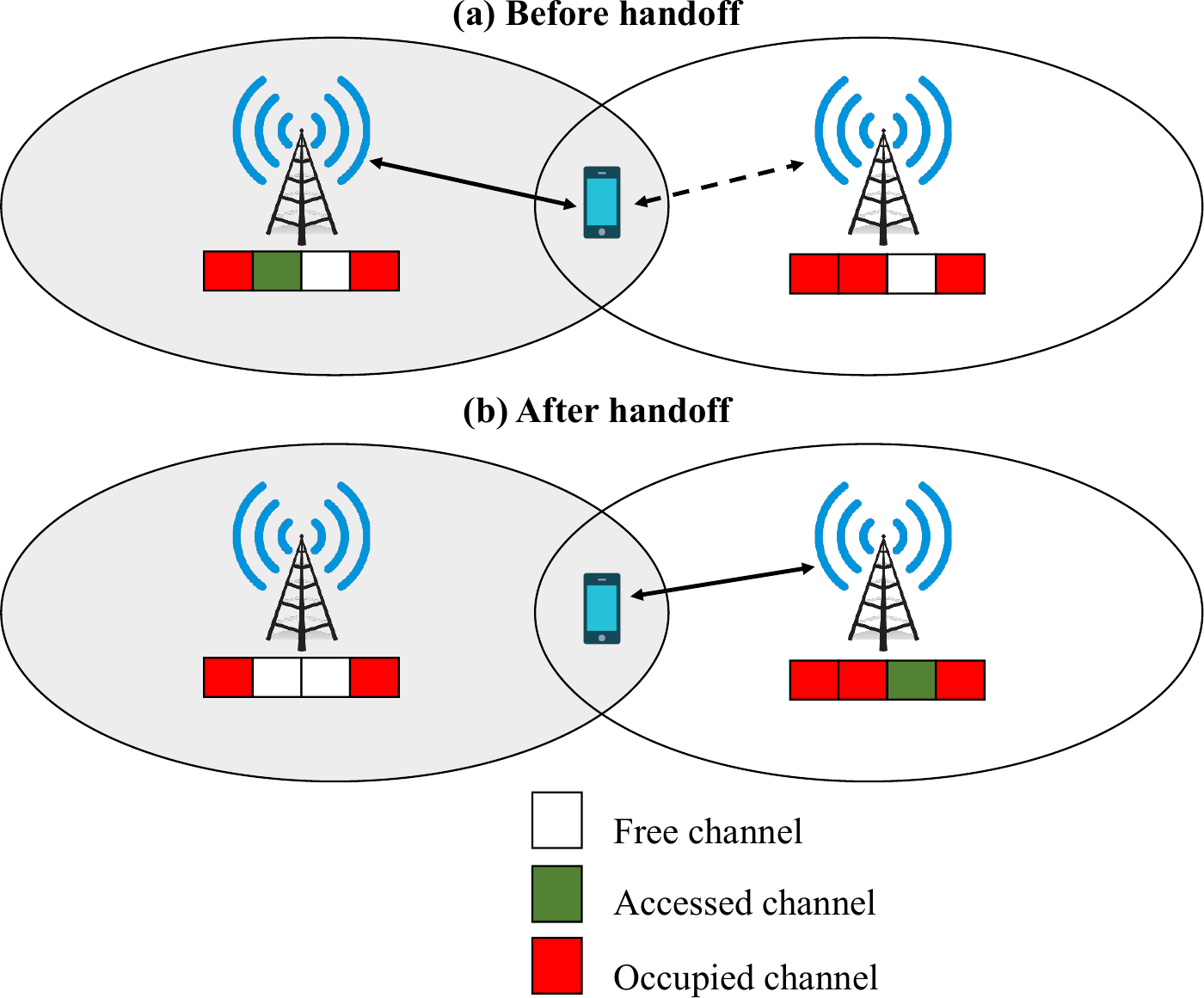}}
	\caption{Illustration of a spectrum handoff process.}
	\label{fig:spectrumhandoff}
\end{figure}

There are situations where a user needs to switch channel or \ac{BS} while transmitting or receiving data as illustrated on Figure \ref{fig:spectrumhandoff}. The handoff, or handover, is done to protect users from interference, to meet users requirement, or when a user changes its geographical location. 
The spectrum handoff, then, needs to provide for both licensed and unlicensed users means to keep their connection while achieving an efficient spectrum utilization.

In the literature, some recent works addressed the spectrum handoff problem with different \ac{ML} solutions to optimize users handovers. Handoff works using \ac{ML} approaches are summarized in Table \ref{tab:spectrum_handoff}.

Authors in \cite{Babjan2023} evaluated  several classification algorithms for adaptive beam selection during handover. Results showed that \ac{k-NN} (see Section \ref{subsub:k-nn}) and Nave Bayes (see Section \ref{subsub:bayes_class}) have a better performance in terms of network latency while Random Forest (see Section \ref{subsub:random_forest}) has a better accuracy. \cite{Iyer2023} compared \ac{SVM} (see Section \ref{subsub:svm}) and \ac{ANN} (see Section \ref{subsub:ANN}) performances for handoff in \acp{CRN}. While \ac{ANN} showed a slightly better accuracy, \ac{SVM} showed to be more suitable due to the need of less data for training. The \ac{SVM} technique is also effective to reduce handoffs when the \acp{SU} have prior knowledge of the environment \cite{Srivastava2023}.

In \cite{Haldorai2017} the mobile users of a \ac{CRN} were modeled as particles which move to the optimal solution, i.e., home locator register or visitor locator register. The proposed \ac{PSO} algorithm identified spectrum holes and maximized the number of \acp{SU} occupying existing primary holes with controlled interference. A \ac{HMM} (see Section \ref{subsub:hidden_markov}) is used in \cite{Jaffar2018} to capture past \ac{CR} users activity to predict future users movements. The best channel is selected based on this prediction reducing both number of handoff and average handoff delay.

References \cite{Oyewobi2019,Shi2019,Koushik2018} proposed \ac{RL}-based strategies (see Section  \ref{sub:reinfo_learning}) to control handoff. The authors in \cite{Oyewobi2019} integrated  algorithms into the channel selection strategy to verify channel occupancy and quality for industrial \ac{IoT} applications. The \ac{RL} algorithms organize a candidate channel list optimizing the ordering and sorting of the channels. This approach reduces the time of sensing stage and the probability of a bad channel quality selection.  In  \cite{Shi2019}, Q-learning (see Section \ref{subsub:sarsa_q_learning}) for  \ac{QoE}-driven approach for \acp{CRN} is considered. The proposed model is based on calculating the \ac{SU} package drop rate to decide if the \ac{SU} continue in the actual channel or it has to perform a handoff due to \ac{PU}'s presence. On the other hand, an implementation of a \ac{CRN} using universal software radio peripheral and GNU radio using Q-learning and transfer learning algorithms is done in \cite{Koushik2018}. Q-learning takes a long time to achieve optimal conditions due to the parameters adjustment. Do deal with this issue, a transfer learning solution is used to accelerate the convergence process. 

\ac{DRL} algorithms (see Section \ref{subsub:approx_techniques}) were explored to develop predictive handoff models. Reference \cite{Cao2020}  used in a  \ac{DQN} to learn the sequence of users handoffs reducing the number of iterations and meeting the network \ac{QoS} requirements. The work in \cite{Alkhateeb2018a} predicted blockage and determined whether a user should or should not change the connected \ac{BS} while keeping its transmission. In \cite{Luo2022} the authors  minimized the transmission latency of \acp{SU} by proposing a priority queuing model for handoff.

\begin{table*}[!t]
	\centering
	\caption{Summary of \ac{ML} works for spectrum handoff.}
	{\footnotesize \begin{tabularx}{\textwidth}{
			|>{\hsize=0.6\hsize}X|
			>{\hsize=1.9\hsize}X|
			>{\hsize=0.5\hsize}X|
		}
		\hline 
		\parbox[c]{2cm}{\centering\vspace*{0.05cm}\textbf{\ac{ML} Approach } }  & \parbox[c]{6cm}{\centering\vspace*{0.05cm}\textbf{Commentary} }  & \parbox[c]{1.7cm}{\centering\vspace*{0.05cm}\textbf{Related Works}} \\ \hline
		\parbox[c]{2cm}{\centering\vspace*{0.05cm}Classification  frameworks}  & Comparison of several classification frameworks 
		performance for beam selection during handoff.          & \parbox[c]{1.5cm}{\centering\vspace*{0.05cm}\cite{Babjan2023} }       \\ \hline
		\parbox[c]{2cm}{\centering\vspace*{0.05cm}\ac{SVM} and {ANN}}  & Comparison of \ac{SVM} and {ANN} algorithms for seamless 
		handoff in \acp{CRN}          &  \parbox[c]{1.5cm}{\centering\vspace*{0.05cm}\cite{Iyer2023}}       \\
		\hline		
		\parbox[c]{2cm}{\centering\vspace*{0.05cm}\ac{SVM}}  & Reducing execution time to complete handoof by proposing a 
		metaheuristic \ac{SVM}-based algorithm.          &  \parbox[c]{1.5cm}{\centering\vspace*{0.05cm}\cite{Srivastava2023}}       \\
		\hline
		\parbox[c]{2cm}{\centering\vspace*{0.05cm}Particle swarm optimization}  & Optimizing visitor location register and home location 
		register mapping for cognitive users.          &  \parbox[c]{1.5cm}{\centering\vspace*{0.05cm}\cite{Haldorai2017}}       \\
		\hline
		\parbox[c]{2cm}{\centering\vspace*{0.05cm}Hidden Markov model} &   Prediction of the next secondary user movement and 
		primary user activity  to reduce handoof delay.           & \parbox[c]{1.5cm}{\centering\vspace*{0.05cm} \cite{Jaffar2018}}     \\
		\hline
		\multirow{3}{*}{\parbox[c]{2cm}{\centering\vspace*{0.05cm}Reinforcement Learning}} &  Evaluation of the channel quality and future 
		occupancy for handoff in industrial Internet of things  networks.    &  \parbox[c]{1.5cm}{\centering\vspace*{0.05cm}\cite{Oyewobi2019}   }  \\\cline{2-3}
		&  Reducing of the packet dropped rate in handoffs for multimedia transmissions over \acl{CRN}s.   & \parbox[c]{1.5cm}{\centering\vspace*{0.05cm}\cite{Shi2019} } \\\cline{2-3}
		& Handoff GNU radio implementation using Q-learning and transfer learning.    & \parbox[c]{1.5cm}{\centering\vspace*{0.05cm}\cite{Koushik2018}}  \\
		\hline
		\multirow{3}{*}{\parbox[c]{2.2cm}{\centering\vspace*{0.05cm}Deep Reinforcement Learning} } &   Predictive and proactive spectrum handoff for hybrid interweave and underlay cognitive radio networks.        & \parbox[c]{1.5cm}{\centering\vspace*{0.05cm} \cite{Cao2020}  }   \\\cline{2-3}
		&  Blockage prediction in handoffs to enable high-mobile millimeter wave applications.     &\parbox[c]{1.5cm}{\centering\vspace*{0.05cm} \cite{Alkhateeb2018a} } \\\cline{2-3}		
		&  \ac{SU} latency handoff reduction by developing a hybrid queuing model.     & \parbox[c]{1.5cm}{\centering\vspace*{0.05cm}\cite{Luo2022}}  \\
		\hline
	\end{tabularx}}
	\label{tab:spectrum_handoff}
\end{table*}   

\section{Beamforming}

In order to meet \ac{5G} and \ac{6G} requirements, cellular networks tend to be denser and heterogeneous (in terms of inter-site distance),
which can lead to high inter-cell interference and reduced performance gains. 
To tackle the issues of dense deployment scenarios,
where many users share the same channel frequency, a combination of massive \ac{MIMO} and \ac{mmWave} can be used to achieve both high spectral efficiency and high capacity 
\cite{Alkhateeb2018,Kao2018,Tao2019,Long2018,Lavdas2022,Aljumaily2019,Elbir2019,Zhang2022a,Liu2022,Lizarraga2019}.
Beamforming refers to spatial filtering applied by the transmitter and/or receiver
in order to favor some spatial directions over others \cite{beamforming:firstpaper}.
Using large antenna arrays overcomes the high pathloss at \ac{mmWave} frequencies,
since the beams are narrow in the direction of the intended user.
In a \ac{CR} networks, for instance, secondary users can take advantage of directional transmission 
to boost their performance while respecting the \ac{QoS} requirements of primary users \cite{Sangdeh2019,Masrour2017}.

\ac{DBF} is a popular choice due to the inherent advantages and cost efficiency of digital processing techniques. These techniques 
include the generation of a large number of beams that can be combined to reduce interference while increasing \ac{SINR} and 
throughput gains. 
The element-wise digital control allows adaptive beamforming to respond to environment changes
by choosing the antenna weights based on the estimated characteristics of the propagation channel and/or the statistics of the received 
data \cite{Bailleul2016}.

Recent works considered \ac{ML} as a tool for beamforming design. The authors of \cite{Luijten2019} used \ac{DNN} (see Section \ref{subsub:ANN}) jointly with adaptive \ac{DBF}
to reconstruct ultrasound images from radio frequency channel data with reduced computational and time complexity. The main idea is to model the \ac{DNN} as a minimum variance beamforming architecture and train the \ac{DNN} to calculate the beamforming weights. Reference \cite{Jee2023} proposed a \ac{DNN} framework for beamforming design in a distributed manner. The authors adopted a unsupervised learning to deal with nonlinear systems complexity. In \cite{Chen2023}, a \ac{DNN} algorithm was proposed to optimize \ac{QoS} constraints for both access point clustering and beamforming design. 

In \cite{Xia2019} the authors considered a  \ac{CNN} (see Section \ref{subsub:CNN} for \ac{CNN} details) to find the best beamforming weights that minimize the downlink power transmission in a \ac{MISO} network. The proposed \ac{CNN} takes advantage of the uplink-dowlink duality to predict the power allocation vector and build the beamforming matrix with low computational delay.
Reference \cite{Kwon2019} considered a two-user \ac{MISO} interference channel network,
in which the users determine which beamforming scheme (maximum ratio transmission or zero forcing)
should be applied to maximize the achievable sum rate. Transmit power and channel vectors are used as input in a \ac{DNN} to obtain the optimal performance strategy. \ac{ANN}, however, leads to a high energy cost for beamforming design as it size increases. To deal with this drawback, authors in \cite{Ge2023} proposed a feature estimation framework based on	spiking neural networks. The energy saving is achieved by coding the spiking neural network input by transforming the \ac{CSI} into binary spike sequences and decoding at the output side.

The work in \cite{Sun2020} used a \ac{DRL} via deep Q-network to predict the balancing
coefficients for selfish and altruistic beamforming to reduce interference and improve
the users efficiency in ultra dense \ac{MIMO} networks. The main idea is to avoid the beamforming matrix prediction by finding the balance coefficients to parameterize the final beamforming vectors, which are based only on large scale channel fading, which reduces the signaling overhead. \ac{DQN}-based beamforming strategies can cause correlation between input and target values. To overcome this issue, a prioritized experience replay is proposed by \cite{Ahmad2022} in \acp{UAV} scenario. The results showed that the beamforming performance can be enhanced due to data training reuse.

In \cite{Diamantaras2019} the authors used \ac{RL} to estimate the channel map parameters
in order to select the optimal relay position. The beamforming weights, which are a function of the relay positions, are then calculated so as to maximize the average \ac{SINR} subject to power constraints. On the other hand, a federated learning approach was explored in \cite{Kim2023a}. The authors proposed a subgradient based algorithm to minimize the \ac{MSE} and optimize the system learning rate. The authors in \cite{Li2023b} showed that \ac{RL} takes to long to converge for large number of elements in reconfigurable intelligent surface. To solve this issue, they proposed a double \ac{DQN} for beamforming optimization. This approach simplified the optimization task and was shown to achieve a good performance.

Previous works on \ac{DBF} also consider \ac{mmWave} frequencies based on \ac{ML} approaches for beam design. Reference \cite{Alkhateeb2018} developed an efficient channel training and a coordinated beamforming design to maximize the system effective achievable rate for high-mobile users in dense deployments. The main idea of that paper is to use uplink training pilots with omni- or quasi-omni beam patterns to predict the best radio frequency beamforming vector.
The gains are achieved for high-speed terminals and a massive number of antennas at the \ac{BS}, when compared to baseline solutions. 
Moreover, the results showed that beamforming adapts to time-varying scenarios and that the \acp{BS} does not require phase 
synchronization for coordination, which makes practical implementations feasible. The work in \cite{Kao2018} proposed a two-stage 
beamforming design. In the first stage, an offline training is performed via a technique that combines clustering with feature selection, 
named \ac{LLC}, which determines a subset of eigen-beams that are used to acquire channel information. The second stage uses the 
initial solution obtained in the first stage
as an input to a Rosembrock search. The Rosembrock search is a numerical beam searching approach,
which is performed online and determines the best beamforming vector. The proposed solution is found to be scalable, operates with reduced time and low power consumption for \ac{mmWave} beamforming. Digital beamforming works are summarized in Table \ref{tab:beamforming_digital}.

\begin{table*}[!t]
	\centering
	\caption{Summary of digital beamforming works with \ac{ML}.}
	{\scriptsize \begin{tabularx}{\linewidth}{
				|>{\hsize=0.7\hsize}X|
				>{\hsize=0.7\hsize}X|
				>{\hsize=2\hsize}X|
				>{\hsize=1.0\hsize}X|
				>{\hsize=0.6\hsize}X|
			}
			\hline
			\parbox[c]{1.2cm}{\centering\vspace*{0.05cm}Work}	 & \parbox[c]{1.5cm}{\centering\vspace*{0.05cm}\ac{ML} Technique} & \centering Solution Approach & \centering Optimization Task & Frequency \\
			\hline
			\parbox[c]{1.2cm}{\centering\vspace*{0.05cm}\cite{Luijten2019}} & \parbox[c]{1.2cm}{\centering\vspace*{0.05cm}\ac{DNN}} & Model the \ac{DNN} as a minimum variance beamforming architecture. &  Minimize variance & \parbox[c]{1.2cm}{\centering\vspace*{0.05cm}sub 6 GHz} \\ 
			\hline
			\parbox[c]{1.2cm}{\centering\vspace*{0.05cm}\cite{Jee2023}}	 & \parbox[c]{1.2cm}{\centering\vspace*{0.05cm}\ac{DNN}} & Applies a unsupervised deep learning with constraint in a distributed way. &  Maximize sum rate & \parbox[c]{1.2cm}{\centering\vspace*{0.05cm}$-$}  \\
			\hline
			\parbox[c]{1.2cm}{\centering\vspace*{0.05cm}\cite{Chen2023}}	 & \parbox[c]{1.2cm}{\centering\vspace*{0.05cm}\ac{DNN}} & Reduces the hyperparameters of loss function, and introduces a learnable safety distance parameter in the loss function. &  Maximize sum rate while minimize the number of access point clustering & \parbox[c]{1.2cm}{\centering\vspace*{0.05cm}$-$ } \\
			\hline
			\parbox[c]{1.2cm}{\centering\vspace*{0.05cm}\cite{Xia2019}}  & \parbox[c]{1.2cm}{\centering\vspace*{0.05cm}\ac{CNN}} & Explores uplink-downlink duality to predict the power allocation vector. & Minimize power subject to \ac{SINR} constraint & \parbox[c]{1.2cm}{\centering\vspace*{0.05cm}sub 6 GHz} \\
			\hline
			\parbox[c]{1.2cm}{\centering\vspace*{0.05cm}\cite{Kwon2019}}  & \parbox[c]{1.2cm}{\centering\vspace*{0.05cm}\ac{DNN}} & Uses transmit power and channel vector as input for the \ac{DNN}. &Maximize sum-rate & \parbox[c]{1.2cm}{\centering\vspace*{0.05cm}$-$} \\ 
			\hline
			\parbox[c]{1.2cm}{\centering\vspace*{0.05cm}\cite{Ge2023}}  & \parbox[c]{1.2cm}{\centering\vspace*{0.05cm}Spiking \ac{NN}} & Reduced \ac{ANN} energy cost for beamforming design. & Maximize sum-rate & \parbox[c]{1.2cm}{\centering\vspace*{0.05cm}$-$} \\
			\hline
			\parbox[c]{1.2cm}{\centering\vspace*{0.05cm}\cite{Sun2020}}  & \parbox[c]{1.2cm}{\centering\vspace*{0.05cm}\ac{DRL}} & The balancing coefficients are used to predict the beamforming vector. & Maximize sum-rate & \parbox[c]{1.2cm}{\centering\vspace*{0.05cm}sub 6 GHz} \\ 
			\hline
			\parbox[c]{1.2cm}{\centering\vspace*{0.05cm}\cite{Ahmad2022}}  & \parbox[c]{1.2cm}{\centering\vspace*{0.05cm}\ac{DQN}} & Introduced prioritized experience replay in \ac{DQN} to enhance beamforing performance for \ac{UAV} networks. & Minimize \ac{MSE} & \parbox[c]{1.2cm}{\centering\vspace*{0.05cm}$-$ } \\
			\hline
			\parbox[c]{1.2cm}{\centering\vspace*{0.05cm}\cite{Diamantaras2019}}  &\parbox[c]{1.2cm}{\centering\vspace*{0.05cm} \ac{RL} }& Cooperative $\epsilon$-greedy \ac{RL} mobile relay positioning. &Maximize average \ac{SINR} subject to power constraints & \parbox[c]{1.2cm}{\centering\vspace*{0.05cm}$-$} \\
			\hline
			\parbox[c]{1.2cm}{\centering\vspace*{0.05cm}\cite{Kim2023a}}  & \parbox[c]{1.2cm}{\centering\vspace*{0.05cm}Federated  learning} & Aplies majorization minimization method to perform jointly beamforming vectors and learning rate optimization. & Minimize \ac{MSE} & \parbox[c]{1.2cm}{\centering\vspace*{0.05cm}$-$} \\ 
			\hline
			\parbox[c]{1.2cm}{\centering\vspace*{0.05cm}\cite{Li2023b} }& \parbox[c]{1.2cm}{\centering\vspace*{0.05cm}\ac{DNN}} & Applies double \ac{DNN} to simplify beamforming optmization task. & Maximize channel capacity & \parbox[c]{1.2cm}{\centering\vspace*{0.05cm}$-$} \\ 
			\hline
			\parbox[c]{1.2cm}{\centering\vspace*{0.05cm}\cite{Alkhateeb2018} } &\parbox[c]{1.2cm}{\centering\vspace*{0.05cm} \ac{DNN}} & Use of uplink training pilots with omni or quasi-omni beam patter to predict the best radio frequency beamforming vector. &Maximize system effective achievable rate &\parbox[c]{1.2cm}{\centering\vspace*{0.05cm} \ac{mmWave}} \\
			\hline
			\parbox[c]{1.2cm}{\centering\vspace*{0.05cm}\cite{Kao2018}}  & \parbox[c]{1.2cm}{\centering\vspace*{0.05cm}\ac{LLC}} & Two-stage process: offline \ac{LLC} and online Rosembrock search. & Maximize spectral efficiency & \parbox[c]{1.2cm}{\centering\vspace*{0.05cm}\ac{mmWave}} \\
			\hline
	\end{tabularx}}
	\label{tab:beamforming_digital}
\end{table*}

Despite the benefits of digital processing, the \ac{DBF} requires a separate radio frequency chain
per antenna element leading to high power consumption and scalability issues.
This is even more critical at \ac{mmWave} frequencies, where a large antenna array is used to reduce path loss effect. The power consumption of the analog-to-digital converter is proportional
to the signal bandwidth \cite{Roh2014,Abbas2017} which makes \ac{DBF} schemes non practical in massive \ac{MIMO} scenarios.

Analog beamforming techniques offer a more efficient solution in terms of power consumption and scalability. However, they may be 
inferior to \ac{DBF} techniques in terms of the achieved \ac{SINR} and overall spectral efficiency.
Consequently, \ac{HBF} techniques have gained attention, because they reduce the number of radio frequency chains while keeping a part of the digital structure to signal processing and offering similar performance as \ac{DBF} techniques \cite{Tao2019,Long2018,Lavdas2022,Aljumaily2019,Elbir2019,Zhang2022a,Liu2022,Lizarraga2019}.

Recognizing their performance, feasibility and cost advantages, many recent works have devised \ac{HBF} with \ac{ML} solutions. In 
\cite{Tao2019}, the authors proposed a \ac{HBF} scheme for multi-user \ac{mmWave} massive \ac{MIMO} systems, based on \ac{DNN} 
which minimizes the sum \ac{MSE} subject to power constraints. By assuming perfect \ac{CSI} at both the transmitter and receiver, the 
authors have integrated the received \ac{CSI} into a hidden layer of the \ac{ANN} (see Section \ref{subsub:ANN} for \ac{ANN} details) to 
calculate the best digital and analog beamformer/combining design. They showed via simulation results that the proposed scheme 
outperforms existing methods in terms of bit error rate.

The work in \cite{Long2018} solved the beam selection problem by a data-driven method that uses the angle and direction of arrival as feature vectors entries. The authors modeled the beam selection task as a multi-class problem using \ac{SVM} (see Section \ref{subsub:svm} for \ac{SVM} details) to maximize the sum rate. To achieve a better performance of \ac{SVM}, the authors designed a method to obtain the optimal parameter of the Gaussian kernel function using McLaughlin classification method. The proposed method achieved near-optimal sum-rate performance while the complexity was reduced via \ac{SVM} approach. The authors in \cite{Lavdas2022} evaluated the \ac{k-NN} (see Section \ref{subsub:k-nn}) performance for adaptive beamforming in \ac{5G} \ac{mmWave}. The proposed framework used the spatial distribution of throughput demand to efficiently generate beamforming configurations to achieve spectral and energy efficiency.

The authors in \cite{Aljumaily2019} assumed a single user \ac{MIMO} system.
The problem is solved in two steps to achieve a spectral efficiency as close as possible to \ac{DBF} approach. In the first step, the digital and analog parts are jointly optimized  in terms of sum-rate via convex optimization. In the second step, a radial basis function network is implemented by a \ac{DNN} for approximation and to improve the designed beamforming to reach a performance close to that achieved by full \ac{DBF}.

In \cite{Elbir2019} the authors proposed a quantized \ac{CNN} to reduce the complexity of the downlink \ac{HBF}. The channel matrix is used as input to the \ac{CNN} resulting in the analog precoders and combiners as outputs that maximize the spectral efficiency. The proposed approach is robust since the input (training) data are contaminated with noise. 

Supervised learning leads to label overhead when training the \ac{ML} algorithm. To overcome this issue, an unsupervised \ac{DNN} beamforming framework can be used. In \cite{Zhang2022a}, the data used for training is obtained from the estimated channel implying the proposed model to learn under imperfect \ac{CSI}, achieving a suboptimal solution. On the other hand, reference \cite{Liu2022} used the loss function to jointly train the antenna selection and hybrid beamforming networks. 

Reference \cite{Lizarraga2019} used \ac{RL} to find the analog beamformer part that maximizes the sum data rate. The digital part is found by using the singular value decomposition of
the equivalent low-dimension channel determined by the analog beamformer. This \ac{HBF} solution had a small performance gap when compared to brute force search. Hybrid  beamforming works are summarized in Table \ref{tab:beamforming_hybrid}.

\begin{table*}[!t]
	\centering
	\caption{Summary of hybrid beamforming works with \ac{ML}.}
	{\scriptsize \begin{tabularx}{\linewidth}{
				|>{\hsize=0.7\hsize}X|
				>{\hsize=0.7\hsize}X|
				>{\hsize=2\hsize}X|
				>{\hsize=1.0\hsize}X|
				>{\hsize=0.6\hsize}X|
			}
			\hline

			\parbox[c]{1.2cm}{\centering\vspace*{0.05cm}Work}	 & \parbox[c]{1.5cm}{\centering\vspace*{0.05cm}\ac{ML} Technique} & \centering Solution Approach & \centering Optimization Task & Frequency \\
            \hline
			\parbox[c]{1.2cm}{\centering\vspace*{0.05cm}\cite{Tao2019}} & \parbox[c]{1.2cm}{\centering\vspace*{0.05cm}\ac{DNN}} & Integrate the \ac{CSI} matrix into a hidden layer of the \ac{ANN} & Minimize \ac{MSE} subject to power constraints& \ac{mmWave} \\
			\hline
			\parbox[c]{1.2cm}{\centering\vspace*{0.05cm}\cite{Long2018}}  & \parbox[c]{1.2cm}{\centering\vspace*{0.05cm}\ac{SVM}} & Solve beam selection problem by data-drive method & Maximize sum-rate & \ac{mmWave} \\
			\hline
			\parbox[c]{1.2cm}{\centering\vspace*{0.05cm}\cite{Lavdas2022}}  & \parbox[c]{1.2cm}{\centering\vspace*{0.05cm} \ac{k-NN}} & Application of  \ac{k-NN} for beamforming configuration selection according to the spatial distribution of throughput demand. & Maximize energy efficiency and spectral efficiency & \ac{mmWave} \\
			\hline
			\parbox[c]{1.2cm}{\centering\vspace*{0.05cm}\cite{Aljumaily2019}}  & \parbox[c]{1.2cm}{\centering\vspace*{0.05cm}\ac{DNN}} & Two-step optimazation: a convex optimizer and \ac{ML} technique & 
			Maximize rate subject to power constraint& \ac{mmWave} \\
			\hline 
			\parbox[c]{1.2cm}{\centering\vspace*{0.05cm}\cite{Elbir2019} } &\parbox[c]{1.2cm}{\centering\vspace*{0.05cm} \ac{CNN}} & Channel matrix with noise contamination as \ac{CNN} input & Maximize rate& \ac{mmWave} \\
			\hline 
			\parbox[c]{1.2cm}{\centering\vspace*{0.05cm}\cite{Zhang2022a} } &\parbox[c]{1.2cm}{\centering\vspace*{0.05cm} \ac{DNN} }& Unsupervised learning approach by using the estimated channel data as training sequence. & Maximize sum rate& \ac{mmWave} \\
			\hline
			\parbox[c]{1.2cm}{\centering\vspace*{0.05cm}\cite{Liu2022}}  & \parbox[c]{1.2cm}{\centering\vspace*{0.05cm}\ac{DNN}} & Unsupervised learning approach by using the loss function data as training sequence to jointly optimize antenna selection and hybrid beamforming. & Maximize rate& \ac{mmWave} \\
			\hline
			\parbox[c]{1.2cm}{\centering\vspace*{0.05cm}\cite{Lizarraga2019}} & \parbox[c]{1.2cm}{\centering\vspace*{0.05cm}\ac{RL}} & Codebook-based analog beamformer and best candidate selected via \ac{RL} algorithm & Maximize average data rate& \ac{mmWave} \\
			\hline
	\end{tabularx}}
	\label{tab:beamforming_hybrid}
\end{table*}

\section{Security and Privacy}

Due to the broadcast nature of wireless channel, wireless systems and networks are vulnerable to various types of attacks. When 
different users and/or \acp{MNO} share the same frequency resource, there is a major concern about the security of information 
exchanged. Security requirements for spectrum sharing networks include \textit{confidentiality}, \textit{integrity}, \textit{availability}, 
\textit{authentication}, \textit{nonrepudiation}, \textit{compliance}, \textit{access control} and \textit{privacy} \cite{Baldini2012,Park2014}.

Spectrum sharing systems are susceptible to various security and privacy concerns, necessitating adaptive, dependable, and scalable 
protection measures. Machine learning techniques are often suggested as solutions to tackle these challenges 
\cite{Wang2022,Ruzomberka2023}. The attackers aim to exploit the sensing or geolocation database \cite{Park2014,Li2019a}, 
including eavesdropping,, 
intentional interference jamming,
spoofing 
and intrusion.

\subsection{Eavesdropping}

An attacker can monitor the frequency channel and try to retrieve confidential users information. The impact of an eavesdropping presence in a spectrum sharing network was investigated in the recent literature. The  secrecy performance of a \ac{5G} network was analyzed in \cite{Soltani2019} while the uplink pilot contamination via eavesdropping in underlay spectrum sharing system was investigated by \cite{Timilsina2018}.

Spectrum sharing strategies need to sacrifice utility to protect shared information from eavesdropper. This privacy-performance tradeoff was analyzed in \cite{Hilli2019} where a precoding matrix for a \ac{MIMO} radar network was designed. The proposed scheme decreases the chance of an eavesdropper to find the radar location at the cost of increasing the interference power at the radar. Reference \cite{Dong2018} designed a scheme to protect \ac{PU} privacy. The \ac{SU} was chosen in order to satisfy \ac{PU} privacy police in detriment to maximize the network performance.

Other works analyzed the  performance optimization of a \ac{CRN} in a presence of an eavesdropping. The total secrecy rate 
maximization of a two-cell \ac{MIMO} \ac{NOMA}-based network was done by \cite{Nandan2018} and    
a \ac{SU} throughput maximization scheme under system constraints was proposed by \cite{Banerjee2019}.

\ac{ML} algorithms were used in recent works at eavesdropping situations in wireless networks. \ac{SVM} (see Section \ref{subsub:svm}) and K-means (see Section \ref{subsub:K_means})  \cite{Hoang2020}, and minimum risk Bayesian classification (see Section \ref{subsub:bayes_class}) \cite{Liu2019} were used to detect possible threats through predictive models. These algorithms focus on identifying whenever an active eavesdropper is attacking the network by 
detecting an abnormal channel behavior. \ac{DNN} (see Section \ref{subsub:ANN}) was used in \cite{Bao2020} to predict the secrecy outage probability from a ground \ac{BS} to an \ac{UAV} user. \ac{MARL} is used in \cite{Kazemi2024} to cooperatively decide a friendly jammer selection to protect data from eavesdroppers.

As discussed in Section \ref{Sec:Allocation}, \ac{DQN} is a powerful tool for resource allocation. Since eavesdropping model is uncertain and dynamic, a channel allocation policy was proven to be effective against eavesdroppers. Reference \cite{Wang2023a} performed a resource allocation scheme based on double \ac{DQN} an actor-critic algorithm. In \cite{Yao2023} the authors designed an algorithm to track attacker nodes and, then, applied the \ac{DQN} strategy to efficiently allocate channel to authorized users in order to avoid eavesdropping. The work also considered a power control scheme to deal with intentional jamming.

These works are summarized in Table \ref{tab:security_eaves}.

\begin{table*}[!t]
	\centering
	\caption{Overview of eavesdropping \ac{ML} works for security in spectrum sharing scenarios.}
{\small     
	\begin{tabularx}{\linewidth}{
			|>{\hsize=0.8\hsize}X|
			>{\hsize=1.4\hsize}X|
			>{\hsize=0.8\hsize}X|
		}
		\hline %
		\parbox[c]{2.5cm}{\centering\vspace*{0.05cm}\textbf{ML Algorithm} } & \parbox[c]{5cm}{\centering\vspace*{0.05cm}\textbf{Security Strategy}} & \parbox[c]{2.5cm}{\centering\vspace*{0.05cm}\textbf{Work}} \\ \hline
		\parbox[c]{2.5cm}{\centering\vspace*{0.05cm}\ac{SVM} and K-means}         &  Identifies abnormal network behavior.    &  
		\parbox[c]{2.5cm}{\centering\vspace*{0.05cm} \cite{Hoang2020} }            \\
		\hline
		\parbox[c]{2.5cm}{\centering\vspace*{0.05cm}Minimum risk Bayesian  }      & Identifies abnormal network behavior.     &     \parbox[c]{2.5cm}{\centering\vspace*{0.05cm}  \cite{Liu2019} }           \\
		\hline
		\parbox[c]{2.5cm}{\centering\vspace*{0.05cm}	\ac{DRL} }      & Beamforming transmission     &    \parbox[c]{2.5cm}{\centering\vspace*{0.05cm} \cite{Bao2020}   }         \\ 
		\hline
		\parbox[c]{2.5cm}{\centering\vspace*{0.05cm}	\ac{MARL}   }    & Intentional jammers to prevent the leakage of information.     &       \parbox[c]{2.5cm}{\centering\vspace*{0.05cm} \cite{Kazemi2024}    }        \\
		\hline
		\parbox[c]{2.5cm}{\centering\vspace*{0.05cm}Double \ac{DQN} }      & Intelligent resource allocation to avoid eavesdroppers.      &    \parbox[c]{2.5cm}{\centering\vspace*{0.05cm}    \cite{Wang2023a}    }        \\ 
		\hline
		\parbox[c]{2.5cm}{\centering\vspace*{0.05cm}\ac{DQN}  }      & Channel allocation policy.      & \parbox[c]{2.5cm}{\centering\vspace*{0.05cm} \cite{Yao2023}}            \\
		\hline
	\end{tabularx} }
	\label{tab:security_eaves}
\end{table*}

\subsection{Jamming}

Users with tight constraints and sensitive requirements are severe harmed by intentional interference jams that corrupt the shared information. When a jamming attack occurs, the channel coefficients suffer from abruptly variance degrading the system performance.

\ac{MDP} combining with \ac{RL} techniques are widely used to model, analyze and  develop different anti-jamming defense strategies. The most used approach is to estimate the jamming pattern via Q-learning (see Section \ref{subsub:sarsa_q_learning}) \cite{Pei2019,Yao2019a,Wang2020} or \ac{DRL} (see Section \ref{subsub:approx_techniques}) \cite{Chen2018,Liu2019a,Han2017a,Xiao2018a,Li2023c,Sharma2023}. Another strategy is to consider frequency and temporal information, known as spectrum waterfall, instead of estimate directly the jamming pattern and parameters using a \ac{DRL} framework \cite{Liu2018}. A\ac{MARL} approach was also explored in literature for solving jamming issues \cite{Yin2022}.

Anti-jamming solutions rely on developing strategies for one or more of following domains: code, power, frequency and space. Reference \cite{Pei2019} proposed a multi-domain  solution where the developed Q-learning framework obtained optimal power and channel selection strategy. To achieve the best anti-jamming strategy, however, it is necessary coordination among the network nodes. The works in \cite{Yao2019a,Wang2020} proposed  collaborative Q-learning methods to obtain the optimal anti-jamming strategy. The algorithms could cope with both malicious jamming and mutual interference among users showed through simulation results \cite{Yao2019a} and testbed \cite{Wang2020}.

However, if the jamming environment changes dynamically, Q-learning algorithms can not process the big amount of information generated by the \ac{MDP}, then a \ac{DRL} algorithm has to be used. In \cite{Chen2018}, a power control anti-jamming strategy was formulated via \ac{DQN} to cope with a large number of \ac{SINR} quantization levels and jamming levels. The proposed method outperformed the benchmark Q-learning improving the transmission and power efficiency. Other works proposed  a channel selection scheme only \cite{Liu2019a} or combined with spatial domain strategy  \cite{Han2017a,Xiao2018a} to avoid intentional jamming. In \cite{Liu2019a}, a \ac{DQN} recognizes the jamming pattern and chooses the best channel available taking into account the switching cost. Reference \cite{Han2017a} used a \ac{DQN} to determine whether a user switch the channel frequency or leave an area with high intentional interference jamming. This work was extended in \cite{Xiao2018a} where the authors designed a faster \ac{DQN} framework based on macro-action technique to observe a large range of feasible \ac{SINR}  levels.

The work in \cite{Li2023c} reduces the \ac{DQN} training time by introducing labels instead of reward in \ac{RL}. In \cite{Sharma2023}, the authors proposed a beamforming and power control strategy applying federated learning with \ac{DQN} to address jamming issue and maximize the achievable rate.

A \ac{MARL} Q-learning anti-jamming strategy was proposed by \cite{Yin2022}. The main goal is to jointly optimizing the channel and power allocation of \ac{UAV} users by reducing the action dimensionality of each agent for a faster convergence of the proposed algorithm. 

These works are summarized in Table \ref{tab:security_jam}.

\begin{table*}[!t]
	\centering
	\caption{Overview of jamming \ac{ML} works for security in spectrum sharing scenarios.}
{\small 	\begin{tabularx}{\linewidth}{
			|>{\hsize=0.8\hsize}X|
			>{\hsize=1.4\hsize}X|
			>{\hsize=0.8\hsize}X|
		}
		\hline %
		\parbox[c]{2.5cm}{\centering\vspace*{0.05cm}\textbf{ML Algorithm} } & \parbox[c]{5cm}{\centering\vspace*{0.05cm}\textbf{Security Strategy}} & \parbox[c]{2.5cm}{\centering\vspace*{0.05cm}\textbf{Work}} \\
		\hline 
		\multirow{2}{*} {\parbox[c]{2.5cm}{\centering\vspace*{0.05cm}Q-Learning}}        & Power control and channel selection scheme.   &\parbox[c]{2.5cm}{\centering\vspace*{0.05cm} \cite{Pei2019}   }               \\ \cline{2-3}
		&  Chooses the best anti-jamming policy.    &   \parbox[c]{2.5cm}{\centering\vspace*{0.05cm}       \cite{Yao2019a,Wang2020}  }   \\ \hline
		\multirow{4}{*} {\parbox[c]{2.5cm}{\centering\vspace*{0.05cm}\ac{DQN}}}        &  Power control scheme.      &    \parbox[c]{2.5cm}{\centering\vspace*{0.05cm}\cite{Chen2018,Yao2023}  }          \\ \cline{2-3}
		& Channel selection scheme.    &      \parbox[c]{2.5cm}{\centering\vspace*{0.05cm}    \cite{Liu2019a,Li2023c}    } \\ \cline{2-3}
		&  Channel selection and spatial avoidance.   &  \parbox[c]{2.5cm}{\centering\vspace*{0.05cm}       \cite{Han2017a,Xiao2018a}   }  \\   \cline{2-3}
		&  Chooses the best anti-jamming policy.    &    \parbox[c]{2.5cm}{\centering\vspace*{0.05cm}      \cite{Liu2018} }\\   \cline{2-3}
		&  Beamforming and power allocation.   &   \parbox[c]{2.5cm}{\centering\vspace*{0.05cm}      \cite{Sharma2023} } \\ 
		\hline  
		\parbox[c]{2.5cm}{\centering\vspace*{0.05cm}\ac{MARL}    }   &  Channel selection and power allocation.   &    \parbox[c]{2.5cm}{\centering\vspace*{0.05cm}      \cite{Yin2022} }    \\ \hline
	\end{tabularx}}
	\label{tab:security_jam}
\end{table*}

\subsection{Spoofing}

The spoofing threat is an identity-based attack where an unauthorized node transmits signals with higher power than legal users, forcing the system to interpret the attacker as an authorized user. 
Consequently, the spoofer is authenticated in the system where it can lunch several attacks such denial-of-service, false packets 
injection, false request messages or sensing data falsification attack.%

Authentication approaches using \ac{ML} frameworks can be used do identify authorized users and recognize spoofing attacks. Reference \cite{Fang2019} discussed how the different categories of \ac{ML} algorithms is used to provide intelligent authentication. The choice of which \ac{ML} algorithm fits better to avoid spoofing depends on the system scenario and the available information. Supervised learning algorithms, such \ac{k-NN} (see Section \ref{subsub:k-nn}) \cite{Pinto2018}  or logistic regression (see Section \ref{subsub:lin_log_regre}) \cite{Wang2018a}  are suitable when the labeled input/output pairs for legitimate users are available or  easy to obtain. When the input/output labeled data are not obtainable, unsupervised algorithms such K-means (see Section \ref{subsub:K_means}) \cite{Pinto2018} can be used for authentication if the legit user has more information available than a spoofer. This 
is a suitable assumption if the number of users are much greater than spoofers. 

Another spoofing threat is related to sensing link disruption and spectrum sensing data falsification attack. At this kind of 
attack, an intruder may try to mimic \ac{PU}'s signal pattern to deceive \acp{SU} \cite{YangLi2016} and/or  intentionally send 
modified sensing reports to the fusion center degrading the \ac{CR} performance \cite{Zhu2018, Luo2022a}. These attacks can 
be identified by \ac{ML} classification algorithms. In \cite{YangLi2016} an unsupervised learning K-means was used to exclude the 
false sensing reports from the trusted  ones. Reference \cite{Zhu2018} focused on identifying malicious \acp{SU} by applying 
supervised learning \ac{SVM} (see Section \ref{subsub:svm}) based on energy values for each trusted \ac{SU}. Authors in 
\cite{Luo2022a} proposed a learning-evaluation-beating framework where a modified \ac{SVM} to recognize the real sensing data 
from the received data at fusion center contaminated with adversarial perturbation.

These works are summarized in Table \ref{tab:security_spoffing}.

\begin{table*}[!t]
	\centering
	\caption{Overview of spoofing \ac{ML} works for security in spectrum sharing scenarios.}
	{\small \begin{tabularx}{\linewidth}{
			|>{\hsize=0.8\hsize}X|
			>{\hsize=1.4\hsize}X|
			>{\hsize=0.8\hsize}X|
		}
		\hline %
		\parbox[c]{2.5cm}{\centering\vspace*{0.05cm}\textbf{ML Algorithm} } & \parbox[c]{5cm}{\centering\vspace*{0.05cm}\textbf{Security Strategy}} & \parbox[c]{2.5cm}{\centering\vspace*{0.05cm}\textbf{Work}} \\
		\hline 
		\parbox[c]{2.5cm}{\centering\vspace*{0.05cm}	K-means and \ac{k-NN}   }      &  Fast variation identification on \ac{RSS} instant samples.    &       \parbox[c]{2.5cm}{\centering\vspace*{0.05cm}\cite{Pinto2018}   }      \\ \hline 
		\parbox[c]{2.5cm}{\centering\vspace*{0.05cm}Logistic regression}           & Anomalies detection on virtual angle of arrival.    &  \parbox[c]{2.5cm}{\centering\vspace*{0.05cm}  \cite{Wang2018a}   }         \\ \hline 
		\parbox[c]{2.5cm}{\centering\vspace*{0.05cm}  K-means }          &   Separates false sensing reports from trusted ones.   &  \parbox[c]{2.5cm}{\centering\vspace*{0.05cm}    \cite{YangLi2016}      }    \\
		\hline 
		\parbox[c]{2.5cm}{\centering\vspace*{0.05cm}  \ac{SVM}  }         & Identifies malicious \acp{SU} based on energy values.     &   \parbox[c]{2.5cm}{\centering\vspace*{0.05cm} \cite{Zhu2018} }           \\
		\hline
		\parbox[c]{2.5cm}{\centering\vspace*{0.05cm} \ac{SVM}     }      & Minimize the perturbation at sensing data.   & \parbox[c]{2.5cm}{\centering\vspace*{0.05cm}   \cite{Luo2022a}    }        \\
		\hline 
	\end{tabularx}}
	\label{tab:security_spoffing}
\end{table*}

\subsection{Intrusion}

Access control techniques prevent unauthorized users from accessing network resources. 
An \ac{IDS} is used to identify whenever an intruder is present in the system by analyzing network activity searching for any abnormal channel behavior \cite{Otoum2019,Verma2019,Otoum2019a} or identifying an authorized user based on some metric \cite{Bassey2019}.

A straightforward strategy is to make usage of a dataset\footnote{KDDCup99, CIDDS-001, UNSW-NB15, NSL-KDD, and others.} that contain intruder attacks signatures which can be use jointly with \ac{ML} algorithms such as learning classifiers or deep learning. The usage of learning classifiers was studied in \cite{Verma2019} where the authors used popular datasets as benchmark to analyze the prominent metrics and validation methods for those classifiers. In \cite{Otoum2019} the dataset was used to train a deep learning algorithm to recognize the possible attackers signatures in the sensor network.

If the intruder activity is not available for the training process, a previous legit dataset can be used to train the \ac{IDS} to learn what is the expected channel behavior as used by \cite{Bassey2019}. The authors observed that each individual device has unique radio frequency fingerprints and used this fact to train the \ac{CNN} (see Section \ref{subsub:CNN}) to identify legal activities. Other suitable strategy is to use Q-learning to capture network behavior and give the \ac{IDS} system a feedback whether an intruder is identified or not \cite{Otoum2019a}.

These works are summarized in Table \ref{tab:security_intrusion}.

\begin{table*}[!t]
	\centering
	\caption{Overview of intrusion \ac{ML} works for security in spectrum sharing scenarios.}
	{\small \begin{tabularx}{\linewidth}{
			|>{\hsize=0.8\hsize}X|
			>{\hsize=1.4\hsize}X|
			>{\hsize=0.8\hsize}X|
		}
		\hline %
		\parbox[c]{2.5cm}{\centering\vspace*{0.05cm}\textbf{ML Algorithm} } & \parbox[c]{5cm}{\centering\vspace*{0.05cm}\textbf{Security Strategy}} & \parbox[c]{2.5cm}{\centering\vspace*{0.05cm}\textbf{Work}} \\
		\hline 
	
		 \parbox[c]{2.5cm}{\centering\vspace*{0.05cm}\ac{ML} classifiers}          & Anomalies detection on network activities.    &   \parbox[c]{2.5cm}{\centering\vspace*{0.05cm} \cite{Verma2019}  }        \\ \hline
	     \parbox[c]{2.5cm}{\centering\vspace*{0.05cm}\ac{DRL} }       &   Anomalies detection on network activities.    &  \parbox[c]{2.5cm}{\centering\vspace*{0.05cm}    \cite{Otoum2019}  }         \\ \hline
	     \parbox[c]{2.5cm}{\centering\vspace*{0.05cm} \ac{CNN}  }      &  Recognizes unique users fingerprints.    &  \parbox[c]{2.5cm}{\centering\vspace*{0.05cm}    \cite{Bassey2019}  }       \\ \hline
	     \parbox[c]{2.5cm}{\centering\vspace*{0.05cm}Q-learning }     &  Anomalies detection on network activities.    &  \parbox[c]{2.5cm}{\centering\vspace*{0.05cm}    \cite{Otoum2019a} }        \\
		\hline
	\end{tabularx}}
	\label{tab:security_intrusion}
\end{table*}

\section{Summary}\label{sub:Chap_FurtherAspects:summary}
In this chapter, we addressed the spectrum handoff mechanism,  beamforming and security aspects for spectrum sharing.

We introduced the handoff task, we presented and summarized recent literature works that used \ac{ML} approach to solve this problem. We surveyed recent literature works that applied digital and hybrid beamforming approaches to address spectrum sharing problem. We also covered security and privacy aspects for spectrum sharing by characterizing eavesdropping, jamming, spoofing and intrusion problems and surveyed recent works that solve these tasks using \ac{ML} algorithms.  

\chapter{Challenges and Future Research}\label{Sec:FutureResearch}
\ac{DSS} brings several advantages to networks and users by optimizing radio resources. As mentioned in Section \ref{Sec:Intro}, the efficient allocation and usage of the shared spectrum increases the network performance by allowing the coexistence of different technologies in the same frequency band.

However, due to heterogeneity of allowing a mix of different wireless systems coexisting in multiple domains, spectrum 
management becomes a complex task. With massive \ac{MIMO}, \ac{mmWave} and new upcoming services, cellular networks 
become denser in space (relays, picocells, femtocells) in addition of an existing macro/microcells structure 
\citep{Ge2016}. Different types of devices will be connected in the \ac{BS} (sensors, smartphones, vehicles), having 
different specifications (e.g., data rate and sensibility). Hence, the spectrum management becomes complicated due to 
the high interference scenario with different requirements.

To attend \ac{5G} demands, \ac{IMT}-2020 expects some key capabilities, also called \acp{KPI}, such as the user 
experienced and peak data rate, area traffic capacity, network energy and spectrum efficiency, connection density, 
latency, and mobility \citep{ITU2015}. These \acp{KPI} allow the emergence of new applications such as \ac{VR}, \ac{AR}, 
\acp{UAV}, smart cities, which have different characteristics and technical requirements. To handle this 
diversity of application and services, three categories were proposed by \ac{ITU}: \ac{MTC}, \ac{eMBB} and \ac{URLLC} 
\citep{Popovski2014,Popovski2018,Abreu2019}.

The support for these services has been investigated by \ac{3GPP} on Release 15 \citep{3GPP2018_URLLC} and the minimum 
technical performance requirements are described in \citep{ITU2017}. Spectrum sharing techniques have to take into 
account which service will be implemented in order to meet their different requirements. For instance, the minimum user 
plan latency is 4 ms for \ac{eMBB} while it is 1ms for \ac{URLLC}. The testing environment is also fundamental to 
establish minimum \acp{KPI} for each service. For example, the downlink average spectral efficiency for an indoor 
hotspot \ac{eMBB} is 9 bits/s/Hz per transmission reception point while in a rural area this value is 3.3 bits/s/Hz per 
transmission reception points \citep{ITU2017}.

Also, we discuss in this section some of the points we believe will drive the research in the upcoming years, regarding 
\ac{ML} and spectrum sharing problems:

\begin{itemize}
  \item \textbf{6G THz frequency bands:} the inclusion of higher frequency bands (THz and sub-THz) will be necessary 
  for \ac{6G} so it can accommodate the requirements in terms of data rates and number of connected users. In addition 
  to the higher complexity of handling the physical effects in such bands (e.g., propagation, fading), the spectrum 
  management considering high frequencies is more challenging due to the need of sensing a much bigger frequency 
  space in order to identify potential white spaces. The need to accomplish this task in a reduced time window also 
  brings another layer of problems to be analyzed since some \ac{ML} methods rely on extensive training models 
  that are time demanding.
  \item \textbf{Dimension reduction for beamforming:} the increasing number of antennas provide by massive \ac{MIMO} arrays and use of \ac{mmWave} (inclusing THz bands) will make the dimensionality of the beamforming problem even higher and even more challenging, regarding the dimensions one will need to handle regarding typical beamforming strategies (e.g. matrix inversions). One expects that \ac{ML} can be part of a solution in this problem, which will affect the capability of the system to use proper radio resources, including spectrum bands. Having in mind that latency will also be a \ac{KPI}, typical \ac{ML} which use long training sequences will not be a straightforward solution and dimensionality reduction will be not only desired but instrumental for some problems at hand. Data information and some structure about them can be required as a way to address the same performance handling much less information.
  \item \textbf{Coexistence of communication and radar systems, and integrated communication and radar systems:} while 
  cellular networks have traditionally been designed to provide connectivity for mobile broadband and \ac{IoT} 
  applications, there is a growing demand for high-precision positioning, localization, and mapping services. 
  Several use cases in the automotive, rail and the emerging urban air mobility transport sector rely on positioning and 
  simultaneous localization and mapping services that require higher accuracy than that provided by current Global 
  Navigation Satellite Systems \citep{Su2022,PeralRosado2020}. 
  As such, we believe that spectrum allocation and management in evolving 6G cellular networks, as well as in radar systems mounted on 
  infrastructure, vehicles, and  uncrewed aerial vehicles, will benefit from \ac{ML}-based dynamic spectrum sharing solutions, 
  as highlighted by the \ac{ITU} in \ac{IMT}-2030~\citep{ITU-R2023}. 
  For example, cellular networks could help mitigate radar-to-radar interference in dense urban environments where  vehicles are 
  equipped with radar systems. 
  Moreover, we expect that integrated communication and sensing systems will be deployed for services requiring the 
  localization and tracking of both passive (unconnected) objects and those connected to the cellular network. 
  One example is the collective perception service, developed by the European Telecommunications Standards 
  Institute (ETSI), which detects vulnerable road users who do not have smartphones and alerts surrounding 
  vehicles~\citep{Caillot2022}. 
  These services will be offered by perceptive cellular networks~\citep{Caillot2022,Jung2022,Pacella2021}, which will coexist with 
  traditional cellular networks and radar systems. 
  We believe that managing spectrum in these complex environments will significantly benefit from the application of \ac{ML} techniques.
  \item \textbf{Dynamic time division duplexing, subband full-duplex and full-duplex systems:} recent advances in large 
  scale antenna systems and self-interference suppression techniques facilitate the deployment of dynamic time division 
  duplexing and, more recently, cross-division duplex systems \citep{Silva2021,Ji2021}. The goal of these systems is to 
  dynamically allocate spectrum resources to spatially and temporally varying uplink and downlink traffic demands in 
  large scale mobile systems. However, these systems may suffer from severe \ac{BS}-to-\ac{BS} (so called cross-link) 
  interference as well as inter-operator interference due to adjacent channel leakage and intermodulation effects 
  \citep{Kim2020}. Also, full-duplex systems have been demonstrated in small scale experiments to improve spectrum 
  utilization \citep{Korpi2016}. We believe that dynamic spectrum management using \ac{ML} techniques will be 
  invaluable for the successive deployment of these systems. 
  \item \textbf{Integrated terrestrial and non-terrestrial networks for integrated terrestrial, airspace and maritime 
  services:} due to the growing interest in advanced air mobility, urban air mobility and maritime connectivity 
  services \citep{Han2021,Saafi2022}, and providing mobile broadband services in commercial aircrafts, integrated 
  terrestrial and non-terrestrial networks attract the interest of the research and standardization communities 
  \citep{Lei2021}. Spectrum management in these systems is highly non-trivial, since depending on the traffic demands, 
  service requirements, mobility pattern, altitude of the flying vehicles, and operational costs, different 
  combinations of terrestrial and non-terrestrial network solutions may be desirable. Therefore, integrated terrestrial 
  and non-terrestrial networks will likely benefit from dynamic spectrum management assisted by \ac{ML} methods. 
  \item \textbf{Real-time \ac{ML}:} some envisaged applications for \ac{6G} will require a strong component of 
  artificial intelligence to achieve promised performance, such as tactile internet, \ac{VR} and \ac{AR} 
  \citep{Lu:2020}. One potential problem for those use cases is related to latency required by \ac{ML} methods, 
  especially regarding training of them. Therefore, some degree of ``real-time'' execution is expected to be available, 
  so aspects of optimization and feature selection, for example, which are important steps for spectrum sharing actions 
  would be ideally performed when the data is available ``on the fly'' and not only performed for a given dataset and 
  then extrapolated for similar (statistically speaking) data.
  \item \textbf{Coexistence with \ac{ML} services:} the increase of \ac{IoT} devices that generated large 
  quantities of data promotes the use of distributed \ac{ML} methods. 
  Specifically, the \ac{IoT} devices exchange model parameters, such as the weights of an \ac{ANN}, to train 
  collectively a \ac{ML} model for prediction and monitoring applications. 
  The use of such distributed \ac{ML} methods increase the privacy when training due to sharing only model parameters 
  instead of the raw data each device generates. 
  The exchange of model parameters happens over the wireless network and, depending on the application, it has to 
  fulfil requirements in terms of communication bandwidth, transmitted power, latency, and reliability. 
  For example, autonomous driving generates large quantities of data that can be used to train collectively a model to 
  improve the cooperative perception of vehicles.
  Such vehicles need to exchange the model parameters near real-time and with high reliability to guarantee the 
  convergence of the \ac{ML} model being trained~\citep{Balkus2022}.
  The training of distributed \ac{ML} models through the wireless network has been termed ``wireless for 
  \ac{ML}''~\citep{Hellstrom2022}, and the demand for \ac{ML} services is projected to grow significantly.
  Specifically, discussions have already begun on a dedicated network slice for \ac{ML} in future-generation cellular 
  networks such as beyond-5G and 6G~\citep{SaadAndBennis2019,ZhangAndXiao2019}. 
  Therefore, the investigation of spectrum sharing in wireless for \ac{ML} becomes relevant not only for 
  local-area networks but also for large-scale cellular networks.  
\end{itemize}


%
\chapter{Conclusions}\label{Sec:Conclusion}
In this survey, we have addressed the \ac{ML} techniques for the problems of spectrum sharing aiming the future 
wireless systems. We have revisited the fundamentals of \ac{ML} techniques which form the basis of the state-of-art of 
tools for spectrum sharing problems, providing some view of their potentials, characteristics and capabilities as well 
as the rationale of  the mathematical modelling of each class of the techniques. The problems of spectrum sharing were 
then split into different categories to ease the identification of the most usual techniques employed for each case and 
when they overlap. Finally, we have also discussed about the main issues regarding spectrum challenges the future 
wireless systems will have to tackle and provided several open research challenges along with their causes and 
guidelines.

As \ac{ML} methods are becoming an ubiquitous tool for the design and optimization of wireless communication 
system we foresee a wider interest in such strategies as the number of parameters grow and also the growing of the 
amount of available data about the behaviour of the users in different scenarios and situations. This \emph{big data} 
picture, associated with new frequency bands and other features of the 6G system (e.g. massive MIMO, holographic 
communications) will increase the need of powerful methods to scrap the (huge) amount of available data and provide 
reliable and robust optimization solutions regarding the several challenges of allocating a scarse and shared resource 
to users in order to allow communication in high data rates. Our findings in this survey provide a roadmap of the 
potential directions one can follow in order to fill some still present gaps in both technological and industrial 
perspectives.

\backmatter  

\printbibliography

\end{document}